\pdfoutput=1
\documentclass[10pt]{IEEEtran}

\usepackage{setspace}
\usepackage{cite}
\usepackage{graphicx}
\usepackage[cmex10]{amsmath}
\usepackage{amssymb}
\usepackage{mathtools}
\usepackage{tikz}
\usepackage{color}
\usepackage{url}
\usepackage{epstopdf}
\usepackage{subfigure}
\usepackage{bm}
\usepackage{framed}

\usetikzlibrary{automata,arrows,positioning,calc,shapes.multipart,chains,shapes,decorations}

\usepackage{texdef2020}
        \newtheorem{theorem}{Theorem}
        
        \newtheorem{lemma}{Lemma}
        \newtheorem{definition}{Definition}
        \newtheorem{corollary}{Corollary}

\newcommand{\agevec}{\mathbf{\age}}
\newcommand{\Tavg}[2][\mathcal{T}]{\left\langle{#2}\right\rangle_{#1}}
\newcommand{\Tcal}{\mathcal{T}}
\newcommand{\age}{\Delta}

\newcommand{\Lcal}{\mathcal{L}}
\newcommand{\Qcal}{\mathcal{Q}}
\newcommand{\R}{\mathbb{R}}

\newcommand{\ql}{q_l}
\newcommand{\laml}[1][l]{\lambda^{(#1)}}

\newcommand{\rowvec}[1]{[\begin{matrix} #1\end{matrix}]}
\newcommand{\rvec}[1]{\smrvec{#1}}
\newcommand{\smrvec}[1]{\setlength\arraycolsep{2pt}\rowvec{#1}}
\newcommand{\piv}{\text{\boldmath{$\pi$}}}

\newcommand{\pibar}{\bar{\pi}}
\newcommand{\vvbar}{\bar{\vv}}

\newcommand{\onev}[1][n]{{\mathbf{1}}_{#1}}
\newcommand{\PAoI}{\age^{(p)}}
\newcommand{\err}[2][X]{\varepsilon_{#1}(#2)}

\newcommand{\qed}{\nobreak \ifvmode \relax \else
      \ifdim\lastskip<1.5em \hskip-\lastskip
      \hskip1.5em plus0em minus0.5em \fi \nobreak
      \vrule height0.75em width0.5em depth0.25em\fi}

\makeatletter
\def\BState{\State\hskip-\ALG@thistlm}
\makeatother

\newcommand{\iid}{{i.i.d.}}
\newcommand{\pr}[1]{\mathbb{P}\left[ #1 \right]}
\newcommand{\imply}{\Rightarrow}
\newcommand{\atick}[1]{-- ++(0,-0.3) ++(0,-0.8) node {$#1$}}
\newcommand{\dtick}[1]{\atick{#1}}

\newcommand{\lyapunovDrift}{\Phi}
\newcommand{\AoI}{\Delta}

\newcommand{\schedulingDecision}{v}

\title{Age of Information: An Introduction and Survey}
\author{Roy D.~Yates, Yin Sun, D.~Richard Brown III, Sanjit K.~Kaul, Eytan Modiano and Sennur Ulukus%
\thanks{Roy D.~Yates is with WINLAB and the ECE Department, Rutgers University, NJ, USA, e-mail: ryates@winlab.rutgers.edu.}% <-this % stops a space
\thanks{Yin~Sun is with the Department of ECE, Auburn University, Auburn, AL 36849, e-mail: yzs0078@auburn.edu.}%
\thanks{D.~Richard~Brown~III is with the Department of Electrical and Computer Engineering, Worcester Polytechnic Institute, Worcester MA 01609 USA, e-mail: drb@wpi.edu.}%
\thanks{Sanjit K.~Kaul is with the Wireless Systems Lab, IIIT-Delhi, India, e-mail: skkaul@iiitd.ac.in.}%
\thanks{Eytan Modiano is with the Laboratory for Information and Decision Systems (LIDS) at the Massachusetts Institute of Technology (MIT), Cambridge, MA, e-mail: modiano@mit.edu.}%
\thanks{Sennur Ulukus is with the Department of Electrical and Computer Engineering, University of Maryland, College Park, MD 20742, e-mail: ulukus@umd.edu.}}
\begin{document}
\maketitle

%\section{Introduction}
% !TEX root = aoisurvey.tex
\begin{abstract}
    We summarize recent contributions in the broad area of age of information (AoI). In particular, we describe the current state of the art in the design and optimization of low-latency cyberphysical systems and applications  in which sources send time-stamped status updates to interested recipients. These applications desire status updates at the recipients to be as timely as possible; however, this is typically constrained by limited system resources. We describe AoI timeliness metrics and present general methods of AoI evaluation analysis that are  applicable to a wide variety of sources and systems. Starting from elementary single-server queues, we apply these AoI methods to a range of increasingly complex systems, including energy harvesting sensors transmitting over noisy channels, parallel server systems, queueing networks,  and various single-hop and multi-hop wireless networks. We also explore how update age is related to MMSE methods of sampling, estimation and control of stochastic processes. The paper concludes with a review of efforts to employ age optimization in cyberphysical applications. 
\end{abstract}
\section{Introduction}\label{sec:introRY}
Low-latency cyberphysical system applications continue to grow in importance. Camera images from vehicles are used to generate point clouds that describe the surroundings. Video streams are  augmented with informative labels. Sensor data needs to be gathered and analyzed to detect anomalies. A remote surgery system needs to update the positions of the surgical tools.  
From a system perspective, these examples  share a common description: a source generates time-stamped status update messages that are transmitted through a network to one or more monitors.  Awareness of the state of the remote sensor or system needs to be as timely as possible. In some cases, a few milliseconds delay  may be too much. 

% The goal of real-time status updating is to ensure that the status of interest, is as timely as possible at each monitor. 
% Ideally, a monitor at time $t$ would have instantaneous knowledge of the state of the system of interest at time $t$, but this is not feasible since network capacity constraints dictate that a status update require a nonzero  and typically random time in the network. Moreover, if sources send update packets too frequently, the network can become congested, causing additional delay in the delivery of the status update. 

% For these low-latency status updating applications,  the networks, protocols and systems will benefit from and perhaps require  a redesign  focused on the timeliness of status updates. 
We have seen that research efforts directed toward low-latency networks are underway. Machine-to-machine communication  and the tactile internet, each requiring  link delays of just a few milliseconds, were key drivers for the 5G cellular standard \cite{Parvez-RGSD-commsurveys2018,Popovski-SNCATB-comm2019,Sachs-AACLRSW-ieeeproc2019}. Edge cloud computing that will eliminate round trip propagation delays on the order of 40~ms for transcontinental routes is another essential ingredient.  However, while new systems supporting low-latency communication are necessary, they are also not sufficient for timely operation. In particular, networks need to regulate congestion. Similarly, edge-cloud processing centers can accumulate backlogged jobs that need to be processed in order to deliver timely updates. 

From these observations, timeliness of status updates has emerged as a new field of network research.  It has been shown, even in the simplest queueing systems, that timely updating is not the same as maximizing the utilization of the  system that delivers these updates, nor the same as ensuring that updates are received with minimum delay \cite{Kaul-YG-infocom2012}. 
While utilization is maximized by sending updates as fast as possible, this strategy will lead to a monitor receiving delayed updates that were backlogged in the communication system. 
In this case, the timeliness of status updates at the receiver can be improved by {\em reducing} the update rate. 
On the other hand,  throttling the update rate will also lead to a monitor having
unnecessarily outdated status information because of a lack of updates.

 %Optimization based on AoI metrics of both the network and the senders' updating policies  has yielded new and even surprising results \cite{Sun-UBYKS-IT2017UpdateorWait,Yates-isit2015}. 
 
This has led to new interest in  {\em Age of Information (AoI)} performance  metrics that describe the timeliness of a monitor's knowledge of an entity or process. AoI is an end-to-end metric that can be used to characterize latency in status updating systems and  applications. An update packet with timestamp $u$ is said to have age $t-u$ at a time $t\ge u$.  An update is said to be {\em fresh} when its timestamp is the current time $t$ and its age is zero. When the monitor's freshest\footnote{One update is fresher than another if its age is less.} received update at time $t$ has time-stamp $u(t)$,  the  age is the random process $\age(t)=t-u(t)$.

While this AoI  survey focuses on recent work on age, we note that data freshness has been a recurring research theme.  This includes 
modeling and maximizing the freshness of query responses from a data warehouse \cite{Karakasidis-VP-iqis2005etl}, network architectures that  limit the ``\emph{degree of staleness}'' of a cache \cite{Yu-BS-sigcomm1999scalable}, distributed QoS routing based on aged and imprecise network state information \cite{Chen1998}, and ad hoc networking mechanisms that avoid propagation of stale route information \cite{Hu-Johnson-pomc2002ensuring} and that balance network congestion against nodes having stale information \cite{Giruka-Singhal-wowmom2005hello}.

Periodic transactions updating real time databases \cite{Song-Liu-ISCSA1990,Xiong-Ramamritham-real1999deriving} was perhaps the earliest use of freshness. In \cite{Song-Liu-ISCSA1990} sensors wrote time-stamped fresh measurements into a real-time database and the age of an update was used to enforce  concurrency of computations based on multiple measurements.  For web-caching, page-refresh policies have been tuned  to maximize the freshness of cached pages \cite{Cho-GM-tods2003effective} using an age metric in which age accumulated once the cached copy became outdated. Also noteworthy is \cite{Ioannidis2009} in which updates from a source are distributed over a graph by a gossip network. This work showed how age in the network is described by the edge expansion  of the graph.

The initial motivation for \cite{Kaul-YG-infocom2012} came from vehicular safety messaging. In particular, \cite{Kaul-GRK-secon2011} looked at minimizing the age of safety messages over a CSMA network of connected cars. For small CSMA contention window sizes, it was observed in simulation that the minimum age could be approached using a gradient descent like algorithm.  Over a random graph of vehicular nodes in a DSRC network, a round robin schedule was shown to lead to an average status-age that is smaller under the condition that nodes' updates piggyback each others'  updates \cite{Kaul-YG-globecom2011piggybacking}.  Note that while both~\cite{Kaul-GRK-secon2011} and~\cite{Kaul-YG-globecom2011piggybacking} used the phrase \emph{system age of information}, the optimization metric was indeed the average age of status updates.

These  simulation studies of vehicular updating \cite{Kaul-GRK-secon2011,Kaul-YG-globecom2011piggybacking} prompted the AoI analysis in single-source single-server queues \cite{Kaul-YG-infocom2012}. In contrast to the prior work \cite{Song-Liu-ISCSA1990,Cho-GM-tods2003effective,Ioannidis2009} based on status update age,  \cite{Kaul-YG-infocom2012} focused on the impact of random service times on the age of delivered updates and showed  that minimizing age required balancing the rate of updates against congestion. The takeaway message  
%in \cite{Kaul-YG-infocom2012} 
was that both the update arrivals  and the service system could be designed, tuned, and even controlled to minimize the age.   

This survey focuses on the large number of  contributions to AoI analysis that followed \cite{Kaul-YG-infocom2012}.
Section~\ref{sec:background} introduces the age process and associated age metrics, and basic methods for the analysis of AoI.  Section~\ref{sec:queues} summarizes AoI results in single-server queues, in order to demonstrate how AoI is influenced by the update arrival rate, the queue discipline, and packet management schemes designed explicitly to optimize freshness. This leads to a review of queueing networks, with a focus on scheduling updates of multiple sources at multiple servers in Section~\ref{sec:queue-networks}. This is followed in Section~\ref{sec:resource-constrained} by the study of energy-constrained updating. Here the emphasis is on energy harvesting systems in which updates by a sensor  are constrained by its harvesting process. In this area, we examine {\em generate-at-will} sources that can generate a fresh update whenever they wish.  Generate-at-will models are further explored in the context of sampling, estimation and control in Section~\ref{sec:sampling}. This is followed by a study of   wireless networks in Section~\ref{sec:wireless} and a discussion of various applications of AoI in Section~\ref{sec:apps}. Finally, the conclusion in Section~\ref{sec:conclusion} discusses potential application areas of AoI.

%\section{AoI Analysis}
% !TEX root = aoisurvey.tex

\section{AoI Metrics and Analysis}
\label{sec:background}
%{\BLUE How about ``AoI Metrics and Analysis"?}
%RY: good sugg
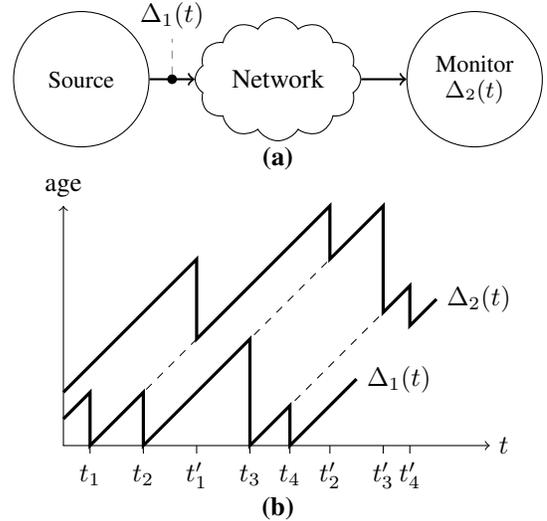
\begin{figure}[t]
\begin{center}
\begin{tikzpicture}[baseline=(current bounding box.center),node distance=0.6cm,
dot/.style={draw,circle,fill=black,minimum size=1mm,inner sep=0pt},
circnode/.style={draw,circle,align=center,minimum size=1.8cm}]
\node [circnode] (newsource) {\small Source};
\node  [cloud, draw,cloud puffs=12,cloud puff arc=120, aspect=1.5%
%, inner ysep=1em
] [right = of newsource](net){\shortstack{Network}%
%\\$[q(t),\xv(t)]$
};
\draw[->,thick] (newsource.east) -- node[dot, pos=0.5, pin={[pin edge={dashed,-}]90:$\age_1(t)$}](x1node){}(net.west);
\node [circnode] (localcache)[right = of net] {\small\shortstack{Monitor\\$\age_2(t)$}};
%Marks a bullet monitor on the link to the Monitor
%\draw[->,thick] (net.east) --  node[dot, pos=0.5, pin={[pin edge={dashed,-}]90:$x_2(t)$}](x1node){}(localcache.west);
\draw[->,thick] (net.east) -- (localcache.west);
\end{tikzpicture}\\
{\bf (a)}\\
\begin{tikzpicture}[baseline=(current bounding box.center),scale=\linewidth/25cm]
\draw [<->] (0,9) node [above] {age} -- (0,0) -- (16,0) node [right] {$t$};
\draw [very thick] (0,2) -- (5,7) -- (5,4)  -- (10,9) 
-- (10,7)  -- (12,9) -- (12,5) -- (13,6) -- (13,4.5) -- (14,5.5) node [right] {$\age_2(t)$};
\draw [ultra thin, dashed] (1,0) to (10,9) to (10,7) to (3,0);
\draw 
(1,0) \atick{t_1} 
(3,0)\atick{t_2}
(5,0) \dtick{t'_1} 
(7,0) \atick{t_3}
(8.5,0) \atick{t_4}
(10,0)\dtick{t_2'}
(12,0) \dtick{t'_3}
(13,0) \dtick{t'_4};
\draw [very thick] (0,1) -- ++(1,1) -- ++(0,-2)  -- ++(2,2) 
-- ++(0,-2)  -- ++(4,4) -- ++(0,-4) -- ++(1.5,1.5)-- ++(0,-1.5)-- ++(2.5,2.5) node [right] {$\age_1(t)$};
\draw [thin, dashed] (7,0) to (12,5);
%\draw [thin, dashed] (8.5,0) to (14,5.5);
\end{tikzpicture}\\
 {\bf (b)} 
\end{center}
\caption{(a) Fresh updates from a source pass through the network to  a destination monitor.  Monitor $1$ (marked by $\bullet$) sees fresh update  packets at the network access link. (b) Since Monitor $1$ sees fresh updates as a point process at times $t_i$, its age process $\age_1(t)$ is reset to zero at times $t_j$.   Since the destination monitor sees updates that are delivered at times $t'_j$ after traveling through the network, its age process $\age_2(t)$ is reset to $\age_2(t'_i)=t'_j-t_j$, which is the age of update $j$ when it is delivered.}
\label{fig:cloudnet}\end{figure}

 As depicted in  Figure~\ref{fig:cloudnet}(a),  the canonical updating model has a source that submits fresh updates to a network that delivers those updates to a destination monitor. In a complex system, there may be additional monitors/observers in the network that serve to track the ages of updates in the network. For example, Figure~\ref{fig:cloudnet}(a) depicts an additional monitor that observes fresh updates as they enter the network. 

These fresh updates are submitted at times $t_1,t_2,\ldots$ and this induces the AoI process $\age_1(t)$ shown in Figure~\ref{fig:cloudnet}(b). Specifically, $\age_1(t)$ is the age of the most recent update seen by a monitor at the input to the network. Because the updates are fresh, $\age_1(t)$ is reset to zero at each $t_i$. However, in the absence of a new update, the age $\age_1(t)$ grows at unit rate.  
If the source  in Fig.~\ref{fig:cloudnet} submits fresh updates as a renewal point process,  the AoI $\age_1(t)$ is simply the age (also known as the backwards excess) \cite{Ross1996stochastic,Gallager2013stochastic} of the renewal process.

These updates pass through a network  and are delivered to the destination monitor at corresponding times $t'_1,t'_2,\ldots$. Consequently, the AoI process $\age_2(t)$ at the destination monitor is reset at time $t'_j$ to $\age_2(t'_j)=t'_j-t_j$, which is the age of the $j$th update when it is delivered. Once again, absent the delivery of a newer update, $\age_2(t)$ grows at unit rate.  Hence the age processes $\age_1(t)$ and $\age_2(t)$ have the characteristic sawtooth patterns shown in Figure~\ref{fig:cloudnet}(b). Furthermore, any other monitor in the network that sees updates arrive some time after they are fresh, will have a sawtooth age process $\age(t)$ resembling that of $\age_2(t)$.

In the rest of this section, we describe three approaches to AoI analysis. We start with with methods that analyze the limiting time-average age by graphical decomposition  of the area under the sawtooth function $\age(t)$. We next introduce the average peak age metric and then the stochastic hybrid systems (SHS) approach to AoI analysis. This is followed by a discussion of nonlinear age penalty functions and functionals of the age process that are intended to capture the role of age in different classes of  applications. 

\subsection{Time-Average Age}
\label{sec:ageprelim}
Initial work on age has focused on applying graphical methods to sawtooth age waveforms $\age(t)$  to evaluate the time-average AoI
\begin{align}\eqnlabel{time-average-age}
\Tavg{\age} = \frac{1}{\mathcal{T}}\int_{0}^{\Tcal} \age(t) dt.
%\label{eqn:averageAge}
\end{align}
in the limit of large $\Tcal$.
While this time average  is often called the AoI, this work employs AoI and age as synonyms that refer to the process $\age(t)$.
%and call $\limty{\Tcal}\Tavg{\age}$ the average AoI or average age.

\begin{figure}[t]
\centering
% %ONECOLUMN
%\begin{tikzpicture}[scale=\linewidth/44cm]
% %TWOCOLUMN
%HELP https://tex.stackexchange.com/questions/331581/tikz-arrow-tips-on-every-subpath
\begin{tikzpicture}[scale=0.25]
\pgfdeclaredecoration{arrows}{draw}{
\state{draw}[width=\pgfdecoratedinputsegmentlength]{%
  \path [every arrow subpath/.try] \pgfextra{%
    \pgfpathmoveto{\pgfpointdecoratedinputsegmentfirst}%
    \pgfpathlineto{\pgfpointdecoratedinputsegmentlast}%
   };
}}
\tikzset{every arrow subpath/.style={|<->|, draw}}
\draw [<-|] (0,11) node [above] {$\age(t)$} -- (0,0) -- (15.5,0);
\draw [|->] (16,0) -- (30,0) node [right] {$t$};
\draw [very thick] (0,2) 
-- ++(6,6) node [above] {$A_0$}
-- ++(0,-3) 
-- ++(5,5)%(10,9) 
node [above] {$A_1$}
-- ++(0,-3)%(10,7)  
-- ++(2,2)%(12,9) 
node [above] {$A_2$} 
-- ++(0,-4)%(12,5) 
-- ++(2,2)%(14,7) 
node [above] {$A_3$} 
-- ++(0,-1.5)%(14,5.5) 
-- ++(0.5,0.5); %(14.5,6);
\draw [fill=lightgray, ultra thin, dashed] (1,0) to ++(10,10) to ++(0,-3) to ++(-7,-7);
\draw 
(1,0) \atick{t_0}
(4,0) \atick{t_1} 
(6,0) \dtick{t'_0} 
(8,0) \atick{t_2}
(9.5,0) \atick{t_3}
(11,0) \dtick{t'_1} 
(13,0) \dtick{t'_2};
\draw (2,2.5) node {$\tilde{Q}_0$} 
(5,2.5) node {$Q_1$} 
(8.5,2.5) node {$Q_2$};
 \path [decoration=arrows, decorate] (1,-3) to node [below] {$Y_1$} ++(3,0) to node [below] {$T_1$}  ++(7,0);
 \path [decoration=arrows, decorate] (1,-5.5) to node [below] {$T_0$} ++(5,0) to node [below] {$D_1$}  ++(5,0);
\draw [thin, dashed] (8,0) to ++(5,5);
\draw [thin, dashed] (9.5,0) to ++(5.5,5.5);
%% right side of plot
\draw [fill=lightgray, ultra thin, dashed] (18,0) to ++(10,10) to ++(0,-7) to ++(-3,-3);
\draw  
(18,0) \atick{t_{n-1}}
(22,0) \dtick{t'_{n-1}}
(25,0) \atick{t_n}
(28,0) \dtick{t'_n};
%\draw [thin] (22,0) -- (28,-0.4);
\draw [very thick] (17,2) -- ++(5,5) -- ++(0,-3) -- ++(6,6) node [above] {$A_n$} -- ++(0,-7) -- ++(2,2); 
\path [decoration=arrows, decorate] (18,-3) to node [below] {$Y_n$} ++(7,0) to node [below] {$T_n$}  ++(3,0);
\path [decoration=arrows, decorate] (18,-5.5) to node [below] {$T_{n-1}$} ++(4,0) to node [below] {$D_n$}  ++(6,0);
%\draw  [|<->|] (21,-3) -- node [below] {$Y_n$} (25,-3);
%\draw  [|<->|] (25,-3) -- node [below] {$T_{n}$} (28,-3);
\draw [<->] (28,2.8) -- node [right] {$T_n$} (28,0); 
\draw (24,2.5) node {$Q_n$};
%\draw[<-] (25,3) to [out=110,in=250] (25,6) node [above] {$Q_{n}$};
\end{tikzpicture}
\caption{ An age sample path: updates from a source  arrive at times $t_0,t_1,\ldots$  and are received at the monitor at times $T_0',T_1.'\ldots$. For the $n$th delivered update, $Y_n$, $T_n$ and $D_n$ are the interarrival, system and inter-departure times, and $A_n$ is the corresponding age peak.}
\label{fig:age}
\vspace{-5mm}
\end{figure}
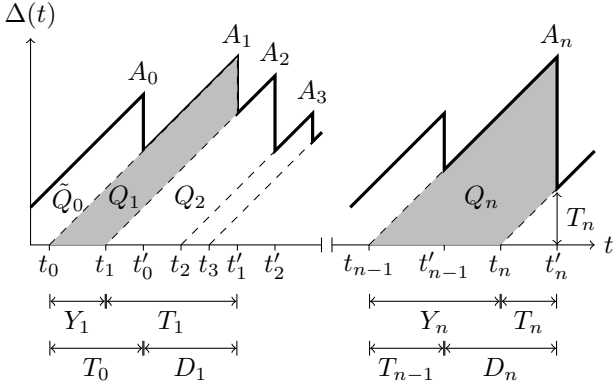

Figure~\ref{fig:age} shows a sawtooth sample path of an age process $\age(t)$ in greater detail. 
% Over an interval $(0,\mathcal{T})$, the time-average age $\Tavg{\age}$ is the area under the age graph normalized by the time interval of observation.
For simplicity of exposition,  the length of the observation interval is chosen to be $\mathcal{T}=t'_n$, as depicted in  Figure~\ref{fig:age}. We decompose the area defined by the integral \eqnref{time-average-age} into the  sum of 
the polygon area $\tilde{Q}_0$, the trapezoidal areas $Q_j$ for $j\ge 1$ ($Q_1$ and $Q_n$ are highlighted in the figure), and the triangular area of width $T_n$ over the time interval $(t_n,t_n')$. 
From Figure~\ref{fig:age}, we see that $Q_n$ can be calculated as the difference between the area of the isosceles triangle whose base connects the points $t_{n-1}$ and $t'_n$ and the area of the isosceles triangle with base connecting the points $t_n$ and $t'_{n}$. 
Defining 
%\begin{equation}
$Y_n=t_n - t_{n-1}$
%\end{equation}
to be the $n$th interarrival time, it follows that
\begin{align}\eqnlabel{Qi}
Q_n = \frac{1}{2}(T_{n} + Y_n)^2 - \frac{1}{2}T_{n}^2 = Y_n T_{n}+Y_n^2/2.
\end{align}
With  $N(\mathcal{T}) = \max\{n|t_n\le \mathcal{T}\}$ denoting the number of updates by time $\mathcal{T}$, 
%
% this decomposition,
% yields the time-average age 
% \begin{align}
% \Tavg{\age} &= \frac{\tilde{Q}}{\mathcal{T}}
%\nonumber\\&\qquad
%+ \frac{(N(\mathcal{T})-1)}{\mathcal{T}}
%\frac{\sum_{j=2}^{N(\mathcal{T})} 
%\paren{X_i T_{i}+X_i^2/2}
% Q_j}{N(\mathcal{T})-1}
% \eqnlabel{delta_tau_2}
%\end{align}
% where $\tilde{Q} = \Qtil_1+T_n^2/2$. We observe that the age contribution $\tilde{Q}$ represents a boundary effect that is finite with probability 1, so  the first term in~\eqnref{delta_tau_2} will vanish as $\Tcal$ grows. 
we will say a status updating system is {\em stationary and ergodic} if  $(Y_n,T_n)$ is a stationary sequence with marginal distribution identical to $(Y,T)$, and $N(\mathcal{T})/\mathcal{T}\to{1}/{\E{Y}}$ and
$\frac{1}{N(\mathcal{T})}\sum_{j=1}^{N(\mathcal{T})} Q_j\to\E{Q}$ with probability $1$ as $\mathcal{T}\to\infty$.
% \begin{align}
%  \frac{N(\mathcal{T})}{\mathcal{T}}\to\frac{1}{\E{Y}}, \quad\text{and}\quad
% \frac{\sum_{j=1}^{N(\mathcal{T})}
% Q_j}{N(\mathcal{T})} \to\E{Q}
% \end{align}

For a stationary ergodic updating system in which $Y$ is the interarrival time between delivered updates and $T$ is the system time of such a delivered packet, it follows that the time-average AoI  
$\age=\limty{\mathcal{T}}\Tavg{\age}$ satisfies
\begin{align}\eqnlabel{sysAge}
\age&=\frac{\E{Q_n}}{\E{Y_n}}=\frac{\E{Y_nT_n} +\E{Y_n^2}/2}{\E{Y_n}}.
\end{align}

We  note this decomposition of the area under the age process is not unique. As can be seen in Fig.~\ref{fig:age}, an alternate approach \cite{Costa-CE-IT2016management} shows 
\begin{align}
Q_n=D_nT_{n-1}+D_n^2/2
\end{align} 
where $D_n =t'_n-t'_{n-1}$ is the $n$th inter-\emph{departure} time. With this decomposition, a stationary ergodic updating system will have average age \begin{equation}
 \age=\frac{\E{Q_n}}{\E{D_n}}
 =\frac{\E{D_nT_{n-1}}+\E{D_n^2/2}}{\E{D_n}}. 
 \eqnlabel{sysAge2}
\end{equation}

Both \eqnref{sysAge} and \eqnref{sysAge2} can be applied to a broad class of  service systems, including both lossless FCFS systems as well as lossy last-come-first-served (LCFS) systems in which updates are preempted and discarded.  Furthermore, they make no specific assumptions regarding other traffic that might share the system with the update packets of interest. 

However, despite their apparent simplicity, exact analysis of the age $\age$ can be challenging.  
%The random variables $Y$ and $T$ typically are  dependent.  
With respect to \eqnref{sysAge}, a large interarrival time $Y_n$ allows the queue to empty, yielding a small waiting time and typically a small system time $T_n$.  That is, $Y_n$ and $T_n$ tend to be negatively correlated and this complicates the evaluation of $\E{T_nY_n}$. 

\subsection{Peak Age}
The challenge in evaluating $\E{TY}$ prompted the introduction in \cite{Costa-CE-isit2014} of peak age of information (PAoI), an alternate  (and generally more tractable) age metric. Referring to Fig.~\ref{fig:age}, we observe that the age process $\age(t)$ reaches a peak
\begin{align}\eqnlabel{peak-age}
A_n &=T_{n-1} +D_n
\end{align}
the instant before the service completion at time $t_n'$.
As an alternative to the (possibly challenging) computation of the average age, \cite{Costa-CE-isit2014} proposed the average peak age of information (PAoI)
\begin{equation}\label{eq_PAoI}
\PAoI =\limty{\Tcal} \frac{1}{N(\Tcal)}\sum_{n=1}^{N(\Tcal)} A_n     
\end{equation}
Under mild ergodicity assumptions, it follows that the PAoI is
\begin{equation}
\PAoI =\E{A}=\E{T_{n-1}}+\E{D_n}.
\end{equation}
Hence PAoI avoids the computation of $\E{TY}$.

Like the average age, the peak age captures the key characteristics of the age process. Specifically, if the system is lightly loaded, then the average inter-departure time $\E{D}$ will be large; conversely as the system load gets heavy, the average system time $\E{T}$ will become large.  Here we also note there is more than one way to calculate PAoI. Inspection of Fig.~\ref{fig:age} reveals that $A_n=Y_n+ T_{n}$. It follows that PAoI is also $\PAoI=\E{Y}+\E{T}$, which is the decomposition in  \cite{Costa-CE-IT2016management}.
% RB note: the literature doesn't have consistent definitions for peak age - some papers look at the average of the peaks and some look and the max/sup of the peaks

For single-server queues, it has been observed  \cite{Inoue-MTT-IT2019} that by defining $t'_{-1}=0$ and $T_{-1}=\age(0)$  that 
\begin{align}
    \age(t)=T_{n-1} +(t-t'_{n-1}),\quad t\in(t'_{n-1},t'_n).
\end{align}
for $n=0,1,2,\ldots$. Thus the sample path of $\age(t)$ is completely determined by the point process $\set{(t_n',T_n):n=0,1,\ldots}$.
Since the departure times $t_n'$ can be reconstructed from the $D_n$ inter-departure sequence and  \eqnref{peak-age} implies $D_n=A_n-T_{n-1}$, the sequence of pairs $(T_{n-1},A_n)$ is also sufficient to reconstruct the age process $\age(t)$. As noted in \cite{Inoue-MTT-IT2019}, this shows that the age peaks $A_n$ are a fundamental characterization of the age process. 

\subsection{Stochastic Hybrid Systems for AoI Analysis}
A stochastic hybrid system (SHS) \cite{Hespanha-2006modelling} was shown in \cite{Yates-Kaul-IT2019}  to provide an alternate approach for average age analysis.
% This approach was then extended in \cite{Yates-IT2020} to analyse higher order moments and the moment generating function of stationary age  processes.
In the SHS approach, the network shown in Fig.~\ref{fig:cloudnet}(a)  has a hybrid state $[q(t),\xv(t)]$ such that $\xv(t)\in \R^{1\times n}$ and $q(t)\in\Qcal=\set{0,\ldots,M}$ is a continuous-time Markov chain. 
For AoI analysis, $q(t)$ describes the discrete state of a network while the real-valued age vector $\xv(t)$ describes the continuous-time evolution of a collection of  age-related processes.   One of the components of $\xv(t)$ is the age $\age(t)$ at a monitor of interest.

The SHS approach was introduced  in \cite{Yates-Kaul-IT2019}, where
it was shown  that age tracking can be implemented as a simplified SHS with non-negative linear reset maps in which the continuous state is a piecewise linear process \cite{Vermes-1980,Davis-1984,Deville-DDZ-siam2016moment}. For finite-state systems,  
this led to a set of age balance equations and simple conditions \cite[Theorem~4]{Yates-Kaul-IT2019}  under which $\E{\xv(t)}$ converges to a fixed point.  
A description of this simplified SHS for AoI analysis now follows.

In the graph representation of the Markov chain $q(t)$, each state $q\in\Qcal$ is a node and each transition $l\in\Lcal$ is a directed edge $(q_l,q'_l)$ with transition rate $\laml$ from state $q_l$ to $q'_l$. 
%The Kronecker delta function $\dq{q_l}$ ensures that transition $l$ occurs only in state $q_l$. 
Associated with each transition $l$  is  a transition reset mapping  $\Amat_l\in\set{0,1}^{n\times n}$  that induces a jump $\xv'=\xv\Amat_l$ in the continuous state $\xv(t)$. 

Unlike an ordinary continuous-time Markov chain, the SHS Markov chain may include self-transitions in which the discrete state is unchanged because a reset occurs in the continuous state. Furthermore, for a given pair of states $q,q'\in\Qcal$, there may be multiple transitions $l$ and $\lhat$ in which $q(t)$ jumps from $q$ to $q'$ but the transition maps $\Amat_l$ and $\Amat_{\lhat}$ are different.

For each state $\qbar$, we denote the respective sets of incoming and outgoing  transitions by 
\begin{align}\eqnlabel{Lcalqbar}
\Lcal'_{\qbar}\!=\!\set{l\in\Lcal: q'_l=\qbar},\ \ 
\Lcal_{\qbar}\!=\!\set{l\in\Lcal: q_l=\qbar}.
\end{align}
%\end{subequations}
Assuming the discrete state Markov chain is ergodic, $q(t)$ has unique stationary probabilities 
%state probability vector $\piv(t)=\rvec{\pi_0(t) &\cdots&\pi_m(t)}$ always  converges to the unique stationary
 $\bar{\piv}=\rvec{\pibar_0&\cdots&\pibar_M}$ satisfying
%\begin{subequations}
\begin{align}
\bar{\pi}_{\qbar}\sum_{l\in\Lcal_{\qbar}}\laml&=\sum_{l\in\Lcal'_{\qbar}}\laml\bar{\pi}_{\ql},\quad\qbar\in\Qcal;\quad 
 \sum_{\qbar\in\Qcal}\bar{\pi}_\qbar=1.
\eqnlabel{AOI-SHS-pi}
\end{align}
%and the normalization constraint $\sum_{\qbar\in\Qcal}\bar{\pi}_\qbar=1$.
%\eqnlabel{AOI-SHS-pi1}
%\end{align}
%\end{subequations}
%If $\piv(t)=\bar{\piv}$,  it is shown \cite{Yates-Kaul-IT2018}  that $\vv(t)=\rvec{\vv_0(t)&\cdots&\vv_m(t)}$ obeys a system of first order differential equations.
% such that for all $\qbar\in\Qcal$,
%\begin{align}
%\dot{\vv}_{\qbar}(t)&=
%\bv_{\qbar}\pibar_{\qbar}+\sum_{l\in\Lcal'_{\qbar}}\laml \vv_{\ql}(t)\Amat_l
%-\vv_{\qbar}(t)\sum_{l\in\Lcal_{\qbar}}\laml.
%%\quad \qbar\in\Qcal.
%\eqnlabel{vv-derivs-pibar}
%\end{align}
%Depending on the  reset maps $\Amat_l$, the differential equation \eqnref{vv-derivs-pibar} may or may not be stable. 
%However,  when \eqnref{vv-derivs-pibar}
%When this system is stable, 
%each $\vv_{\qbar}(t)=\E{\xv(t)\dqt{\qbar}}$ converges to a limit $\vvbar_{\qbar}$ as $t\goes\infty$. 
%In this case,
%\begin{align}
%\E{\xv}\equiv\limty{t}\E{\xv(t)}
%&=\!\limty{t}\sum_{\qbar\in\Qcal} \E{\xv(t)\delta_{\qbar,q(t)}}\nn
%&=\sum_{\qbar\in\Qcal} \bar{\vv}_{\qbar}
%\end{align}
%is the vector of average ages at the set of observers. 
The next theorem derives the limiting average age vector $\E{\xv}=\limty{t}\E{\xv(t)}$.
\begin{theorem}\thmlabel{AOI-SHS}
\cite[Theorem~4]{Yates-Kaul-IT2019}
If the discrete-state Markov chain $q(t)$ is ergodic with stationary distribution $\bar{\piv}>0$ and there exists a non-negative vector $\vvbar=\rvec{\vvbar_0&\cdots\vvbar_M}$ % an age-of-information SHS $\Acal$ has average age 
such that 
%\begin{subequations}
\begin{align}
\bar{\vv}_{\qbar}\sum_{l\in\Lcal_{\qbar}}\laml &=\onev[]\bar{\pi}_{\qbar}+ \sum_{l\in\Lcal'_{\qbar}}\laml \bar{\vv}_{\ql}\Amat_l,\quad \qbar\in\Qcal,\eqnlabel{AOI-SHS-v}
\end{align}
then
%the differential equation \eqnref{vv-derivs-pibar} is stable and 
the average age vector is %given by 
$\E{\xv}=
%\limty{t}\E{\xv(t)}=
\sum_{\qbar\in\Qcal} \vvbar_{\qbar}$.
%\begin{equation}
%\E{\xv}=\limty{t}\E{\xv(t)}=\sum_{\qbar\in\Qcal} \vvbar_{\qbar}.
%\eqnlabel{age-vsum}
%\end{equation}
%\end{subequations}
\end{theorem}
In \cite[Theorem~1]{Yates-IT2020}, this theorem is extended to provide the stationary age moments $\limty{t} \E{\rvec{x_1^m(t) \cdots x_n^m(t)}}$ and staionary age MGF $\limty{t} \E{\rvec{e^{sx_1(t)} \cdots e^{sx_n(t)}}}$ of the age process $\xv(t)$.

\subsection{Nonlinear Age Functions} \label{sec:nonlinear}
%The AoI $\Delta(t)$ measures information ageing as the time difference between data generation and data usage, which grow linearly in time with a unit slope. 
%In practice, some information sources (e.g., vehicle location, stock price) vary quickly over time, while others (e.g., temperature, interest-rate) change slowly. Consider the example of autonomous driving: The location information of motor vehicles collected 0.5 seconds ago could already be quite stale for making control decisions\footnote{A car will travel 15 meters during 0.5 seconds at the speed of 70 mph.}, but the engine temperature measured a few minutes ago is still valid for engine health monitoring. In addition, 

Although the AoI $\age(t)$ grows over time with a unit rate, the performance degradation caused by information aging may not be a linear function of time. For instance, consider the problem of estimating the state of a Gaussian Linear Time-Invariant (LTI) system: If the system is stable, the state estimation error is a sub-linear function of $\age(t)$ that converges to a finite constant as $\age(t)$ tends to infinite \cite{Ornee2019,Ornee2020ArXiv}; if the system is unstable, the state estimation error increases exponentially as $\age(t)$ grows \cite{Champati2019,Markus2019}. 

% feedback control systems, the state estimation error 
%
%in the feedback control problem for stabilizing a pendulum in its inverted position, a slight deviation from the equilibrium position will cause a gravitation torque on the pendulum, which will accelerate it away from the equilibrium. Hence, the state estimation error in the inverted pendulum system grows 
%
%When the position of the pendulum, the estimation error in such systems an exponential function of the AoI. 
%From these examples, one can observe that  information freshness should be evaluated based on (i) the time-varying pattern of the source and (ii) how valuable the fresh data is in the specific application.

One approach for characterizing the nonlinear behavior of information aging is to define \emph{freshness} and \emph{staleness} as nonlinear functions of the AoI \cite{SunInfocom2016,SunJournal2017,Kosta2017,SunNonlinear2019,KostaTCOM2020}. Specifically, dissatisfaction with information staleness (or eagerness for data refreshment) can be represented by a penalty function $p(\age(t))$ of the AoI $\age(t)$, where the function $p: [0,\infty) \mapsto \mathbb{R}$ is \emph{non-decreasing}. This non-decreasing requirement on $p(\cdot)$ complies with the observations that stale data is usually less desirable than fresh data \cite{Shapiro1999,Cho-GM-tods2003effective,Even:2007,Heinrich:2009,Ioannidis2009,Altman2011,Razniewski:2016}. 
This information staleness model is quite general, as it allows $p(\cdot)$ to be non-convex or  discontinuous. 

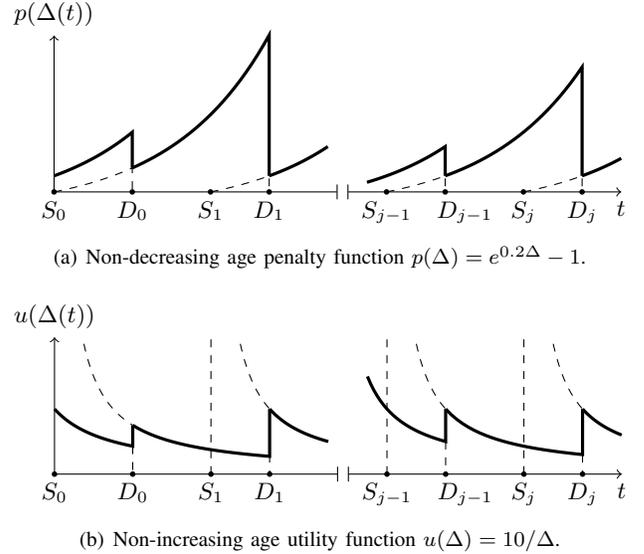
\begin{figure}[t]
\centering
\subfigure[][Non-decreasing age penalty function $p(\age)=e^{0.2\age}-1$.]
{
\begin{tikzpicture}[scale=0.26]
\draw [<-|] (0,8) node [above] {\small $p(\age(t))$} -- (0,0) -- (14.5,0);
\draw [|->] (15,0) -- (29,0) node [below] {\small$t$};
%\draw (-3,9) node [right] {\small$p(\age(t));%=e^{0.2\age(t)}-1$};
\fill
(8,0)  circle[radius=4pt]
(17,0)  circle[radius=4pt]
(24,0)  circle[radius=4pt];
\draw
(0,0) node [below] {\small$S_0$}
(8,0) node [below] {\small$S_1$}
(17,0) node [below] {\small$S_{j-1}$}
(24,0) node [below] {\small$S_{j}$};
\fill
(0,0)  circle[radius=4pt]
(4,0)  circle[radius=4pt]
(11,0)  circle[radius=4pt]
(20,0)  circle[radius=4pt]
(27,0)  circle[radius=4pt];
\draw
(4,0) node [below] {\small$D_0$}
(11,0) node [below] {\small$D_1$}
(21,0) node [below] {\small$D_{j-1}$}
(27,0) node [below] {\small$D_{j}$};
\draw[ very thick, domain=0:4] plot (\x, {exp(0.2*(\x+3))-1}) -- (4, {exp(0.2*(4))-1});
\draw[ very thick, domain=4:11] plot (\x, {exp(0.2*(\x))-1}) -- (11, {exp(0.2*(11-8))-1});
\draw[ thin,dashed,  domain=0:4] plot (\x, {exp(0.2*(\x))-1}) -- (4,0);
\draw[ very thick, domain=11:14] plot (\x, {exp(0.2*(\x-8))-1});
\draw[ thin,dashed,  domain=8:11] plot (\x, {exp(0.2*(\x-8))-1}) -- (11,0);
\draw[ very thick, domain=16:20] plot (\x, {exp(0.2*(\x-14))-1}) -- (20, {exp(0.2*(20-17))-1});
\draw[ very thick, domain=20:27] plot (\x, {exp(0.2*(\x-17))-1}) -- (27, {exp(0.2*(27-24))-1});
\draw[ thin,dashed,  domain=17:20] plot (\x, {exp(0.2*(\x-17))-1}) -- (20,0);
\draw[ very thick, domain=27:29] plot (\x, {exp(0.2*(\x-24))-1});
\draw[ thin,dashed,  domain=24:27] plot (\x, {exp(0.2*(\x-24))-1}) -- (27,0);
\end{tikzpicture}
}

\subfigure[][Non-increasing age utility function $u(\age)=10/\age$.]
{
\begin{tikzpicture}[scale=0.26]
\draw [<-|] (0,7) node [above] {\small $u(\age(t))$} -- (0,0) -- (14.5,0);
\draw [|->] (15,0) -- (29,0) node [below] {\small$t$};
%\draw (0.2,2) node [left] { $Y_0$};
%\draw (-3,9) node [right] {\small$u(\age(t))=10/\age(t)$};
\fill
(8,0)  circle[radius=4pt]
(17,0)  circle[radius=4pt]
(24,0)  circle[radius=4pt];
\draw
(0,0) node [below] {\small$S_0$}
(8,0) node [below] {\small$S_1$}
(17,0) node [below] {\small$S_{j-1}$}
(24,0) node [below] {\small$S_{j}$};
\fill
(0,0)  circle[radius=4pt]
(4,0)  circle[radius=4pt]
(11,0)  circle[radius=4pt]
(20,0)  circle[radius=4pt]
(27,0)  circle[radius=4pt];
\draw
(4,0) node [below] {\small$D_0$}
(11,0) node [below] {\small$D_1$}
(21,0) node [below] {\small$D_{j-1}$}
(27,0) node [below] {\small$D_{j}$};
\draw[ very thick, domain=0:4] plot (\x, {10/(\x+3))}) -- (4, {10/4});
\draw[ thin, dashed] (4, {10/7}) -- (4,0);
\draw[ thin, dashed, domain=1.5:4] plot (\x, {10/(\x)});
\draw[ very thick, domain=4:11] plot (\x, {10/(\x)}) -- (11, {10/(11-8)});
\draw[ thin, dashed] (11, {10/10}) -- (11,0);
\draw[ thin, dashed, domain=9.5:11] plot (\x, {10/(\x-8)});
\draw[ thin, dashed] (8,0)--(8,7);
\draw[ thin, dashed] (17,0)--(17,7);
\draw[ thin, dashed] (24,0)--(24,7);
\draw[ very thick, domain=11:14] plot (\x, {10/(\x-8)});
\draw[ very thick, domain=16:20] plot (\x, {10/(\x-14)}) -- (20, {10/(20-17)});
\draw[ thin, dashed] (20, {10/6}) -- (20,0);
\draw[ thin, dashed, domain=18.5:20] plot (\x, {10/(\x-17)});
\draw[ very thick, domain=20:27] plot (\x, {10/(\x-17)}) -- (27, {10/(27-24)});
\draw[ thin, dashed, domain=25.5:27] plot (\x, {10/(\x-24)});
\draw[ very thick, domain=27:29] plot (\x, {10/(\x-24)});
\draw[ thin, dashed] (27, {10/10}) -- (27,0);
\end{tikzpicture}
}
\caption{Two examples of non-linear age functions, where $S_i$ and $D_i$ are the generation time and delivery time of the $i$-th sample, respectively.}
\label{fig:age2}
\vspace{-5mm}
\end{figure}

Similarly, information freshness can be characterized by a \emph{non-increasing} utility function $u(\cdot)$ of the AoI $\Delta$ \cite{Even:2007,Ioannidis2009}. One simple choice is $u(\cdot)=-p(\cdot)$. 
Note that because the AoI $\Delta(t)$ is a function of time $t$,  $p(\Delta(t))$ and $u(\Delta(t))$ are both time-varying, as illustrated in Fig.~\ref{fig:age2}. 
In practice, one can choose $p(\cdot)$ and $u(\cdot)$ based on the information source and the application under consideration, as illustrated in the examples below. In addition to these examples, we note that additional usage cases of $p(\cdot)$ and $u(\cdot)$ can be found in \cite{Cho-GM-tods2003effective,Even:2007,Heinrich:2009,Ioannidis2009,Altman2011,Razniewski:2016} and that other information freshness metrics that cannot be expressed as functions of    $\age(t)$ were discussed in \cite{Costa-CE-IT2016management,Bedewy-SS-isit2016,Bedewy-SS-isit2017,BedewyJournal2017,BedewyJournal20172,Sun-UBK-aoi2018,maatouk2020status}. 

\subsubsection*{Auto-correlation Function}
%\begin{itemize}
%1. \emph{Auto-correlation Function of the Source:} 
The auto-correlation function $\mathbb{E}[X_t^* X_{t-\age(t)}]$ of a source signal $X_t$ can be used to evaluate the freshness of the sample $X_{t-\age(t)}$ \cite{Kosta2017}. For some stationary sources, $|\mathbb{E}[X_t^* X_{t-\age(t)}]|$ is a non-negative, non-increasing function of the AoI $\age(t)$ that can be considered as an age utility function $u(\age(t))$.
For example, in stationary ergodic Gauss-Markov block fading channels, the impact of channel aging can be characterized by the auto-correlation function of fading channel coefficients. When the AoI $\age(t)$ is  small, the auto-correlation function and the data rate both decay with respect to $\age(t)$ \cite{Truong2013}.

\subsubsection*{Real-time Signal Estimation Error} 
Consider a status-update system, where samples of a Markov source $X_t$ are forwarded to a remote estimator. The estimator uses causally received samples to reconstruct an estimate $\hat X_t$. Let $S_i$ and $D_i$ denote the generation time and delivery time of the $i$-th sample, respectively. If the sampling times $\{S_i\}_{i=1,2,\ldots}$ are independent of the observed source $\{X_t, t\geq 0\}$, %then 
%$\{\age(t), t\geq 0\}$ is independent of $\{X_t, t\geq 0\}$. In this case, 
one can show that the mean-squared estimation error at time $t$ is an age penalty function $p(\age(t))$ \cite{Yates-Kaul-IT2019,SunISIT2017,SunTIT2020,Ornee2019}. %For example, if the source is a Wiener process, the estimation error is exactly the age $\age(t)$ \cite{YatesTIT2018,SunISIT2017,SunTIT2020ZZJN19a}. 
However, if the sampling times $\{S_i\}_{i=1,2,\ldots}$ are chosen based on causal knowledge of the source, the estimation error is not necessarily a  function of $\age(t)$ \cite{SunISIT2017,SunTIT2020,Ornee2019}.

The above result can be generalized to the state estimation error of feedback control systems \cite{Champati2019,Markus2019}. Consider a single-loop feedback control system, where a plant and a controller are governed by a discrete-time Gaussian LTI system
\begin{align}
X_{t+1} = A X_t + B U_t + N_t,
\end{align}
where $X_t \in \mathbb{R}^n$ is the state of the system at time slot $t$, 
%$n$ is the system dimension, 
$U_t\in \mathbb{R}^m$  represents the control input,  $N_t \in \mathbb{R}^n$ is an exogenous zero-mean Gaussian noise vector with covariance matrix $\Sigma$. 
The constant matrices $ A\in \mathbb{R}^{n\times n}$  and $B\in \mathbb{R}^{n\times m}$ are the system and input matrices, respectively, where $(A,B)$ is assumed to be controllable, and the exogenous noise $N_t$ is \iid{} across time. 
Samples of the state process $X_t $ are forwarded to the controller, which determines $U_t$ at time $t$ based on the samples that have been delivered by time $t$. Under some assumptions, the state estimation error can be proven to be independent of the adopted control policy \cite{SoleymaniArXiv2018}. 
Furthermore, if the sampling times $\{S_i\}_{i=1,2,\ldots}$ are independent of the state process $X_t$,  the state estimation error is an age penalty function $p(\age(t)) = \sum_{r=0}^{\age(t)-1} \text{tr} [A^r\Sigma (A^T)^r]$ \cite{Champati2019,Markus2019}. If the  sampling time $\{S_i\}_{i=1,2,\ldots}$ are determined based on history values of the system state, the state estimation error is not necessarily a function of the AoI.
%redundant
%, where  $A$ is the system matrix and  $\Sigma$ is the covariance of the exogenous noise.

\subsubsection*{Mutual Information based Freshness Metrics} 
Let ${W}_t = \{(X_{S_i},S_i): D_i \leq t\}$
denote the samples that have been delivered to the receiver by time $t$. One can use the mutual information \cite{SunSPAWC2018}
\begin{align}
I(X_t; {W}_t) = H(X_t) - H(X_t|{W}_t),
\end{align}
i.e., information that the received samples ${W}_t$ carry about the current source value $X_t$, to evaluate the freshness of ${W}_t$. %\footnote{After this mutual information based freshness metric appeared in an earlier version of this paper \cite{SunSPAWC2018}, the authors heard from Jing Yang (personal communication) that she had developed a version of this metric.}
%\footnote{In this metric, the knowledge implied by the sampling times $\{{S_i}: D_i \leq t\}$ is neglected. One interesting future research direction is to investigate how to exploit timing information to improve data freshness.}
If $I(X_t; {W}_t)$ is close to $H(X_t)$, the sample ${W}_t$ contains a lot of information about $X_t$ and is considered to be fresh; if $I(X_t; {W}_t)$ is near $0$, ${W}_t$ provides little information about $X_t$ and is deemed to be obsolete. 

One way to interpret  $I(X_t; {W}_t)$ is to consider how helpful the received samples ${W}_t$ are for inferring $X_t$. 
By using the Shannon code lengths \cite[Section 5.4]{Cover}, the expected minimum number of bits $L$ required to specify $X_t$ satisfies 
$H(X_t) \leq L < H(X_t) + 1$,
where $L$ can be interpreted as the expected minimum number of binary tests that are needed to infer $X_t$. 
On the other hand, with the knowledge of ${W}_t$, the expected minimum number of bits $L'$ that are required to specify $X_t$ satisfies
$H(X_t| {W}_t) \leq L' < H(X_t| {W}_t) + 1$.
If $X_t$ is a random vector consisting of a large number of symbols (e.g., $X_t$ represents an image containing many pixels or the coefficients of MIMO-OFDM channels), the one bit of overhead is insignificant. 
Hence, $I(X_t; {W}_t)$ is approximately the reduction in the description cost for inferring $X_t$ without and with the knowledge of ${W}_t$. 

If $X_t$ is a stationary Markov chain, the following\ lemma is a consequence of the data processing inequality \cite[Theorem 2.8.1]{Cover}:

\begin{lemma}\label{lem1}\cite{SunSPAWC2018}
If $X_t$ is a stationary (continuous-time or discrete-time) Markov chain and the sampling times $\{S_i\}_{i=1,2,\ldots}$ are independent of $\{X_t, t\geq0\}$, then the mutual information
\begin{align}\label{eq_lem1}
I(X_t; {W}_t) = I(X_t; X_{t-\age(t)})
\end{align}
is a non-negative non-increasing function $u(\age(t))$ of $\age(t)$.
\end{lemma}
%\begin{proof}
%See Appendix \ref{app_lem1}. 
%\end{proof}
%If the sampling times $S_i$ are independent of $\{X_t, t \geq 0\}$, then the condition ``$\age(t)$ is independent of $X_t$ and ${W}_t$'' is satisfied. 

Lemma \ref{lem1} provides an intuitive interpretation of ``information aging.'' The information $I(X_t; {W}_t)$ that is preserved in ${W}_t$ for {inferring} the current source value $X_t$ decreases as the AoI $\age(t)$ grows. We note that Lemma \ref{lem1} can be generalized to the case that $X_t$ is a stationary discrete-time Markov chain with memory $k$. In this case, each sample $V_{t} = (X_{t}, X_{t-1}, \ldots, X_{t-k+1})$ should contain the source values at $k$ successive time instants. Let
 $W_t = \{(V_{S_i},S_i): D_i \leq t\}$, then $V_{t-\age(t)}$ is a sufficient statistic of ${W}_t$ for inferring $X_t$ and $I(X_t; {W}_t) = I(X_t; V_{t-\age(t)})$ is  a non-negative non-increasing function of $\age(t)$. 

%If the $S_i$'s are independent of $\{X_t, t\geq0\}$, the sampling times $\{{S_i}:D_i \leq t\}$ of delivered samples do not carry any information about the current source value  $X_t$. However, 
Lemma \ref{lem1} relies on a condition that the sampling times $\{S_i\}_{i=1,2,\ldots}$ are independent of $\{X_t, t\geq0\}$. 
If the sampling times $\{S_i\}_{i=1,2,\ldots}$ are determined using causal knowledge of $X_t$, $I(X_t; {W}_t)$ is not necessarily a function of the AoI. One interesting future research direction is how to choose the sampling time $\{S_i\}_{i=1,2,\ldots}$ based on the signal and utilize the timing information in $\{S_i\}_{i=1,2,\ldots}$ to improve information freshness.

Similarly, one can also use the conditional entropy $H(X_t| {W}_t)$ to represent the staleness of ${W}_t$ \cite{Soleymani2016-1,Soleymani2016-2,Soleymani2016-3}. In particular, 
$H(X_t| {W}_t)$ can be interpreted as the uncertainty about the  current source value $X_t$ after receiving the samples ${W}_t$. 
 If the sampling times  $\{S_i\}_{i=1,2,\ldots}$ are independent of $\{X_t, t \geq 0\}$ and $X_t$ is a stationary Markov chain, 
 \begin{IEEEeqnarray}{rCl}
     H(X_t| {W}_t)&= H(X_t| \{X_{S_i}\!:\!D_i\leq t\})=H(X_t| X_{{t-\age(t)}})\IEEEeqnarraynumspace
 \end{IEEEeqnarray}  
 is a non-decreasing penalty function $p(\age(t))$ of the AoI $\age(t)$. %In this result, the Markov chain $X_t$ is required to be time-homogeneous, but not necessarily stationary.
%(e.g., the conditional distribution $\Pr[X_{t+s}|X_t]$ is invariant in $t$, but $ X_0=x$ is a constant). 
If (i) $\{S_i\}_{i=1,2,\ldots}$ are determined based on causal knowledge of $X_t$ or (ii) $X_t$ is not a Markov chain, $H(X_t| {W}_t)$ is not necessarily a function of the AoI. 

\subsubsection*{Age Violation Probability} 
If $p(\Delta(t))$ is chosen as the indicator function \cite{BedewyJournal2017} 
\begin{equation}\label{functional311}
p(\Delta(t)) = 1_{\{\age(t)>d\}} = \left\{\begin{array}{l l } 1, \text{ if } \age(t)>d;\\
0, \text{ if } \age(t)\leq d,\\
\end{array}\right.
\end{equation}
then the fraction of time such that $\age(t)$ exceeds a threshold $d$ is given by  
\begin{equation}\label{eq_violation}
\Pr\{\Delta(t)>d \} = \limty{\Tcal}\frac{1}{\Tcal}\int_{0}^{\Tcal} 1_{\{\age(t)>d\}} dt. 
\end{equation}
Therefore, the age violation probability is a time-average of the indicator age penalty function in \eqref{functional311}. 

\subsubsection*{Soft Updates} 
\begin{figure}[t]
	\centerline{\includegraphics[width=3in]{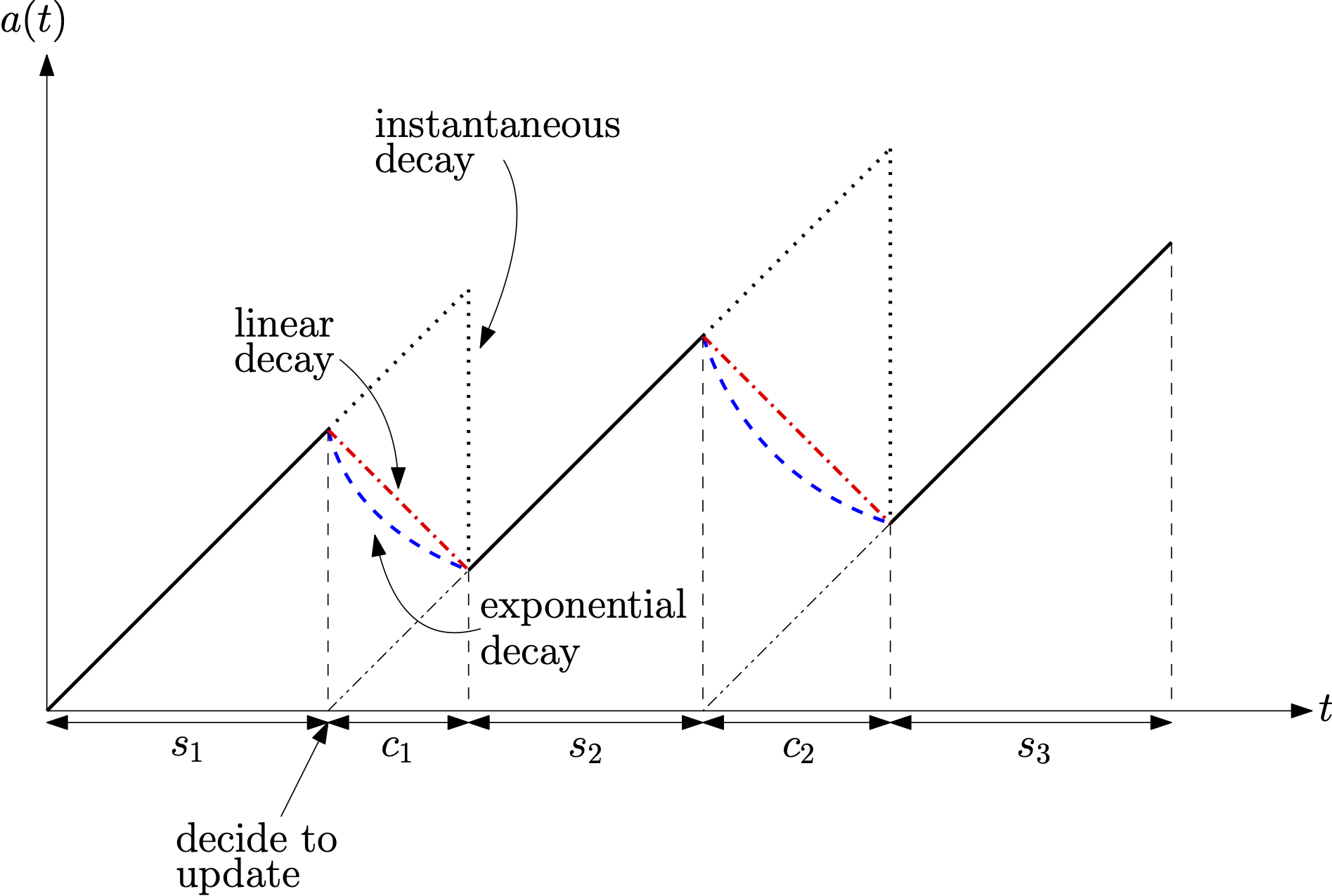}}
	\caption{Hard and soft updates: Hard updates take effect after a service time but cause instantaneous decay. Soft updates start taking effect right away but gradually over time giving rise to exponential and linear decay depending on the model.}
	\label{System_Model}
	\vspace{-5mm}
\end{figure}

Another instance where nonlinear age metric appears is in {soft updates} \cite{Bastopcu18, Bastopcu19b}. This setting models human interactions and social media interactions where an update consists of viewing and digesting many small pieces of information posted, that are of varying importance, relevance and interest to the monitor. Most of the AoI literature considers \emph{hard} updates, which are contained in information packets. A hard update \emph{takes effect} and reduces the age instantaneously
%to the age of the packet itself 
at the time of update's arrival at the monitor. This is denoted as \emph{instantaneous decay} in Fig.~\ref{System_Model}. The time for the update to take effect (denoted by $c_1$ for the first update) is the service time. 
% Essentially, this is the time for the update packet to \emph{travel} from the transmitter to the receiver, and when it arrives, it drops the age instantaneously. In contrast, in soft updates, the update begins reducing the age at the time of information source making a decision to update. However, the drop in age is not instantaneous, rather it is \emph{gradual} over time.
Essentially, this is the time for the transmitter to deliver the update packet to the monitor, and when it arrives, it drops the age instantaneously. In contrast, in soft updates, the update gradually reduces the age while the source is delivering the update. 
%However, the drop in age is not instantaneous, rather it is \emph{gradual} over the delivery time.
Depending on the model used, this gradual decrease may yield nonlinear instantaneous age functions.  

References \cite{Bastopcu18, Bastopcu19b} consider two models for the age function $a(t)$ of the soft update process: In the first model, the rate of decrease in age is proportional to the current age:
%\begin{align} \label{age-model1}
$da(t)/dt = -\alpha a(t)$,
%\end{align}
where $\alpha$ is a fixed constant. This is motivated by the fact that new information is most valuable when the current information is most aged, i.e., when the new information is most innovative. This model leads to an exponential decay in the age %(denoted by \emph{exponential decay} 
in Fig.~\ref{System_Model}. Note also that, the exponential decay in the age is consistent with information dissemination in human interactions as well as in social media feeds, where the most important information is conveyed/displayed first, reducing the age faster initially, and less important information is conveyed/displayed next, reducing the age slower subsequently. In the second model, the rate of decrease in age is not a function of the current age, rather it is constant:
%\begin{align} \label{age-model2}
%\frac{da(t)}{dt} = -\alpha
%\end{align}
$da(t)/dt = -\alpha$.
%where again $\alpha$ is a fixed constant. 
In this case, the age decreases linearly, as shown in Fig.~\ref{System_Model}.

\subsection{Functionals of  Age Processes}

More generally, one can use a non-decreasing functional\footnote{Recall that a functional is a mapping from functions to real numbers.
A functional $f$ is \emph{non-decreasing} if  $f(\{\Delta_1(t): t\ge0\}) \leq f(\{\Delta_2(t): t\ge0\})$   whenever $\Delta_1(t)\leq \Delta_2(t)$ for all $t\ge0$ \cite[p. 281]{StochasticOrderBook}.} $f(\{\Delta(t): t\ge0\})$ of the age process $\{\Delta(t): t\ge 0\}$ to represent the level of dissatisfaction for having aged information at the monitor, which is called an \emph{age penalty functional} \cite{BedewyJournal2017}. 
%The time-average age in \eqnref{time-average-age} and the peak age in \eqref{eq_PAoI} are two examples of age penalty functionals. 
Examples of age penalty functionals include the time-average age in \eqnref{time-average-age} and  the \emph{time-average age penalty}, defined by
\begin{equation}\label{functional3}
f_{\text{avg-penalty}}(\{\Delta(t): t\ge 0\})=\limty{\Tcal} \frac{1}{\Tcal}\int_{0}^{\Tcal} p(\Delta(t)) dt,
\end{equation}
where $p$ : $[0,\infty)\to \mathbb{R}$ can be any non-decreasing age penalty function. 

We note that the peak age in \eqref{eq_PAoI} is not an age penalty functional.%
% In fact, one can increase an age process to create some additional low peaks which do not exist in the original age process. Because of these additional low peaks, the average of the peaks in \eqref{eq_PAoI} may drop. Therefore, the peak age may reduce as the age process grows.
\footnote{The claim that ``the peak age is a non-decreasing functional of the age process" in \cite{Bedewy-SS-isit2016,Bedewy-SS-isit2017,BedewyJournal2017,BedewyJournal20172,Sun-UBK-aoi2018,maatouk2020status} is wrong. In fact, one can increase an age process $\age_1(t)$ to create another age process $\age_2(t)$ satisfying: (i) all peaks of $\age_1(t)$ are also peaks of $\age_2(t)$ and (ii) $\age_2(t)$ has some additional smaller peaks that do not exist in $\age_1(t)$. Even though $\age_2(t)\ge \age_1(t)$,  these additional low peaks can be chosen so that the average of peaks in $\age_2(t)$ is smaller than that in $\age_1(t)$. Therefore, the average peak age in \eqref{eq_PAoI} may drop even though the age process is increased.} Similarly, the peak age violation probability, i.e., the probability that the peak age exceeds a threshold $d$
\begin{equation}
\Pr[A_n>d]=\limty{\Tcal} \frac{1}{N(\Tcal)}\sum_{n=1}^{N(\Tcal)} 1_{\{A_n>d\}},
\end{equation}
and the peak age penalty
\begin{equation}
\limty{\Tcal} \frac{1}{N(\Tcal)}\sum_{n=1}^{N(\Tcal)} p(A_n) 
\end{equation}
with $p(\cdot)$ being a non-decreasing penalty function,
are not  age penalty functionals.

%\section{Updates through Queues}
% !TEX root = aoisurvey.tex
%\section{Updates through Queues}
\section{Age in Elementary Queues}
\label{sec:queues}
Over the past few years, there has been a considerable effort to understand AoI in simple queueing systems. In fact, this survey could be devoted solely to this subject. In this section, we  examine AoI when the network in Fig.~\ref{fig:cloudnet}(a) is an elementary queue and  sources submit updates  as a stochastic process, independent of the queue state. This model includes the M/M/1, M/D/1 and D/M/1 queues and versions of those queues that incorporate  preemption (in service or in waiting)  or blocking of new arrivals. 

While space considerations preclude an in-depth discussion of the entire literature of AoI in queues, there are a number of other notable contributions.  
For a single updating source, distributional properties of the age process were analyzed for the D/G/1 queue under first-come-first-served (FCFS) service \cite{Champati-AG-aoi2018}.  General queueing systems in the form of G/G/1/1 queues were analyzed in \cite{soysal18, soysal19}. Non-i.i.d.~service times modeled as a Gilbert-Elliott Markov chain with \emph{good} and \emph{bad} serving states were studied in \cite{Buyukates20a}. Packet deadlines were   found to improve AoI \cite{Kam-KNWE-isit2016deadline}. Age-optimal preemption policies were identified for updates with deterministic service times \cite{Wang-FY-spawc2018}.  AoI was evaluated in the presence of packet erasures at the M/M/1 queue output \cite{Chen-Huang-isit2016} and for memoryless arrivals to a two-state Markov-modulated service process \cite{Huang-Qian-globecom2017}. 

In Section~\ref{sec:single-server-queue}, we examine average age for a single source sending updates through a queue. We focus on arrival and services processes that are either deterministic or memoryless in order to characterize elementary properties of the average age. This is followed by Section~\ref{sect:minAchAge}, which uses zero-wait systems to derive age lower bounds, and Section~\ref{sec:multi-source-queue}. which examines age in queues that serve multiple sources. 

\subsection{Age in Single-source Single-server Queues}
\label{sec:single-server-queue}

This review is based chiefly on AoI results in  \cite{Kaul-YG-infocom2012,Costa-CE-IT2016management,Inoue-MTT-IT2019}. 
We start with variations on non-preemptive and preemptive single server queues, for which the representation  in \cite{Inoue-MTT-IT2019} of the age process $\age(t)$ by the point process $\set{(t_n',T_n):n=0,1,\ldots}$ has led to a panoply of results, including the extension to distributional results for the stationary age $\age(t)$ and the peak age $A_n$  and also generalization to GI/GI/1 queues.

% For each system, we find the server utilization that minimizes the average age $\age$. Numerical performance comparisons between queueing disciplines then follow.

Throughout this discussion, each  server has expected service time $\E{S}$ and each service system has \iid{} interarrival times with expected value $\E{Y}$. For consistency of presentation, $\lambda=1/\E{Y}$ is the arrival rate, $\mu=1/\E{S}$ is the service rate, and the system has offered load $\rho=\lambda/\mu$. Numerical comparisons will be presented in terms of the load $\rho$ with $1/\mu=1$.

For the FCFS M/M/1 queue
% in which update interarrival times $Y_j$ are independent and identically distributed (iid) exponential random variables with $\E{Y}=1/\lambda$ and  service times are \iid{} exponentials with  average service time $1/\mu$. In terms of the 
with offered load $\rho$, it was shown  \cite{Kaul-YG-infocom2012} using \eqnref{sysAge} that the average age is
\begin{align}
\age_{\text{M/M/1}} = \frac{1}{\mu}\paren{1+\frac{1}{\rho} +\frac{\rho^2}{1-\rho}}.
\eqnlabel{sysAgeMM1fcfs}
\end{align}
For fixed service rate $\mu$, the age-optimal utilization $\rho^{*}$ satisfies $\rho^{4} - 2\rho^3 + \rho^2 - 2\rho + 1 = 0$ and thus $\rho^*\approx 0.53$. The server is idle $\approx 47\%$ of the time. The optimal age is achieved by choosing a $\lambda$ that biases the server towards being busy only slightly more than being idle.
Note that we would want $\rho$ close to $1$ if we wanted to maximize the throughput, which is the number of packets delivered to the monitors every second. If we instead wanted to minimize packet delay, that is minimize the system time of a packet, we would want $\rho$ to be close to $0$. Analysis of the M/D/1 queue \cite{Inoue-MTT-IT2019} and the D/M/1 queue \cite{Yates-Kaul-IT2019} yielded 
\begin{align}
\age_{\text{M/D/1}}&=\frac{1}{\mu}
\paren{\frac{1}{2(1-\rho)}+\frac{1}{2}+\frac{(1-\rho)\exp(\rho)}{\rho}},\\
\age_{\text{D/M/1}}&=\frac{1}{\mu}\paren{\frac{1}{2\rho}+\frac{1}{1-\gamma(\rho)}},
\end{align}
where, in terms of the Lambert-W function, $\mathcal{W}(\cdot)$,
\begin{align}
%\beta = e^{-\mu(1-\beta)D} = -\rho \mathcal{W}\left(\frac{-e^{\frac{-1}{\rho}}}{\rho}\right)\label{eqn:beta},
\gamma(\rho) = -\rho \mathcal{W}\left({-\rho^{-1}e^{(-1/\rho)}}\right).
\label{eqn:lambert}
\end{align}

\begin{figure}[t]
\begin{center}
	\includegraphics{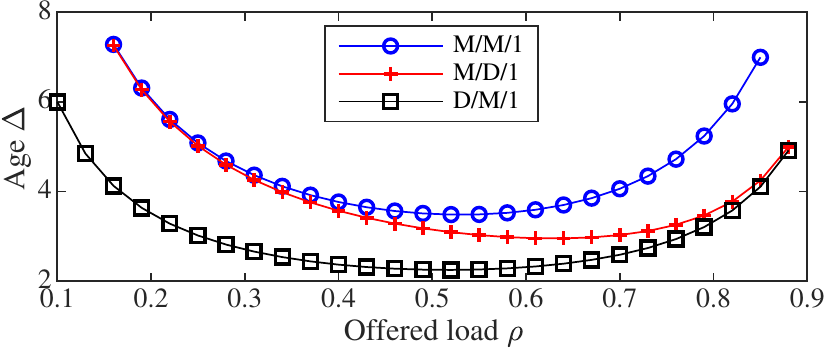}
	\caption{\small Time-average age as a function of offered load $\rho=\lambda/mu$ for the FCFS  M/M/1, M/D/1 and D/M/1 queues.
	The expected service time is $1/\mu = 1$.}
	\label{fig:compMM1MD1}
\end{center}
\vspace{-5mm}
\end{figure}

Fig.~\ref{fig:compMM1MD1} presents age comparisons of the M/M/1, M/D/1, and D/M/1 queues from \cite{Kaul-YG-infocom2012}. For each queue there is an age-minimizing offered load.
Among these FCFS queues, we observe that for each value of system load, D/M/1 is better than M/D/1, which is better than M/M/1. At low load, randomness in the interarrivals  dominates the average status-age.  At high load, M/D/1 and D/M/1 substantially outperform M/M/1  because the determinism in either arrivals or service helps to reduce the average queue length.  For each queue, we observe a unique value of $\rho$ that minimizes the average age.  
% For example, we recall that the age is minimized at a utilization $\rho^* = 0.53$ for the M/M/1 queue. We also note that the D/M/1  queue with optimal load $\rho^*=0.515$ achieves an average age  not very far from the $2/\mu$ lower bound for all memoryless service systems.  For the M/D/1 queue, we observe that $\rho=0.625$ minimizes the age $\age$. Just as in the M/M/1 case, there is an optimal update rate  for the given service facility.

What these FCFS queues make apparent is that the arrival rate can be optimized to balance update frequency against the possibility of congestion. This prompted the study of lossy queues that may discard an arriving update while the server was busy or replace an  older waiting update with a fresher arrival \cite{Kaul-YG-ciss2012,Costa-CE-isit2014,Costa-CE-IT2016management}. These strategies, identified as {\em packet management} \cite{Costa-CE-isit2014,Costa-CE-IT2016management}, include  the M/M/1/1 queue that  blocks and clears a new arrival while the server is busy, the M/M/1/2 queue that will queue one waiting packet but blocks an arrival when the waiting space is occupied, and the $\text{M/M/1/2}^*$ queue that will preempt a waiting packet with a fresh arrival.\footnote{While  Kendall notation $A$/$S$/$c$ is consistently used to signify the \emph{A}rrival process, the \emph{S}ervice  time, and the number of servers $c$, there is no consensus on a fourth entry for these systems. Here we (mostly) follow  
\cite{Costa-CE-IT2016management}, with the fourth entry classifying how arrivals access the servers: $\cdot$/$\cdot$/$c$ is a $c$ server system with an unbounded queue; $\cdot$/$\cdot$/$c$/$m$ indicates a system capacity of $m$ updates (i.e.~an FCFS waiting room of size $m-c$) with new arrivals blocked when the waiting room is full, and $\cdot$/$\cdot$/$c$/$m^*$ with $m=c+1$, indicates a single packet waiting room with preemption in waiting.  We then add the convention $\cdot$/$\cdot$/$c^*$ to signify that a new arrival preempts the oldest update in service. (Since preempted packets are discarded, the waiting room becomes irrelevant.)
% In  \cite{Costa-CE-IT2016management}, the fourth entry indicated the maximum number of updates in the system; the queue with a waiting room for one update was called M/M/1/2 and the M/M/1/2 queue supporting preemption in waiting was called $\text{M/M/1/2}^*$.  
Note that in \cite{Yates-Kaul-IT2019}, the  $\text{M/M/1}^*$ and $\text{M/M/1/2}^*$ queues were called LCFS-S and LCFS-W, with S and W denoting preemption, S in Service and W in Waiting. In both \cite{Costa-CE-IT2016management,Yates-Kaul-IT2019}, it was assumed that obsolete updates were discarded. In \cite{Inoue-MTT-IT2019}, the fourth entry was the size of the waiting room, LCFS designated queues in which a new arrival moved in front of any waiting updates and the prefixes P and NP indicated whether the service was preemptive (P)  or Non-Preemptive (NP), i.e. does the new arrival go immediately into service or simply to the head of the waiting line. \cite{Inoue-MTT-IT2019} also used suffixes (C) and (D) to indicate whether the queue was work Conserving or whether obsolete updates were Discarded. Thus the $\text{M/M/1/2}^*$ queue in \cite{Costa-CE-IT2016management} was the M/M/1 LCFS-W queue in \cite{Yates-Kaul-IT2019} and was the  M/M/1/1 NP-LCFS (D) queue  in \cite{Inoue-MTT-IT2019}.}

Another system  in this category is the LCFS queue with preemption in service that permits a new arrival to preempt an update in service.
Extending the notation introduced in \cite{Costa-CE-IT2016management}, we  call this an $\text{M/M/1}^*$ queue. 
These systems were shown \cite{Costa-CE-IT2016management} to achieve average ages
\begin{subequations}\eqnlabel{MMages}
\begin{IEEEeqnarray}{rCl}
\age_{\text{M/M/1}^*}&=&\frac{1}{\mu}\paren{1+\frac{1}{\rho}},\eqnlabel{MM1p-age}\\
\age_{\text{M/M/1/1}}&=&\frac{1}{\mu}\paren{1+\frac{1}{\rho}+\frac{\rho}{1+\rho}},\\
\age_{\text{M/M/1/2}^*}&=& 
    \frac{1}{\mu}\paren{1+\frac{1}{\rho}+\frac{\rho^2(1+3\rho+\rho^2)}{(1+\rho+\rho^2)(1+\rho)^2}}
    %\bracket{1+\frac{\rho}{(1+\rho)^2}}},
    \\
\age_{\text{M/M/1/2}}&= &
    \frac{1}{\mu}\paren{1+\frac{1}{\rho}+\frac{2\rho^2}{1+\rho+\rho^2}}.
\end{IEEEeqnarray}
\end{subequations}
From \eqnref{MMages}, simple algebra will verify that 
\begin{subequations}
\begin{align}
\age_{\text{M/M/1}^*}&\le \age_{\text{M/M/1/1}},\\
\age_{\text{M/M/1}^*}&\le \age_{\text{M/M/1/2}^*}\le \age_{\text{M/M/1/2}}.
\end{align}
\end{subequations}
However, the age performance of the M/M/1/1 system is less easy to classify. Specifically, as $\rho$ increases, the relative performance of the M/M/1/1 system improves.  In terms of the golden ratio $\rho^*=(1+\sqrt{5})/2$, we have that 
\begin{subequations}
\begin{align}
 \rho\le 1/\rho^*&\imply 
     \age_{\text{M/M/1/2}}\le \age_{\text{M/M/1/1}}; \\   \frac{1}{\rho^*}\le\rho\le\rho^*&\imply 
    \age_{\text{M/M/1/2}^*}\le \age_{\text{M/M/1/1}}\le \age_{\text{M/M/1/2}};\\
     \rho^*\le\rho &\imply 
    \age_{\text{M/M/1/1}}\le \age_{\text{M/M/1/2}^*}.
\end{align}
\end{subequations}

\begin{figure}[t]
\begin{center}
	\includegraphics{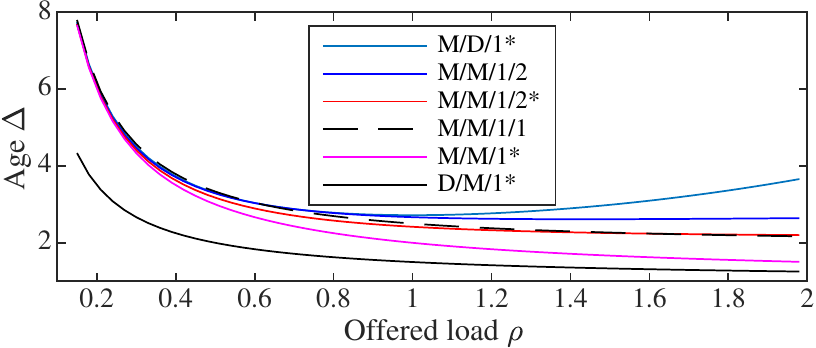}
	\caption{\small Time-average age as a function of offered load $\rho$ for the FCFS  M/M/1/2 and M/M/1/1 blocking queues, the  LCFS $\text{M/D/1}^*$, $\text{M/M/1}^*$,  and $\text{D/M/1}^*$ queues supporting preemption in service, and the $\text{M/M/1/2}^*$ queue supporting preemption in waiting.  The expected service time is $1/\mu=1$.}
	\label{fig:MM1preempt}
\end{center}
\vspace{-5mm}
\end{figure}

Figure~\ref{fig:MM1preempt} compares the average age for these queues. At low load, all of these queues achieve essentially the same average age  $(1+1/\rho)/\mu$ as the M/M/1 and M/D/1 FCFS queues evaluated in Fig.~\ref{fig:age}.  When the queue is almost always empty, the LCFS ability for the freshest update to jump ahead of older update packets is negated. However, at high loads, packet management ensures that that the $\text{M/M/1/2}^*$, M/M/1/1 and $\text{M/M/1}^*$ queues have average age that  decreases with offered load.\footnote{Because congestion is avoided by blocking packets, $\age_{\text{M/M/1/2}}$ avoids blowing up as $\rho\to\infty$. However,  it  achieves a minimum age of $\age=2.61$ at $\rho=1.427$ and then becomes an increasing function for $\rho>1.427$. For large $\rho$, the M/M/1/2 queue admits its next update too quickly.}
 In fact, as the arrival rate $\lambda\to\infty$, the age $\age_{\text{M/M/1}^*}$ will approach the $2/\mu$  lower bound for exponential service systems because  bombarding the server with new update packets ensures that a fresh status update packet will enter the waiting room the instant before each service completion.

One can conclude for memoryless service systems that preemption of old updates by new always helps. However, the comparisons are muddier when we compare buffering vs.~discarding. This is particularly true when we compare $\age_{\text{M/M/1}}$, which buffers every update, against the ages $\age_{\text{M/M/1/1}}$ and $\age_{\text{M/M/1/2}}$. 

We also note, however, that the apparent superiority of preemption in service is somewhat misleading; this property holds for memoryless service times, but not in general. For example,  the $\text{M/D/1}^*$ and $\text{D/M/1}^*$ queues, both of which support preemption in service, have average ages  \cite[pp.~8318]{Inoue-MTT-IT2019}
\begin{align}\eqnlabel{md-dm-preemptive}
\age_{\text{M/D/1}^*}&=\frac{1}{\mu}\frac{\exp(\rho)}{\rho},\qquad
\age_{\text{D/M/1}^*}=\frac{1}{\mu}
\paren{1+\frac{1}{2\rho}}.
\end{align}
In \eqnref{md-dm-preemptive}, and also in Fig.~\ref{fig:age2}, we see that the average age $\age_{\text{D/M/1}^*}$ of the D/M/1 preemptive server is monotonically decreasing in the offered load $\rho$. This is because no matter how high the arrival rate $\lambda$ is (and thus how fast packets are being preempted), the departure rate is $\mu$ as long as an update is in service. By contrast, in the preemptive $\text{M/D/1}^*$ queue, $\age_{\text{M/D/1}^*}$ has a minimum at $\rho=1$ and increases without bound for $\rho>1$. With deterministic unit-time service and arrival rate $\lambda=\rho$,  an update completes service with probability $e^{-\rho}$. As $\rho$ becomes large, too many updates are preempted, and the system thrashes, with updates being preempted before  they can complete service and be delivered to the monitor.
\subsection{Zero-wait updates}
\label{sect:minAchAge}
When the update generator (source) can neither observe nor control the state of the packet update queue, the optimal load $\rho^*$ strikes a balance between overloading the queue and leaving the queue idle. Here we derive lower bounds to the age $\age$ by considering a system in which the update generator observes the state of the packet update queue so that a new status update arrives just as the previous update packet departs the queue. Since each delivered update packet is as fresh as possible, the average age for this system is a lower bound to the age for any queue in which updates are generated as a stochastic process independent of the current queue state.

In this zero-wait systems, the update service times $S_n$ are \iid{} with first and second moments $\E{S}$ and $\E{S^2}$. Referring to the age function $\age(t)$ in Figure~\ref{fig:age},  $t_n = t'_{n-1}$. This implies update $n$ has interarrival time $Y_n=S_{n-1}$, zero waiting time, and system time $T_n=S_n$. Further, $\E{YT} = \E{Y_nT_n} = \E{S_{n-1}S_n}= (\E{S})^2$. From \Eqnref{sysAge},
the average age becomes
\begin{align}
\age^* &= \frac{1}{\E{S}}\left[\frac{E[S^2]}{2} + (\E{S})^2\right]\eqnlabel{minAchAge_Exp}.
\end{align}
It follows that for a system with memoryless service times with $\E{S}=1/\mu$, the minimum average age is 
\begin{equation}
    \age_{\cdot/M/1}^*=2/\mu.
\end{equation}
Moreover, non-negativity of the variance of $S$ implies $\E{S^2}\ge (\E{S})^2$,  Thus, for all service time distributions with $\E{S}=1/\mu$, \eqnref{minAchAge_Exp} yields  the lower bound  
\begin{align}
\age^*  
  &\ge \frac{3\E{S}}{2}=\frac{3}{2\mu}.\eqnlabel{minAchAge_Exp_LB}
\end{align}
This lower bound is tight as it 
is achievable when the service times are deterministic. 
\subsection{Multiple Sources at a Single-server Queue}
\label{sec:multi-source-queue}
When updates have stochastic service times, AoI analysis of multiple updating sources sharing a simple queue has proven challenging and there have been relatively few contributions\footnote{We will see in Sections~\ref{sec:queue-networks} and~\ref{sec:wireless} that there has been much more interest in using complex scheduling to support multiple sources.}
\cite{Yates-Kaul-isit2012, Huang-Modiano-isit2015,Najm-Telatar-aoi2018,Yates-Kaul-IT2019,Moltafet-LC-comm2020}. In these papers, each source $i$ generates updates as an independent Poisson process of rate $\lambda_i$ and the service time $S$ of an update has expected value $1/\mu$. Thus source $i$ has offered load $\rho_i=
\lambda_i/\mu$ and the total offered load is $\rho=\sum_{i=1}^N\rho_i$.

Extending the single-source age analysis in \cite{Najm-YS-isit2017},   reference \cite{Najm-Telatar-aoi2018} derived the average age and average peak age of each source. With $P_{\lambda}=\E{e^{-\lambda S}}$ denoting  the Laplace transform of the service time $S$ at  $\lambda=\lambda_1+\cdots+\lambda_N$, \cite{Najm-Telatar-aoi2018} showed that user $i$ in the M/G/1/1 queue has average age $\age_i$ and average peak age $\PAoI_i$ given by
\begin{align}
    \age_i&=\frac{1}{\lambda_iP_{\lambda}},\qquad
    \PAoI_i =\frac{1}{\lambda_iP_{\lambda}}
    +\frac{\E{Se^{-\lambda S}}}{P_{\lambda}}.
\end{align}

The first age analysis of the multi-source FCFS M/M/1 queue appeared in \cite{Yates-Kaul-isit2012}, which propagated to  \cite{Yates-Kaul-IT2019}. Unfortunately this analysis had an error, as observed in \cite{Moltafet-LC-comm2020}. In a corrected analysis using SHS \cite{Kaul-Yates-ciss2020}, it was shown that with $\rho_{-i}\equiv\rho-\rho_i$ and 
\begin{align}
\mathcal{E}_i \equiv \frac{1 + \rho - \sqrt{(1+\rho)^2 - 4\rho_{-i}}}{2\rho_{-i}},
\end{align}
source $i$ has average age
\begin{align}
\age_i = \frac{1}{\mu}\left[\frac{1-\rho}{(\rho - \rho_{-i} \mathcal{E}_i) (1 - \rho \mathcal{E}_i)} + \frac{1}{1-\rho} + \frac{\rho_{-i}}{\rho_i}\right].\label{eqn:agetwouser}
\end{align}
This result can be shown to be numerically identical to the independently derived result in \cite[Theorem~1]{Moltafet-LC-comm2020}. Also, this corrected average age is sufficiently close to the original claim in \cite{Yates-Kaul-isit2012, Yates-Kaul-IT2019} that the qualitative comparisons between FCFS and LCFS service for multiple sources in \cite{Yates-Kaul-IT2019} remain valid.

Heterogeneous users sharing a single queue have been analyzed in~\cite{Najm-NT-IT2020,Kaul-Yates-isit2018priority,MaatoukISIT2019, Huang-Modiano-isit2015}. The users are modeled as having different priorities in~\cite{Najm-NT-IT2020,Kaul-Yates-isit2018priority,MaatoukISIT2019}. Work~\cite{Najm-NT-IT2020} further considers different queueing disciplines, specifically, FCFS for the lower priority stream and LCFS with preemption allowed in service for the higher priority stream. In \cite{Huang-Modiano-isit2015} users' updates may have different service time distributions.
%\bigskip

%\subsection{Age in Multiple-Source Queues}

\section{Age in Queueing Networks}
\label{sec:queue-networks}
We now examine updates from one or more sources traversing a network of queues. Our starting point is the single-source, single-hop parallel-server network depicted in Fig.~\ref{Fig:sysMod1}, consisting of one queue with buffer size $B\geq0$ feeding  $c$ servers. If $B$ is finite, each packet arriving to a full buffer is either dropped or replaces an existing packet in the system. If $B=0$, the system can keep at most $c$ packets that are being processed by the servers. For this class of systems, the challenge of AoI analysis  is out-of-order packet delivery; a packet in service can be rendered obsolete by a delivery of another server. 

Section~\ref{sec:SimpleScheduler} considers elementary versions of the parallel server system, specifically, the M/M/2 with $c=2$ servers and $B=\infty$ buffer space, the M/M/$\infty$ queue with $c=\infty$ servers in which every update immediately goes into service, and the M/M/$c^*$ system with a scheduler that will preempt the oldest update in service if a new update arrives when all servers are busy. This is followed by Section~\ref{sec:parallel-scheduling}, which examines of age-optimal scheduling for the parallel server system, and Section~\ref{sec:other-queue-networks}, which presents AoI results for update scheduling in other network of queues settings.

%Needs a better heading:
\subsection{The Parallel-Server Queue}
\label{sec:SimpleScheduler}
The observation in \cite{Kam-KE-isit2014diversity} that a single server queue is not representative of networks in which packets may be delivered via multiple paths prompted  AoI analysis of a queue with $c$ parallel servers depicted in Fig.~\ref{Fig:sysMod1}. The initial work 
\cite{Kam-KE-isit2013random,Kam-KNE-IT2016diversity} addressed the extreme cases, namely the M/M/2  and M/M/$\infty$ queues.  Since the M/M/2 queue has infinite buffers, the M/M/2 average age is similar to M/M/1 in that it is subject to congestion induced by waiting updates.  The M/M/2 queue age performance is also penalized by obsolete updates remaining in service. Nevertheless, it was found M/M/2 service still could reduce the AoI by an approximate factor of 2 relative to M/M/1 \cite{Kam-KNE-IT2016diversity}. 

The M/M/$\infty$ age analysis in \cite{Kam-KNE-IT2016diversity} is complex, and average age does not reduce to a simple formula. However, with arrival rate $\lambda$ and per-server rate $\mu$ service, the  exact AoI $\age_{\text{M/M/$\infty$}}$ was found to be subject to the reasonably tight lower and upper bounds
\begin{align}
\frac{1}{\mu}\paren{\frac{1}{\rho}+\frac{1+\rho+\rho^2}{(1+\rho)^3}}
\le  \age_{\text{M/M/$\infty$}}\le
\frac{1}{\mu}\paren{1+\frac{1}{\rho}}\eqnlabel{MMinfty-bounds}
\end{align}
% \begin{IEEEeqnarray}{rCl}
%     \age_{\text{LB},\infty}&=&
%  \frac{1}{\mu}\paren{\frac{1}{\rho}+\frac{1+\rho+\rho^2}{(1+\rho)^3}},\\
%  \age_{\text{UB},\infty}&=&\frac{1}{\mu}\paren{1+\frac{1}{\rho}}.
%  \end{IEEEeqnarray}
 It is not surprising that the right side upper bound  in \eqnref{MMinfty-bounds}
 equals the age $\age_{\text{M/M/1}^*}$ in \eqnref{MM1p-age}  since the monitor in the M/M/$\infty$ system  can choose to mimic the $\text{M/M/1}^*$ update delivery process by discarding all service completions except those by the freshest update in the system.  
 
 In \cite{Yates-isit2018}, the $\text{M/M/$c$}^*$ preemptive parallel server system was analyzed using SHS. In this system,  a fresh update arriving when all servers are busy preempts the oldest update in service.  For this system, the average age was found to be
\begin{IEEEeqnarray}{rCl}
    \age_{\text{M/M/$c$}^*}
&=&\frac{1}{\mu}\bracket{\frac{1}{c}\prod_{i=1}^{c-1}\frac{\rho}{i+\rho}+\frac{1}{\rho}+\frac{1}{\rho}\sum_{l=1}^{c-1}\prod_{i=1}^l\frac{\rho}{i+\rho}}.\IEEEeqnarraynumspace
\end{IEEEeqnarray}
When $\rho=\lambda/\mu\ll c$, one can expect that M/M/$c^*$ system will approximate the infinite server system.
Once again, since the M/M/$\infty$ system can mimic this system by discarding service completions other than those of the $c$ most recent arrivals, $\age_{\text{M/M/$\infty$}}\le\age_{\text{M/M/$c$}^*}$.

% Properties of PAoI were also studied for various M/M/1 queues that support preemption of updates in service or discarding of updates that find the server busy \cite{Costa-CE-IT2016management,Kavitha-AS-arxiv2018}.  

\subsection{Scheduling for Parallel Servers}
\label{sec:parallel-scheduling}
%In the sequel, we will introduce optimal scheduling policies that 

%The comparison among different scheduling policies naturally leads to the following question: Can we identify the optimal scheduling policy for minimizing the AoI in queueing networks? This question was answered in by using sampling-path and stochastic ordering techniques. 

%When update transmission times over network links are exponentially distributed,  were used \cite{Bedewy-SS-isit2016,Bedewy-SS-isit2017,Bedewy-SS-arxiv2017} to show that a preemptive Last-Generated, First-Served (LGFS) policy results in smaller age processes at all nodes of the network than any other causal policy. 
% {\BLUE The scheduling results in this subsection involve single-server/multi-server, single-hop/multi-hop, single-user/multi-user. Hence, the old title ``Scheduling in Queues" is better. Fig. 6 only covers part of the results in this subsection.}

\begin{figure}
\centering 
\includegraphics[width=0.45\textwidth]{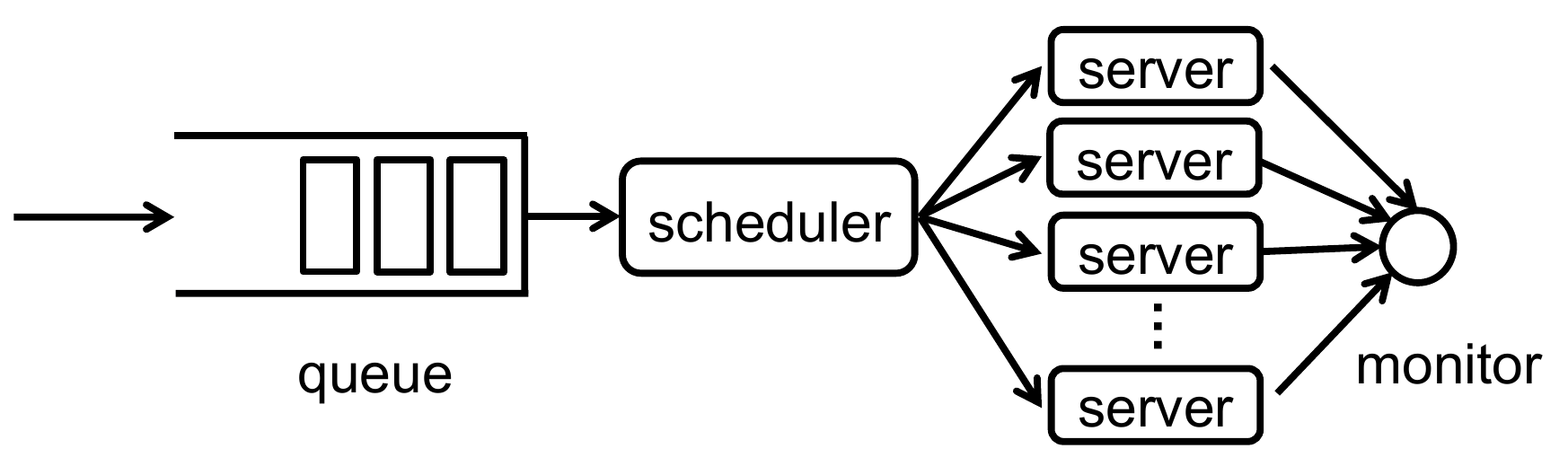} 
\vspace{-0mm}
\caption{Status updates in a single-hop, multi-server queueing network.}
% work--efficiency ordering holds for any priorities of the jobs.
\label{Fig:sysMod1} 
\vspace{-5mm}
\end{figure} 

We now discuss how to schedule update transmissions  to minimize AoI in the parallel server system of Fig.~\ref{Fig:sysMod1}.   In \cite{Bedewy-SS-isit2016,Bedewy-SS-isit2017,BedewyJournal2017,BedewyJournal20172,Sun-UBK-aoi2018,maatouk2020status}, (near) age-optimal scheduling results were established by using sample-path arguments. These results hold for out-of-order packet arrivals and quite general AoI metrics (e.g., age penalty functions and functionals), for which AoI analysis is  challenge. Therefore, AoI analysis and age-optimal scheduling provide complementary perspectives on status updates systems. 

%We will see that the LCFS policy, and more generally, a scheduling policy named Last-Generated, First-Served (LGFS), can minimize the AoI in several queueing networks. 

%Let us first consider the preemptive LGFS policy, in which a fresh incoming packet  preempts an old packet under service. 

%The following definitions of stochastic orders \cite{} are needed to compare two age processes:   
%A random variable ${X}$ is said to be \emph{stochastically smaller} than another random variable ${Y}$, denoted by ${X}\leq_{\text{st}}{Y}$, if $\mathbb{P}({X}>x) \leq \mathbb{P}({Y}>x)$ for all~$x\in \mathbb{R}$.
%A set $\mathcal{U} \subseteq \mathbb{R}^n$ is called \emph{upper}, if $\bm{y} \in \mathcal{U}$ whenever $\bm{y}\geq \bm{x}$ and $\bm{x} \in \mathcal{U}$. 
%A random vector $\bm{X}$ is said to be \emph{stochastically smaller} than another random vector $\bm{Y}$, denoted by $\bm{X}\leq_{\text{st}}\bm{Y}$, if $\mathbb{P}(\bm{X}\in \mathcal{U}) \leq \mathbb{P}(\bm{Y}\in \mathcal{U})$ for all upper sets ~$\mathcal{U}\subseteq \mathbb{R}^n$. %If $\bm{X}\leq_{\text{st}}\bm{Y}$ and $\bm{X}\geq_{\text{st}}\bm{Y}$, then $\bm{X}$ and $\bm{Y}$ follow the same distribution, denoted by $\bm{X}=_{\text{st}}\bm{Y}$. 

Let $\pi$ represent a scheduling policy that  determines (i) the packets being sent by the servers over time and (ii) packet droppings and replacements when the queue buffer is full. %Let $\Pi$ denote the set of \emph{online} policies in which the scheduling decisions are made based on the history and current states of the system.
Let $\Pi$ denote the set of \emph{causal} policies, in which the scheduling decisions are made based on the state history and current state of the system. 

The age processes of different scheduling policies are compared in terms of stochastic ordering \cite{StochasticOrderBook}: An age process $\{\Delta_1(t): t\ge 0 \}$ is \emph{stochastically smaller} (denoted $\leq_{\text{st}}$) than another age process $\{\Delta_2(t): t\ge 0 \}$ 
%(denoted by $\{\Delta_1(t): t\ge 0 \}\leq_{\text{st}}\{\Delta_2(t): t\ge 0\}$) 
if, and only if, 
\begin{align}\label{ch02thm1eq201}
\!\!\E{f(\{\Delta_{1}(t): t\ge 0\})}\leq\E{f(\{\Delta_{2}(t): t\ge 0\})}\!\!
\end{align}
for all non-decreasing functionals $f$, provided the expectations in \eqref{ch02thm1eq201} exist. Suppose that packet $i$ is generated at the source node at time $S_{i}$, arrives at the queue at time $C_{i}$, and is delivered to the monitor at time $D_i$ such that $S_i\leq C_i\leq D_i$. The sequences $\{S_1, S_2,\ldots\}$ and $\{C_1,C_2,\ldots\}$ are \emph{arbitrarily given}. Hence, {out-of-order arrivals}, e.g., $S_i < S_{i+1}$ but $C_i > C_{i+1}$, may occur. Let $\mathcal{I}=\{S_{i}, C_{i},~ i=1,2,\ldots\}$ denote the packet generation and arrival times, and $[\{\Delta_{\pi}(t): t\ge 0\}\vert\mathcal{I}]$ denote the age process of policy $\pi$ for given packet generation/arrival times $\mathcal{I}$.

We consider a scheduling policy named \emph{Last-Generated, First-Served (LGFS)} \cite{Bedewy-SS-isit2016}, in which the last generated
packet is served first, with ties broken arbitrarily. In the Preemptive LGFS (P-LGFS) policy, a fresh  packet can preempt an old packet that is under service. The preempted packets can be dropped or stored back to the queue; whether the preempted packets are dropped or stored back to the queue does not affect the age performance of the P-LGFS policy. In the Non-Preemptive LGFS (NP-LGFS) policy, each server must complete sending the current packet before starting to serve a fresher packet; in order to reduce the AoI, the freshest packet should be kept in the queue when packet dropping/replacement occurs. 

\begin{figure}[t]
\centering 
\includegraphics[width=0.3\textwidth]{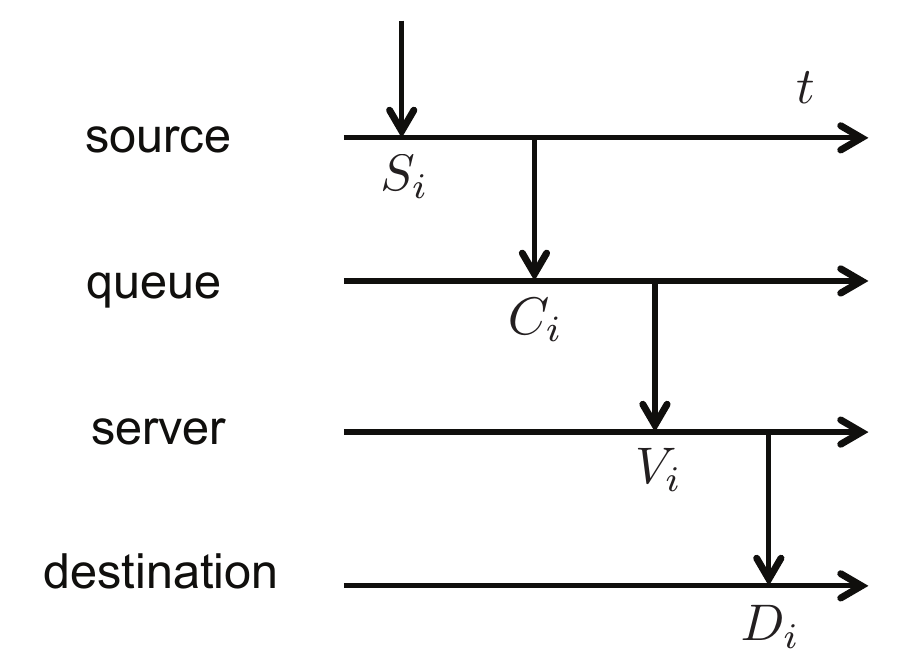} \caption{An illustration of the packet arrival times $S_{i}$, $C_{i}$, $V_{i}$, and $D_{i}$. A packet $i$ arrives at the source at time $S_i$, at the queue at time $C_i$, at a server in the queue at time $V_i$, and at the monitor at time $D_i$.}
% work--efficiency ordering holds for any priorities of the jobs.
\label{fig_times1} 
\end{figure} 

%We first present a scheduling result for \emph{i.i.d.} exponential service times. In particular, 
%If the service times are exponentially distributed, the preemptive LGFS policy is age-optimal, as stated below. 
\begin{theorem}\label{ch02thm1}\cite{Bedewy-SS-isit2016,BedewyJournal2017}
If the packet service times are exponentially distributed, \iid{} across servers and time, then for all $c\geq1$, $B\geq0$, $\mathcal{I}$, and $\pi\in\Pi$
\begin{equation}\label{ch02thm1eq10}
[\{\Delta_{\text{P-LGFS}}(t): t\ge 0\}\vert\mathcal{I}]\leq_{\text{st}}[\{\Delta_{\pi}(t): t\ge 0\}\vert\mathcal{I}].
\end{equation}
In other words, for all $c\geq1$, $B\geq0$, packet generation times $\mathcal{I}$, non-decreasing functionals $f$, and policies $\pi\in\Pi$,
\begin{IEEEeqnarray}{rCl}
\label{ch02thm1eq20}
\E{f(\{\Delta_{\text{P-LGFS}}(t)\!:\! t\ge 0\})\vert\mathcal{I}}&\le&
%\min_{\pi\in\Pi}
\E{f(\{\Delta_{\pi}(t): t\ge 0\})\vert\mathcal{I}},\IEEEeqnarraynumspace
\end{IEEEeqnarray}
provided the expectations in \eqref{ch02thm1eq20} exist.
\end{theorem}

According to Theorem \ref{ch02thm1}, if the service times are i.i.d. exponentially distributed, then for \emph{arbitrarily given} packet generation times $(S_1, S_2, \ldots)$, packet arrival times $(C_{1}, C_{2}, \ldots)$, buffer size $B$, and number of servers $c$, the P-LGFS policy results in a smaller  age process than \emph{any} other causal policy. 
In addition, \eqref{ch02thm1eq20} implies that the P-LGFS policy also minimizes \emph{any} non-decreasing functional of the age process, including the time-average age \eqnref{time-average-age} and time-average age penalty  \eqref{functional3}.

If the packets arrive at the queue in the same order of their generation times, i.e., $(S_{i}- S_{i+1})(C_{i} - C_{i+1}) \geq 0$ for all $i$, then the LGFS policy becomes the LCFS policy. Hence, Theorem \ref{ch02thm1} suggests that the P-LCFS policy is age-optimal for in-order arrivals, which agrees with the AoI analysis above. However, the P-LGFS policy and the P-LCFS policy may not be optimal for minimizing the peak age in \eqref{eq_PAoI}, which is not an age penalty functional.

A weaker version of Theorem \ref{ch02thm1} is to consider the mixture over the realizations of the generation and arrival times in $\mathcal{I}$. In this case, it follows from \eqref{ch02thm1eq10} that
\begin{align}
&\{{\Delta}_{\text{P-LGFS}}(t): t\geq 0\} \leq_{\text{st}} \{{\Delta}_\pi(t): t\geq 0\}.
\end{align} 
Hence, the condition on given realization of $\mathcal{I}$ in \eqref{ch02thm1eq10} and \eqref{ch02thm1eq20} can be removed. 

Next, we will show that the NP-LGFS policy is near age-optimal for a class of New-Better-than-Used service time distributions. 
A non-negative random variable $X$ is said to be \emph{New-Better-than-Used (NBU)} \cite{StochasticOrderBook} if for all $t,\tau\geq0$
\begin{equation}\label{NBU_Inequality}
\bar{F}(\tau +t)\leq \bar{F}(\tau)\bar{F}(t),
\end{equation} 
where $\bar{F}(t) = \Pr[X>t]$.
Examples of NBU distributions include constant, exponential, shifted exponential, geometric,  gamma, and negative binomial distributions.

\begin{figure}[t]
\centering 
\includegraphics[width=0.3\textwidth]{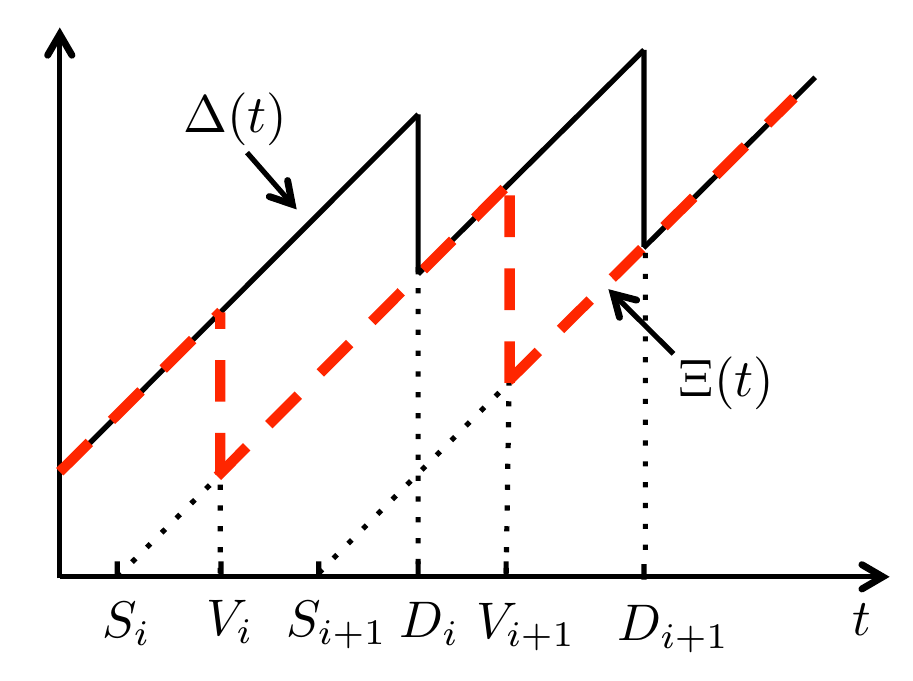} \caption{Age of Served Information $\Xi (t)$ as a lower bound of the age $\age(t)$.}
% work--efficiency ordering holds for any priorities of the jobs.
\label{fig_times2} 
\end{figure}

To show the NP-LGFS policy can come close to age-optimal, we need to construct a lower bound of the age  $\Delta (t) $, as shown below: 
 Let $V_{i}$ denote the time that packet $i$ is assigned to a server, i.e., the service starting time of packet $i$, as  illustrated in Fig.~\ref{fig_times1}. By definition, one can get
$S_{i}\leq C_{i}\leq V_{i}\leq D_{i}$. Consider now an alternative age metric called  \emph{Age of Served Information} \cite{Sun-UBK-aoi2018}, which is defined as
 % by the packets that have started service up to time $t$: 
%The {Age of Served Information} is defined as
\begin{align}\label{eq_age_served}
\Xi (t) \!=\! t\! -\! \max\{S_{i}\!:\! V_{i} \leq t\}.
\end{align}
Age of served information is the time difference between the current time $t$ and the generation time of the freshest packet that has \emph{started} service by time $t$. That is, $\Xi(t)$ is the age process seen by an observer  positioned at the entrance to the server. Since $\age(t)$ is the age process seen by an observer positioned at the exit of the server, 
% Recall that the age $\age(t)$ is the time difference between the current time $t$ and the generation time of the freshest packet that has \emph{completed} service by time $t$. 
$\Xi (t)\leq \age(t)$, 
as shown in Fig.~\ref{fig_times2}. Let $\Pi_{\text{NP}}\subset{\Pi}$ denote the  set of non-preemptive, {causal} policies.

%(i) The buffer size $B$ is no less than $c$, and (ii)  if the buffer is full when a packet arrives, the oldest packet (among the incoming packet and the packets in the queue) is dropped. These two conditions ensure that the freshest $c$ arrived packets are maintained in the queue until they are (i) assigned to the servers or (ii) replaced by fresher incoming packets. 

\begin{theorem}\label{ch02thmNBU1}\cite{BedewyJournal2017}
If  (i) the packet service times are NBU, {i.i.d.} across servers and time, and (ii) the freshest packet is kept in the queue when packet dropping/replacement occurs, 
then for all $c\geq1$, $B\geq 1$, $\mathcal{I}$, and $\pi\in\Pi_{\text{NP}}$
\begin{equation}\label{ch02thm1NBUeq1}
[\{\Xi _{\text{NP-LGFS}}(t): t\ge 0\}\vert\mathcal{I}]\leq_{\text{st}}[\{\Delta_{\pi}(t): t\ge 0\}\vert\mathcal{I}],
\end{equation}
or equivalently, for all $c\geq1$, $B\geq 1$,  $\mathcal{I}$, and non-decreasing functional $f$
\begin{align}\label{ch02thm1NBUeq2}
&\E{f(\{\Xi _{\text{NP-LGFS}}(t): t\ge 0\})\vert\mathcal{I}}\nn
&\qquad\leq \min_{\pi\in\Pi_{\text{NP}}}\E{f(\{\Delta_{\pi}(t): t\ge 0\})\vert\mathcal{I}} \nn
&\qquad\leq \E{f(\{\Delta_{\text{NP-LGFS}}(t): t\ge 0\})\vert\mathcal{I}},
\end{align}
provided the expectations in \eqref{ch02thm1NBUeq2} exist. 
\end{theorem}

Because $S_i=D_i-V_{i}$ is the service time of packet $i$, by choosing $f$ as the time-average age in \eqnref{time-average-age},  it follows from  \eqref{ch02thm1NBUeq2} that
\begin{align}\label{ch02thm1NBUeq3}
&\min_{\pi\in\Pi_{\text{NP}}}\! \limsup_{\Tcal\rightarrow \infty} \frac{1}{\Tcal} \E{\int_{0}^{\Tcal} \Delta_{\pi}(t) dt\bigg| \mathcal{I}} \!
\nn
&\qquad\leq \limsup_{\Tcal\rightarrow \infty} \frac{1}{\Tcal} \E{\int_{0}^{\Tcal} \Delta_{\text{NP-LGFS}}(t) dt\bigg| \mathcal{I}} \nn
&\qquad\leq \min_{\pi\in\Pi_{\text{NP}}}\! \limsup_{\Tcal\rightarrow \infty} \frac{1}{\Tcal} \E{\int_{0}^{\Tcal} \Delta_{\pi}(t) dt\bigg| \mathcal{I}}+ \E{S},
\end{align}
where $\E{S}$ is the mean service time of a packet.
According to  \eqref{ch02thm1NBUeq3}, if the service times are \iid~NBU and the queue can store at least one packet ($B\geq 1$), then the expected time-average age of the NP-LGFS policy is within a small additive gap from the optimum, where the gap $\E{S}$ is invariant of the packet generation and arrival times $\mathcal{I}$, the number of servers $c$, and the buffer size $B$. 

Age-optimal scheduling results for single-source single-server queues can be also obtained from Theorems \ref{ch02thm1}-\ref{ch02thmNBU1} by choosing $c=1$. 

\subsection{Scheduling for Multiple Hops and Multiple Sources}
\label{sec:other-queue-networks}

The scheduling results in Theorems \ref{ch02thm1}-\ref{ch02thmNBU1} have been extended to a few other network settings. In \cite{Bedewy-SS-isit2017,BedewyJournal20172}, the scheduling of a single packet flow in multi-hop queueing networks was studied. If service times are i.i.d. exponentially distributed, the P-LGFS policy is optimal for minimizing the age processes at all nodes of the network. In addition, the NP-LGFS policy is near age-optimal for i.i.d. NBU service times. 

%Multiple Sources:
Age-optimal scheduling of multiple flows with synchronized arrivals in a single-hop queue was investigated in \cite{Sun-UBK-aoi2018}. The authors first proposed a scheduling policy named Preemptive, Maximum Age First, Last Generated First Served (P-MAF-LGFS), in which the last generated packet from the flow with the maximum age is served first among all packets of all flows, with ties broken arbitrarily. When the packet service times are {i.i.d.} exponentially distributed and {the queue has one server},  the P-MAF-LGFS policy  minimizes the stochastic process $\{p_t(\mathbf\age(t)): t\ge 0\}$ in terms of stochastic ordering, where $p_t(\cdot)$ is a time-dependent, symmetric, and non-decreasing penalty function of the age vector $\mathbf\Delta(t)=[\Delta_1(t),\ldots,\Delta_N(t)]$ of the flows. 

In addition, a Non-Preemptive, Maximum Age of Served Information First, Last Generated First Served (NP-MASIF-LGFS) policy was introduced in \cite{Sun-UBK-aoi2018}. In the NP-MASIF-LGFS policy, the last generated packet from the flow with the maximum age of served information is served first among all packets of all flows, with ties broken arbitrarily.
When the packet service times are {i.i.d.} NBU and the queue has multiple servers, the NP-MASIF-LGFS policy is within a small additive gap from the optimum for minimizing the total time-average age. 
If multiple servers are idle, these servers are assigned to process different flows. Therefore, the behavior of the NP-MASIF-LGFS policy is similar to the maximum age matching approach \cite{Tripathi-Moharir-globecom2017} for orthogonal channel systems.

Motivated by \cite{Bruce2005}, a notion of lexicographic age optimality, or simply lex-age-optimality, was introduced in \cite{maatouk2020status} for scheduling multiple flows with diverse priority levels. A lex-age-optimal scheduling policy first minimizes the AoI of high-priority flows, and then, within the set of optimal policies for high-priority flows, achieves the minimum AoI metrics for low-priority flows. When the packet service times are {i.i.d.} exponentially distributed, 
a scheduling policy named Preemptive Priority, Maximum Age First, Last-Generated, First-Served (PP-MAF-LGFS) 
 was shown to be lex-age-optimal in a {single-hop, single-server queue}. In the PP-MAF-LGFS policy, the system will serve an informative packet\footnote{A packet is \emph{informative} if it is fresher than any delivered packet.} that is selected as follows: (i) among all flows with informative packets, pick the class of flows with the highest priority; (ii) next, among the flows from the selected priority class, pick the flow with the maximum age, with ties broken arbitrarily; (iii) finally, among the informative packets from the selected flow, pick the last generated informative packet, with ties broken arbitrarily. The scheduling result in \cite{maatouk2020status} complements the AoI analysis of preemptive priority service systems for multiple sources \cite{Najm-NT-IT2020,Kaul-Yates-isit2018priority,MaatoukISIT2019}.

\section{Resource Constrained Updating}
\label{sec:resource-constrained}
%%%%%Begin-Sennur

%\subsection{AoI in Energy Harvesting Systems: Introduction}

In this section, we focus on resource constrained updating, where the ability to make an update is further constrained due to external factors. A prominent example of this is the setting of energy harvesting transmitters. When transmitters (e.g., sensors) rely on energy harvested from nature to transmit their status updates, they cannot transmit continuously; otherwise they may run out of energy and risk having overly stale status updates at the monitor. Therefore, the fundamental question is  how to manage the harvested energy to send timely status updates.

We will start with an overview of the area and then go on to summarize  \cite{jing-age-online, arafa-age-online-finite, jing-age-erasures-infinite-jour, arafa-age-erasure-no-fb, arafa-age-erasure-fb} that focus on online generate-at-will policies with various battery capacities (one unit battery, arbitrary but finite size battery, and infinite battery) over noiseless and erasure channels in Sections~\ref{sec_prfct} and Sections~\ref{sec_erase}, respectively. In these sections, there is no server and the randomness is purely due to uncertain energy arrivals. In Section~\ref{sec_servers}, we summarize works \cite{Yates-isit2015, farazi-K-B-ISIT2018-aoi-eh-preempt, Farazi-KB-aoi2018}, where there is an additional server and also the updates arrive exogenously as opposed to being generated at-will. Finally, in Section~\ref{sec_advanced}, we summarize works that considered more advanced settings, including the use of ARQ/HARQ feedback, reinforcement learning, energy harvesting via wireless energy transfer, two-way data exchange with power splitting, UAV-assisted systems, cognitive radio systems, intermittent sensing and transmission, and caching systems with energy harvesting.

\subsection{Overview}
There have been a number of works studying AoI with energy harvesting under various assumptions \cite{Yates-isit2015, elif_age_eh, farazi-K-B-ISIT2018-aoi-eh-preempt, Farazi-KB-aoi2018,  liu-age-eh-sensing, arafa-age-2hop, arafa-aoi-eh-2hop-inf-battery, arafa-age-var-serv, elif-age-Emax, jing-age-online, jing-age-error-infinite-no-fb, jing-age-error-infinite-w-fb, jing-age-erasures-infinite-jour, baknina-age-coding, ZhengTWC2019, baknina-updt-info, arafa-age-rbr, arafa-age-sgl, arafa-age-online-finite, elif-age-online-threshold, bacinoglu-aoi-eh-finite-gnrl-pnlty, arafa-age-erasure-no-fb, arafa-age-erasure-fb, krikidis-aoi-wpt, chen-aoi-eh-mac-het,Leng-Yener-2019TCCN, stamatakis-aoi-eh-alarm, rafiee-aoi-eh-drop, ozel-aoi-eh-sensing, dong-aoi-mse, Tunc-Panwar-2019}. With the exception of \cite{arafa-age-var-serv,krikidis-aoi-wpt,Tunc-Panwar-2019}, an underlying assumption in these works is that energy expenditure is normalized, i.e., sending one status update consumes one energy unit. References \cite{Yates-isit2015, elif_age_eh} consider a sensor with infinite battery, with \cite{Yates-isit2015} focusing on online policies under stochastic service times, and \cite{elif_age_eh} focusing on both offline and online policies with zero service times, i.e., with updates being transmitted instantly.\footnote{In fact, most studies on AoI with energy harvesting sensors focus on the zero service time model in which transmission times are negligible relative to the large inter-transmission times induced by energy constraints.} Reference \cite{liu-age-eh-sensing} studies the effect of sensing costs on AoI with an infinite battery sensor transmitting through a noisy channel. Using a harvest-then-use protocol, \cite{liu-age-eh-sensing} presents a steady state analysis of AoI under both deterministic and stochastic energy arrivals. The offline policy in \cite{elif_age_eh} is extended to non-zero, but fixed, service times in \cite{arafa-age-2hop} for both single and multi-hop settings, and in \cite{arafa-age-var-serv} for energy-controlled variable service times.

The online policy in \cite{elif_age_eh} is analyzed through a dynamic programming approach in a discretized time setting, and is shown to have a threshold structure, i.e., an update is sent only if the age grows above a certain threshold and energy is available for transmission. Motivated by such results for the infinite battery case, \cite{elif-age-Emax} then studies the performance of online threshold policies for the finite battery case under zero service times. Reference \cite{jing-age-online} proves the optimality of online threshold policies under zero service times for the special case of a unit-sized battery, via tools from renewal theory. It also shows the optimality of best-effort online policies, where updates are sent over uniformly-spaced time intervals if energy is available, for the infinite battery case. Reference \cite{arafa-aoi-eh-2hop-inf-battery} shows that such a best-effort policy is optimal in the online case of multihop networks, thereby extending the offline work in \cite{arafa-age-2hop}. Best-effort is also shown to be optimal, for the infinite battery case, when updates are subject to erasures, with and without erasure feedback, in \cite{jing-age-error-infinite-no-fb, jing-age-error-infinite-w-fb, jing-age-erasures-infinite-jour}.

Under the same system model of \cite{jing-age-error-infinite-no-fb}, reference \cite{baknina-age-coding} analyzes  the best-effort online policy as well as the save-and-transmit online policy in which the sensor saves some energy in its battery before attempting transmission, for the purpose of coding to combat channel erasures. A slightly different system model is considered in \cite{Farazi-KB-aoi2018}, in which status updates' arrival times are exogenous, i.e., their measurement times are not controlled by the sensor. With a finite battery, and stochastic service times, reference \cite{Farazi-KB-aoi2018} employs tools from stochastic hybrid systems to analyze the long-term average AoI. The work in \cite{farazi-K-B-ISIT2018-aoi-eh-preempt} considers a similar queuing framework as in \cite{Farazi-KB-aoi2018} and studies the value of preemption in service on AoI. Reference \cite{zheng-aoi-eh-queue} also considers a similar approach as in \cite{Farazi-KB-aoi2018, farazi-K-B-ISIT2018-aoi-eh-preempt} under general energy and data buffer sizes. An interesting approach is followed in \cite{baknina-updt-info} where the idea of sending extra information, on top of the measurement status updates, is introduced and analyzed for unit batteries and zero service times.

Optimality of threshold policies for finite batteries with online energy arrivals has been shown in \cite{arafa-age-rbr, arafa-age-sgl, arafa-age-online-finite} using tools from renewal theory and a Lagrangian framework, which provides closed-form solutions of the optimal thresholds. This has also been shown independently and concurrently in \cite{elif-age-online-threshold} using tools from optimal stopping theory. Reference \cite{bacinoglu-aoi-eh-finite-gnrl-pnlty} shows the optimality of threshold policies under general age-penalty functions. Online policies for unit batteries with update erasures also have been shown to have a threshold structure in \cite{arafa-age-erasure-no-fb, arafa-age-erasure-fb}.

Other frameworks that combine AoI with energy harvesting include wireless power transfer \cite{krikidis-aoi-wpt}, multiple access channels \cite{chen-aoi-eh-mac-het}, cognitive radio systems \cite{Leng-Yener-2019TCCN}, monitoring with priority \cite{stamatakis-aoi-eh-alarm}, operational and sensing costs \cite{rafiee-aoi-eh-drop, ozel-aoi-eh-sensing}, and trade-offs between AoI and distortion \cite{dong-aoi-mse}.

%\subsection{Status Updating over Energy Harvesting Noiseless Channels} \label{sec_prfct}
\subsection{Energy Harvesting Noiseless Channels} \label{sec_prfct}
\begin{figure}[t]
\center
\includegraphics[scale=.4]{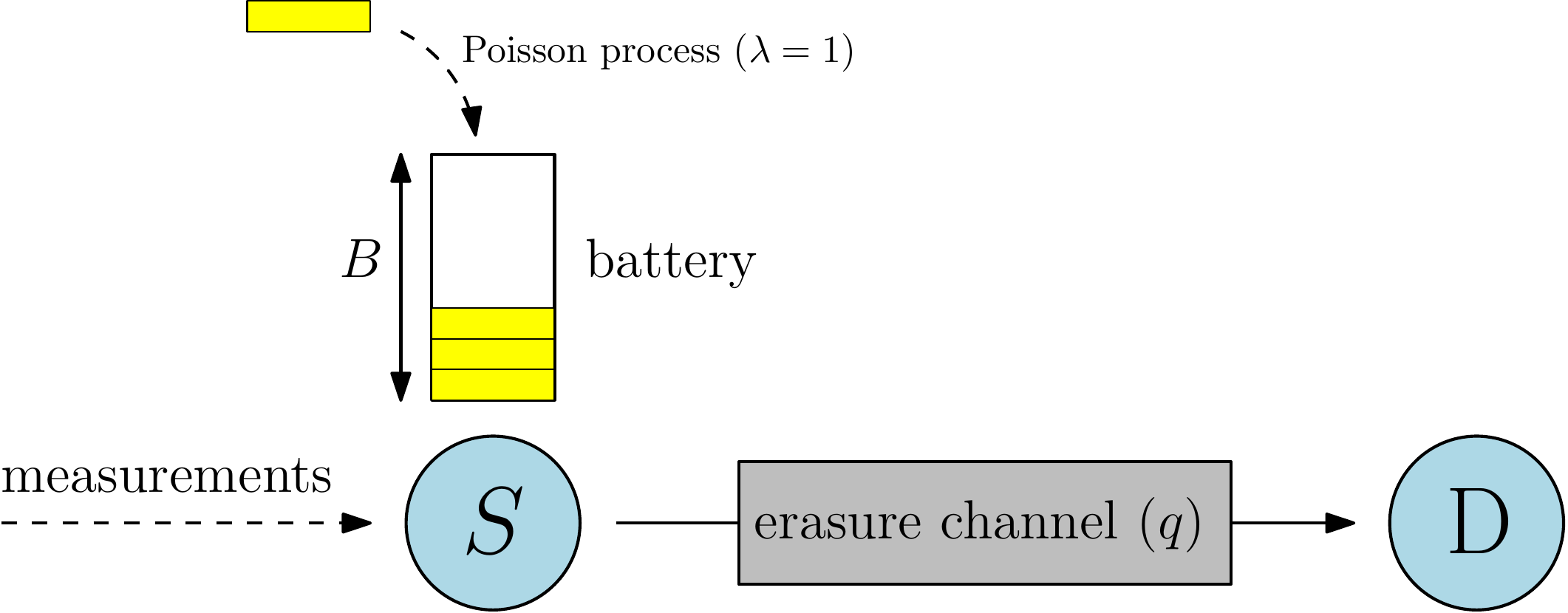}
\caption{System model for status updating over erasure channels. Updates packets are erased with probability $q$. The channel is noiseless when $q=0$.}
\label{fig_sys_mod_erase}
\end{figure}

In this section, we discuss the results reported in \cite{jing-age-online, arafa-age-online-finite}. In these works,  the channel is noiseless with packet erasure probability $q=0$ and energy arrives in units according to a Poisson process with normalized rate $\lambda=1$ arrival per unit time; see Fig.~\ref{fig_sys_mod_erase}. The energy expenditure is  normalized in the sense that one update transmission consumes one energy unit. In addition, the transmission time of an update is negligible, i.e, zero.\footnote{Normalized arrival rates and zero transmission times are without loss of generality. Extensions to non-normalized arrival rates and  fixed nonzero transmission times can be directly derived, at the expense of increased AoI as the arrival rate decreases and/or the transmission time increases.} Hence, updates are sent as a point process $s_1,s_2,\ldots$ such that $s_i$ is the time at which the sensor acquires (and transmits) the $i$th measurement update. 

At time $s_i^-$, the instant before  transmission $i$, the sensor must have an energy unit. Thus, with $\mathcal{E}(t)$ denoting the energy in the battery at time $t$, we have the  energy causality constraint
\begin{align} \label{eq_en_caus}
\mathcal{E}\left(s_i^-\right)\geq1,\quad\forall i.
\end{align}
We assume that the system starts with an empty battery at time $0$. The battery evolves  over time as
\begin{align} \label{eq_battery_inc}
\mathcal{E}\left(s_i^-\right)=\min\left\{\mathcal{E}\left(s_{i-1}^-\right)-1+\mathcal{A}\left(x_i\right),B\right\},
\end{align}
where $x_i\triangleq s_i-s_{i-1}$,  $\mathcal{A}(x_i)$ denotes the number of energy arrivals in $[s_{i-1},s_i)$, and $B$ is the battery capacity. Note that $\mathcal{A}(x_i)$ is a Poisson random variable with expected value  $x_i$. We denote by $\mathcal{F}$, the set of feasible transmission times $\{s_i\}$ described by (\ref{eq_en_caus}) and (\ref{eq_battery_inc}) in addition to an empty battery at time 0, i.e., $\mathcal{E}(0)=0$.

% \begin{figure}[t]
% \center
% \includegraphics[scale=0.5]{images/figs-eh/sys_mod_prfct}
% \caption{System model for status updating over perfect (zero-error) channels including a sensor reporting to a destination.}
% \label{fig_sys_mod_prfct}
% \end{figure}

\begin{figure}[t]
\center
\includegraphics[scale=.4]{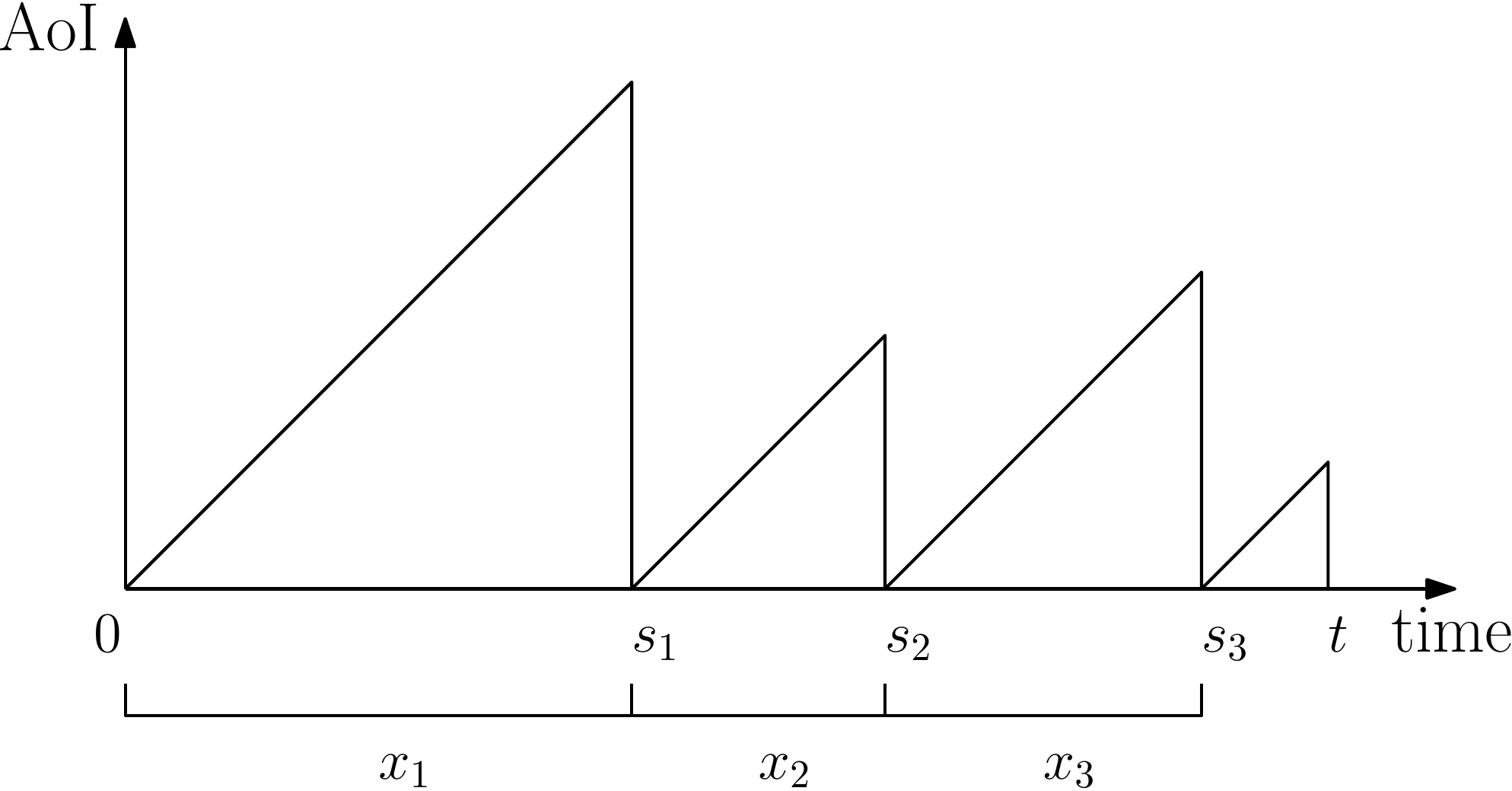}
\caption{Example of the age evolution versus time with $n(t)=3$.}
\label{fig_age_xmpl}
\end{figure}

Let $n(t)$ denote the total number of updates sent by time $t$. We are interested in minimizing the average AoI represented by the area under the age evolution curve vs.~time, see Fig.~\ref{fig_age_xmpl} for a possible sample path with $n(t)=3$. At time $t$, this area is given by
\begin{align} \label{eq_aoi}
Q(t)\equiv
%triangleq
\frac{1}{2}\sum_{i=1}^{n(t)}x_i^2+\frac{1}{2}\left(t-s_{n(t)}\right)^2.
\end{align}
The goal is to choose a set of feasible transmission times $\{s_1,s_2,s_3,\dots\}\in\mathcal{F}$ such that the long-term average AoI is minimized. Equivalently, one can optimize the inter-update times $\{x_1,x_2,x_3\dots\}$ to do so. Therefore, the goal is to characterize the optimal long-term average AoI $\age^*(B)$ as a function of the battery size $B$
%, $\rho(B)$ 
by solving
\begin{align} \label{opt_main_inc}
\age^*(B)\equiv
%\triangleq
\min_{\{x_i\}\in\mathcal{F}}\limsup_{T\rightarrow\infty}\frac{1}{T}\E{Q(T)}.
\end{align}
% Next, we present the solutions for $B=\infty$ and $B<\infty$ separately over the next subsections.

% \subsubsection*{The case $B=\infty$} %\label{sec_B_inf_prfct}

When $B=\infty$, i.e., the battery size is infinite, no energy overflow will happen. Let us define the following policy:
\begin{definition}[Best-Effort Uniform Updating \cite{jing-age-online}] The sensor is scheduled to send a new update at $s_n=n$, $n=1,2,3,\dots$. The sensor performs the task as scheduled if $\mathcal{E}(s_n^-)\geq1$. Otherwise, it stays silent until the next scheduled sampling time. \label{def_bu}
\end{definition}

Clearly, the best-effort uniform (BU)  updating policy is always feasible. One of the main results of \cite{jing-age-online} is showing that it is also optimal for $B=\infty$, i.e., proving the following theorem:
\begin{theorem}[\hspace{-.0025in}\cite{jing-age-online}]
The best-effort uniform updating policy is optimal for $B=\infty$, with $\age^*(\infty)=1/2$.
\end{theorem}

%In order to minimize the long-term average AoI 
When $B$ is finite, the status update policy should try to prevent battery overflows, since wasted energy leads to performance degradation. On the other hand, owing to the nature of AoI, one should also try to send updates as uniformly as possible (as seen in the $B=\infty$ case). The optimal policy would then strike a balance between these objectives.

One main attribute of the optimal policy's behavior is that it has a {\it renewal} structure. In particular, if we define $\left(\mathcal{E}(t),\age(t)\right)$ as the state of the system at time $t$, where $\age(t)$ is the AoI, then we have the following result:
\begin{theorem}[\hspace{-.0025in}\cite{arafa-age-online-finite}] \label{thm_inc_rnwl}
For  $B<\infty$, the optimal  update policy for problem (\ref{opt_main_inc}) is a renewal policy for which visits to state $(0,0)$ form a renewal process.
\end{theorem}

Using the result of the theorem, reference \cite{arafa-age-online-finite} shows that the optimal renewal-type status update  policy has a {\em multi threshold} structure: a new status update is transmitted only if the AoI grows above a certain threshold that depends on the amount of energy available in the battery. Such thresholds are found via a Lagrangian approach in closed-form. Further, it is shown that the thresholds are monotonically decreasing in the energy available. That is, the sensor is less eager to send an update when it has relatively lower energy available in its battery than it is when it has relatively higher energy available.

\subsection{Energy Harvesting Erasure Channels} \label{sec_erase}

In this section, we discuss the results reported in \cite{jing-age-erasures-infinite-jour, arafa-age-erasure-no-fb, arafa-age-erasure-fb}. The system model is similar to that described in Section~\ref{sec_prfct}, except {\it status updates are subject to erasures.} Specifically, the communication channel between the sensor and the destination is modeled as a time-invariant noisy channel, in which each update transmission gets erased with probability $q\in(0,1)$, independently from other transmissions, see Fig.~\ref{fig_sys_mod_erase}. We differentiate between two main cases in our treatment:
\begin{enumerate}
\item {\it No updating feedback.} In this case, the sensor has no knowledge of whether an update is successful. It can only use the up-to-date energy arrival profile and status updating decisions as well as the statistical information, such as the energy arrival rate and the erasure probability of the channel, to decide the upcoming updating time points.
\item {\it Perfect updating feedback}. In this case, the sensor receives an instantaneous, error-free, feedback when an update is transmitted. Therefore, it can decide when to update next based on the feedback information, along with the information it uses for the no feedback case.
\end{enumerate}

Since each update transmission is not necessarily successful, we denote by $\{l_i\}$ the set of update transmission times, and by $\{s_i\}$ the times of that are {\it successful}. Therefore, in general,  $\{s_i\}\subseteq\{l_i\}$. The energy causality constraint in (\ref{eq_en_caus}) now becomes
\begin{align}
\mathcal{E}\left(l_i^-\right)\geq1,\quad\forall i,
\end{align}
and the battery evolution in (\ref{eq_battery_inc}) becomes
\begin{align}
\mathcal{E}\left(l_i^-\right)=\min\{\mathcal{E}\left(l_{i-1}^-\right)-1+\mathcal{A}(x_i),B\},\quad\forall i,
\end{align}
where $x_i\triangleq l_i-l_{i-1}$ now denotes the inter-update attempt delay. We assume $s_0=l_0=0$ without loss of generality, i.e., the system starts with fresh information at time $0$. We denote by $\mathcal{F}_q$, the set of feasible transmission times $\{l_i\}$ described by (\ref{eq_en_caus}) and (\ref{eq_battery_inc}) in addition to an empty battery at time 0, i.e., $\mathcal{E}(0)=0$.\footnote{In \cite{jing-age-erasures-infinite-jour}, it is assumed that $\mathcal{E}(0)=1$ to simplify the analysis. For $\mathcal{E}(0)=0$, the same results would follow after slightly modifying the proofs. We set $\mathcal{E}(0)=0$ for consistency.}

\begin{figure}[t]
\center
\includegraphics[scale=1]{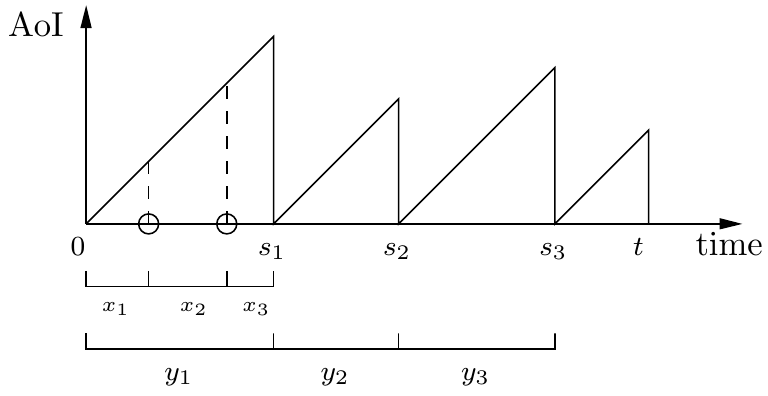}
\caption{Age evolution versus time with $n(t)=3$ successful updates. Circles denote failed attempts. In this example, the first update is successfully received after three update attempts.}
\label{fig_age_xmpl_erasure}
\end{figure}

Let us denote by $y_i\triangleq s_i-s_{i-1}$ the {\it successful} inter-update delay, and by $n(t)$ denote the number of updates that are {\it successfully} received by time $t$. We are interested in the average AoI given by the area under the age evolution curve, see Fig.~\ref{fig_age_xmpl_erasure}, which is given by
\begin{align} \label{eq_aoi_area}
Q(t)=\frac{1}{2}\sum_{i=1}^{n(t)}y_i^2+\frac{1}{2}\left(t-s_{n(t)}\right)^2.
\end{align}
The goal is to choose a set of feasible transmission times $\{l_1,l_2,l_3,\dots\}\in\mathcal{F}_q$, or equivalently $\{x_1,x_2,x_3,\dots\}$, such that the long-term average AoI is minimized. Therefore, the goal is to characterize the optimal long-term average AoI $\rho_q^\omega(B)$ as a function of the battery size $B$  by solving
\begin{align} \label{opt_main_erase}
\age_q^\omega(B)\equiv\min_{\{x_i\}\in\mathcal{F}_q}\limsup_{T\rightarrow\infty}\frac{1}{T}\E{Q(T)},
\end{align}
where the superscript $\omega\equiv\text{noFB}$ in the case without updating feedback, and $\omega\equiv\text{wFB}$ in the case with perfect feedback. In the following, we present the solution of (\ref{opt_main_erase}) for $B=\infty$ followed by the special case of $B=1$, in view of the two feedback availability cases.

\subsubsection*{The case $B=\infty$}

For the case $B=\infty$, without updating feedback, \cite{jing-age-erasures-infinite-jour} shows that the best-effort uniform (BU) updating policy, as per Definition~\ref{def_bu}, is optimal. While for the scenario with perfect feedback, \cite{jing-age-erasures-infinite-jour} proposes a {\it best-effort uniform with retransmission (BUR) updating policy} and shows its optimality. Optimality in both cases is shown by first evaluating a lower bound on the long-term average AoI, and then showing that it can be achieved by best-effort uniform updating.

Reference \cite{jing-age-erasures-infinite-jour} proposes a novel {\it virtual policy} based approach to prove its results. Specifically, for both BU and BUR updating policies, a sequence of virtual policies defined by a time parameter $T_0$  is constructed. These policies are strictly suboptimal to their original counterparts, but eventually converge to them as $T_0\to\infty$. Leveraging the virtual policies, the effects of battery outages and updating errors could be decoupled in the performance analysis. Finally, the long-term average AoI under the virtual policies is shown to converge to the corresponding lower bound, which implies the optimality of the original policy.

%We highlight the main results over the next two subsections.

\paragraph{Updating Without Feedback} For this case, we introduce the following virtual policy:

\begin{definition}[BU-ER$_{T_0}$ \cite{jing-age-erasures-infinite-jour}] \label{def:bu-er}
The sensor performs BU updating until the battery level after sending an update becomes zero for the first time, or until time $T_0^+$, in which case the sensor depletes its battery. After that, when the battery level becomes higher than or equal to 1 after a successful update for the first time, the sensor reduces the battery level to one, and then repeats the process.
\end{definition}

Observe that as $T_0\rightarrow\infty$, the BU-ER$_{T_0}$ updating policy becomes a BU updating policy. One can then show the following:
\begin{theorem}[\hspace{-.0025in}\cite{jing-age-erasures-infinite-jour}]
As $T_0\rightarrow\infty$, the BU-ER$_{T_0}$ updating policy becomes AoI-optimal, with \begin{align}
\age_q^{\text{noFB}}(\infty)=\frac{1+q}{2(1-q)}.
\end{align}
\end{theorem}

\paragraph{Updating With Perfect Feedback} With perfect updating feedback, the sensor has the choice to retransmit the update immediately or wait and update later, thus leading to optimal solutions different from the no feedback case. Let us define the following {\it BUR updating policy:}

\begin{definition}[BUR Updating \cite{jing-age-erasures-infinite-jour}]
The sensor is scheduled to send new updates at $s_n={n}/(1-q)$, $n=1,2,3,\dots$. The sensor keeps sending updates at $s_n$ until an update is successful or until it runs out of battery. Otherwise, the sensor keeps silent until the next scheduled status update time.
\end{definition}

Now let us introduce the following virtual policy:
\begin{definition}[BUR Energy Removal (BUR-ER$_{T_0}$) \!\!\! \cite{jing-age-erasures-infinite-jour}]
The sensor performs BUR updating policy until the battery level after sending an update becomes zero for the first time, or until time $T_0^+$, in which case the sensor depletes its battery after a successful update at $T_0$. After that, when the battery level becomes higher than or equal to 1 after a successful update for the first time, the sensor reduces the battery level to 1, and then repeats the process.
\end{definition}

Observe that as $T_0\rightarrow\infty$, the BUR-ER$_{T_0}$ updating policy becomes a BUR updating policy. One can then show the following:
\begin{theorem}[\hspace{-.0025in}\cite{jing-age-erasures-infinite-jour}] \label{2lemma:upperbound}
As $T_0\rightarrow\infty$, the BUR-ER$_{T_0}$ updating policy becomes AoI-optimal, with \begin{align}
\age_q^{\text{wFB}}(\infty)=\frac{1}{2(1-q)}. 
\end{align}
\end{theorem}

\subsubsection*{The case $B=1$}

We now focus on the special finite battery case of $B=1$ in which one update completely depletes the battery. Similar to the finite battery case analysis of Section~\ref{sec_prfct}, it will be shown that an {\it erasure-dependent threshold policy} is optimal for the case without feedback. For the case with perfect feedback, the focus will be on a class of policies denoted {\it threshold-greedy policies.} We have the following structural result:

\begin{theorem}[\hspace{-.0025in}\cite{arafa-age-erasure-no-fb, arafa-age-erasure-fb}]
The optimal status updating policy that solves problem (\ref{opt_main_erase}) for $B=1$, and for both cases with and without updating feedback, is a renewal policy in the sense that the \emph{actual} update times sequence $\{s_i\}$ forms a renewal process, and the \emph{actual} inter-update times $y_i$'s are i.i.d.
\end{theorem}

The renewal structure in the above theorem greatly reduces the complexity of the problem. Now we only need to look at one renewal interval and optimize the updating policy over it. How this is done depends on whether there is feedback, as we discuss next.

\paragraph{Updating Without Feedback} Reference \cite{arafa-age-erasure-no-fb} shows that the optimal policy in this case is given by an {erasure-dependent} threshold policy in which a new status update is sent only if the AoI grows above a certain threshold that depends on the value of the erasure probability $q$. It is also shown that the optimal threshold is non-increasing in $q$, which is quite intuitive, since the sensor should be more eager to send new updates if the erasure probability is high, so that when the update is eventually received successfully the AoI would not be large.
\paragraph{Updating With Perfect Feedback}
%\paragraph adds a : at the end
Reference \cite{arafa-age-erasure-fb} focuses on an intuitive class of policies in which the first update attempt has a threshold structure, and the subsequent attempts, if the first is not successful, follow a greedy structure. This class is intuitive because if the first update is unsuccessful, then the AoI has already grown to a relatively high value, which urges the sensor to transmit its subsequent updates as soon as energy is available. It is then shown that this class of {\em threshold-greedy policies} represent an
equilibrium 
%fixed-point 
solution of the problem in the sense that if the first update attempt is threshold-based then the following attempts should be greedy, and conversely if the second update attempt onwards are all greedy then the first one should be threshold-based. The optimal threshold-greedy policy is then fully characterized.

While many useful takeaway points can be drawn from the above discussions in Sections~\ref{sec_prfct} and \ref{sec_erase}, where the focus has been on generate-at-will policies, a crucial one is that {\it greedy status updating, whenever energy is available, is not always optimal.} Rather, it is optimal to evenly spread out the status updates over time, to the extent allowed by  energy availability and energy causality. This is achieved by best effort-based policies when $B=\infty$ and threshold-based policies when $B<\infty$.

%\subsection{Status Updating over Energy Harvesting Channels with Servers} \label{sec_servers}
\subsection{Energy Harvesting Channels with Servers} \label{sec_servers}
While the previously discussed studies largely focused on a generate-at-will source and zero service delays, several papers including the original work in \cite{Yates-isit2015} consider a setting with stochastic service delays. In \cite{Yates-isit2015}, the energy harvesting process $H(t)$ is assumed to be ergodic with rate $\eta$, the source is assumed to have infinite battery capacity.
%RY this sentence gives the wrong impression:
%and the source receives new samples at rate $\lambda$. 
The source is also assumed to know the state of the server and can time its updates relative to the  service completions. A $\beta$-minimum update policy was developed to minimize the average age. This policy avoids transmitting an update immediately after an update with fast service completion since the payoff from this subsequent update will be small whereas the cost is fixed. On the other hand, if an update has a slow service completion, a subsequent update is transmitted immediately since the payoff in that case is higher. This policy, which counterintuitively leaves the server idle for periods of time even if the sensor has sufficient energy to transmit an update, was shown to outperform ``best effort'' and ``fixed delay'' policies.

A similar setting was considered in \cite{Farazi-KB-aoi2018}, where the average age was characterized as a function of the information and energy arrival rates, $\lambda$ and $\eta$, respectively, as well as the battery capacity $B$. In this study, the source was assumed to always submit new updates immediately to the server. New updates enter service if the server is idle and has sufficient energy to service the packet. If the server is busy or does not have sufficient energy, the update is dropped. This paper leveraged SHS 
%framework first used for the analysis of average age in \cite{Yates-Kaul-IT2019} 
to determine the average AoI for two cases corresponding to whether the server is able or unable to harvest energy during service. If the server is unable to harvest energy while a packet is in service, it was shown that the average age satisfied
\begin{align}
\Delta & = \begin{cases}
 \frac{2B\rho^2 + (2B+2)\rho + B+2}{\mu[B\rho^2+(B+1)\rho]} & \beta = \rho, \\[2pt]
 \frac{(2\rho^2 + 2\rho + 1)\beta^{B+2} -  (2\beta^2 + 2\beta + 1)\rho^{B+2}}{\mu [(\rho^2+\rho)\beta^{B+2} - (\beta^2+\beta)\rho^{B+2}]} & \beta \neq\rho, 
  \end{cases}
\end{align}
where $\beta = \eta/\mu$ is the normalized energy arrival rate, $\rho = \lambda/\mu$ is the normalized packet arrival rate, $\mu$ is the service rate, and $B \in \{1,2,\dots \}$ is the battery capacity where one unit of battery capacity is required to service an update. Note the somewhat surprising result that the average age in this setting is invariant to exchanging $\beta$ and $\rho$ even though energy and packets are handled in different manners by the server. 

A subsequent study in \cite{farazi-K-B-ISIT2018-aoi-eh-preempt} extended this work to consider the effect of preemption in service under the assumption that the energy of any update in service is lost if the update is preempted. Preemption was shown to decrease average age only in energy-rich operating regimes, i.e., regimes in which the server typically has a full battery and energy lost to preempted packets is inconsequential. In energy-starved operating regimes, preemption was shown to increase average age since preemption led a higher probability of battery depletion. Another study in \cite{ZhengTWC2019} extended these results to queues of arbitrary length, FCFS and LCFS queue disciplines, and nonlinear age penalty functions.

%\subsection{Status Updating in Advanced Energy Harvesting Settings} \label{sec_advanced}
\subsection{Advanced Energy Harvesting Settings} \label{sec_advanced}
In addition to the studies detailed in Sections~\ref{sec_prfct}, \ref{sec_erase}, \ref{sec_servers}, several other contributions have considered the interplay of age, energy, and damaged or lost updates. The case of  erasures without feedback was also considered in \cite{Gu-CZLV2019} where a truncated automatic repeat request (TARQ) scheme was developed to retransmit the current status update until a time threshold is exceeded or a new update is available. The proposed TARQ scheme was shown to achieve a lower average AoI than classical ARQ. 

In \cite{Ceran-GG-wcnc2018,CeranTWC2019}, optimal transmission policies were derived for a setting where a generate-at-will source subject to an energy constraint transmits updates to a monitor. Some updates are damaged or lost and the monitor provides an ACK/NACK feedback bit for each packet to inform the source of successful and unsuccessful updates. For a monitor using classic ARQ, fresh updates are always transmitted after a failed update since the probability of a failed update is assumed to be constant. For a monitor using Hybrid-ARQ, however, it can be optimal to retransmit a failed update since the receiver can combine packets so that each retransmission has a lower error probability. In unknown environments, \cite{CeranTWC2019} went on to use reinforcement learning techniques.

The tradeoff between transmit energy and error probability was explored in the context of age of information in \cite{krikidis-aoi-wpt} and \cite{Tunc-Panwar-2019}. While their settings are slightly different, both papers explore the fundamental tradeoff between using more energy in each transmission to improve the probability of successfully delivering an update and reducing age and the potential for depleting the battery and consequently increasing age. 

In \cite{rafiee-aoi-eh-drop}, a setting was considered in which a sensor node must decide when to sleep to save energy but miss updates or wake to receive updates and use energy. An age-threshold based power ON-OFF scheme was developed to minimize age subject to energy harvesting and battery capacity constraints.

Further, in addition to the previously discussed settings with exogenous sources of harvested energy, another line of research has considered systems in which source nodes harvest energy through wireless energy transfer (WET) from an access point with a stable energy supply. In \cite{Abd-Elmagid-Dhillon-vt2019}, a time-slotted system is considered. In each time slot, based on the available energies at the source nodes, the AoI values of different processes at the destination node, and the channel state information, the system must determine whether the time slot should be used for WET or for an update from a source node. A finite-state finite-action Markov decision process (MDP) is formulated and deep reinforcement learning techniques are used to find a solution. A similar setting, except with two-way data exchange, was considered in \cite{Hu-Dong-JCN-2019}. In this work, the downlink is assumed to use power splitting between WET and data and various tradeoffs between uplink and downlink AoI are analyzed.

Like \cite{Abd-Elmagid-Dhillon-vt2019}, several studies have developed online policies for resource-constrained AoI using a MDP framework. In \cite{Gindullina-Badia-Gunduz-2020}, a setting is considered where multiple sensors monitor the same process. Each sensor has a different age distribution and energy cost and the goal is to choose sensors to minimize the age at the destination subject to an energy constraint. In \cite{Tunc-Panwar-2019}, the source selects a transmission mode with an associated energy cost and error probability. Age only decreases if the transmission is received without error and the goal is to choose the transmission modes to minimize age subject to an energy constraint. A similar study exploring the tradeoff between frequent low-energy updates with high error probability and less frequent high-energy updates with low error probability was also considered in \cite{krikidis-aoi-wpt}. In \cite{Leng-Yener-2019TCCN}, a cognitive radio setting is assumed and a secondary user must decide whether to use energy to sense the channel or send updates. In all of these studies, the MDP framework is used to develop online policies to minimize age subject to energy constraints.

\section{Sampling, Estimation and Control}
\label{sec:sampling}

\begin{figure}[t]
\centering
\includegraphics[width=0.45\textwidth]{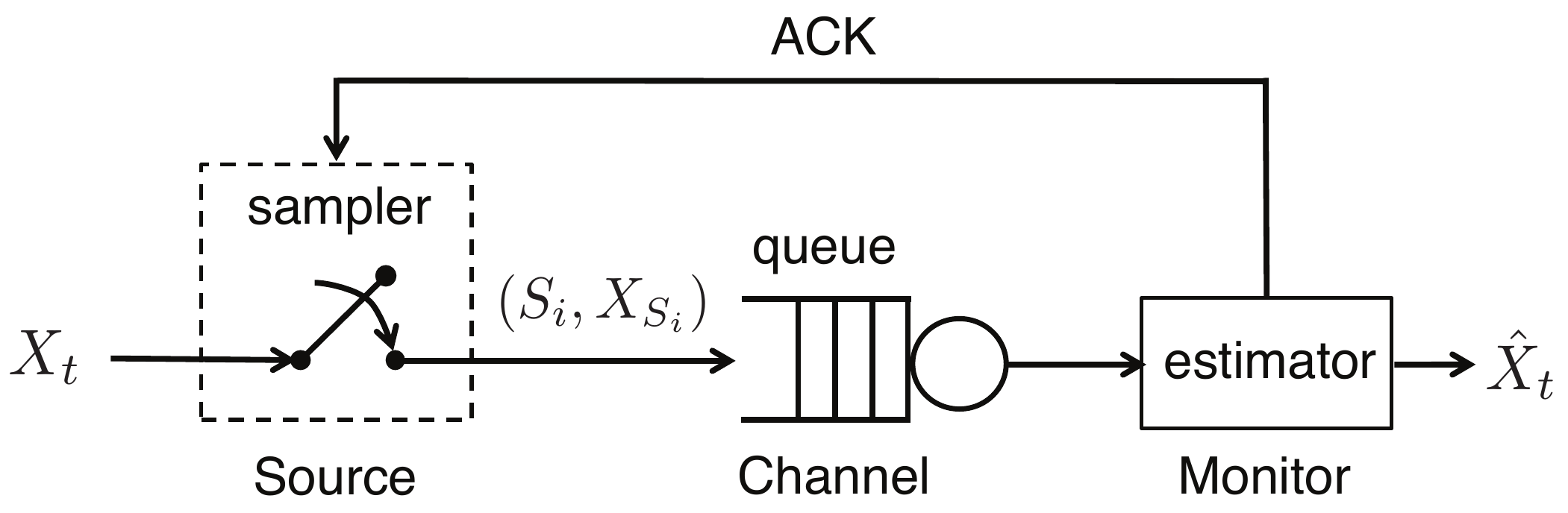}   
\caption{A status update system with a sampler and an estimator.}\vspace{-0.0cm}
\label{fig_sampling_model}
\end{figure}  

One method for reducing the AoI is to design a \emph{sampling} policy that progressively determines when to generate the update packets at the source. Let us consider the status update system illustrated in Fig. \ref{fig_sampling_model}, where samples of a signal $X_t$ are taken at the source and sent one-by-one to the monitor through a FCFS queue with \emph{i.i.d.} service times $Y_i$. Because the source is able to choose when to sample the $X_t$ process and generate an update, this is  called the ``generate-at-will'' sampling model \cite{SunInfocom2016}. Once a sample/update is delivered, an acknowledgement (ACK) is fed back to the sampler with no delay. By these ACKs, the sampler has access to the idle/busy state of the server in real-time. 

Here we examine optimal sampling policies from age and estimation error perspectives. We start in Section~\ref{sec:sampling-example} with an example to show how sampling policies can be counterintuitive and then move on  to  explore sampling for AoI minimization in Section~\ref{sec:sampling-AoI}. In Section~\ref{sec:sampling-estimation}, AoI  minimization is then compared against sampling approaches that aim to minimize signal reconstruction error at the monitor.    
%We want to find the optimal sampling policy that can minimize the AoI by  the sampling times $\{S_i\}_{i=1,2,\ldots}$ over time. 
%In Fig. \ref{fig_sampling_model}, the source can decide when to generate status update packets. This results in a \emph{sampling}  problem 
%
%the transmitter is able to generate status samples at any time. 
%\subsubsection{When to Take A Sample? A Counter-Intuitive Example}
%Consider the status update system in Fig. \ref{fig_sampling_model}, where samples of a signal $X_t$ are taken at the source and sent to the monitor through a FCFS queue. 
\subsection{Introduction}
\label{sec:sampling-example}
\begin{figure}[t]
\centering
%{\subfigure[][The zero-wait sampling policy.]{\resizebox{0.45\textwidth}{!}{\includegraphics{./matlab_SY/example1}}}}
  %  \hspace{0.1\textwidth}
\includegraphics[width=0.45\textwidth]{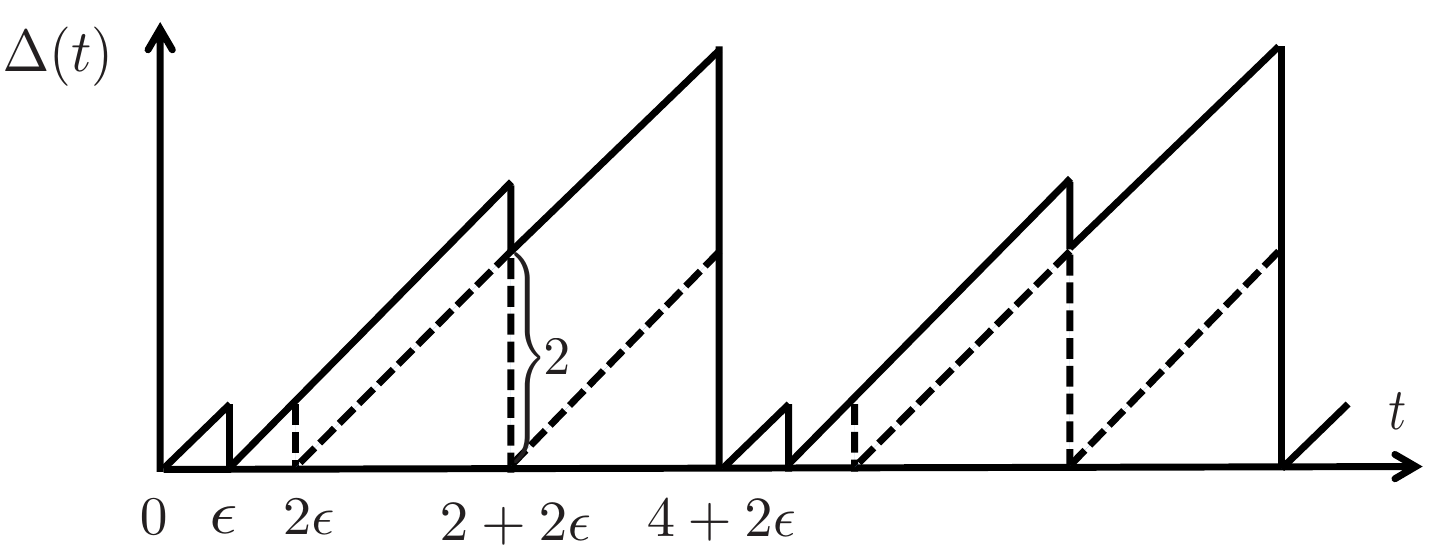}   
\caption{Time evolution of the AoI $\age(t)$ (solid line) for the $\epsilon$-wait policy in the example.}\vspace{-0.0cm}
\label{fig_example}
\end{figure}

In the event of queueing, the sampled packets would need to wait in the queue for their transmission opportunity  and would become stale during the waiting time. Hence, it is better to suspend sampling when the channel is busy, and reactivate it when the channel becomes idle. A reasonable sampling policy is the \emph{zero-wait} policy that submits a new sample once the previous sample is delivered. The zero-wait policy appears to be quite good, as it simultaneously achieves the maximum throughput and the minimum delay: Because the server is busy at all time, the maximum possible throughput is achieved; meanwhile, since the waiting time in the queue is zero, the delay is equal to the mean service time, which is the minimum possible delay. Surprisingly, this zero-wait policy does \textbf{not} always minimize the AoI. The following example reveals the reason behind this counter-intuitive phenomenon:

\begin{quotation}\em
\noindent \textbf{Example:}
Suppose that the service times of the samples are {i.i.d.}~across the samples and are equal to either 0 or 2 with probability 0.5. If Sample $1$ is generated at time $t=0$ and has a zero service time, it will be delivered to the receiver at time $t=0$. The question is when to take the 2nd sample?

Under the zero-wait sampling policy, Sample $2$ is also generated at time $t=0$ and sent out. After Sample $1$ is delivered at time $t=0$, Sample $2$ cannot bring any new information to the receiver, because both samples are taken at the same time $t=0$, i.e., they are exactly the same sample. On the other hand, Sample $2$ will occupy the channel by 1 second on average. Hence, taking the 2nd sample at time $t=0$ is not  a good strategy.

For comparison, consider a $\epsilon$-wait sampling policy that waits for $\epsilon$ seconds after each sample with a zero service time, and does not wait after each sample with a service time of 2 seconds. 
The time-evolution of the AoI $\Delta(t)$ in the $\epsilon$-wait sampling policy is plotted in Fig. \ref{fig_example}. One can compute the time-average age of the $\epsilon$-wait sampling policy, which is given by
% \begin{align}
% &(\epsilon^2/2 + \epsilon^2/2 + 2\epsilon + 4^2/2) / (4 + 2\epsilon) \nn
% \qquad=& (\epsilon^2 + 2\epsilon + 8) / (4 + 2\epsilon)~\text{seconds.}\nonumber
% \end{align} 
\begin{align}
\frac{\epsilon^2/2 + \epsilon^2/2 + 2\epsilon + 4^2/2}{4 + 2\epsilon} 
= \frac{\epsilon^2 + 2\epsilon + 8}{4 + 2\epsilon}~\text{seconds.}\nonumber
\end{align} 
If the waiting time is $\epsilon=0.5$ seconds, the time-average age of the $\epsilon$-wait sampling policy is 1.85 seconds. If the waiting time is $\epsilon=0$, the $\epsilon$-wait sampling policy becomes the zero-wait sampling policy, whose time-average age is 2 seconds. 
Hence, the zero-wait sampling policy is not optimal!
\end{quotation}

In fact, the zero-wait sampling policy can be far from the optimum if (i) the goal is to minimize a nonlinear age function that grows quickly with respect to the AoI, and/or (ii) when the service times follow a heavy-tail distribution \cite{SunJournal2017,SunNonlinear2019}.
This example points out a key difference between data communication systems and status update systems: 
In data communication, all packets are equally important; however,
in status updating, a sample packet is useful only if it carries fresh information to the monitor. 

%\subsubsection*{Age-optimal Sampler Design}

\subsection{Sampling for AoI Minimization}
\label{sec:sampling-AoI}
Let $\pi=(S_1,S_2,\ldots)$ represent a sampling policy where $S_i$ is the generation time of sample $i$, and $\Pi$ denote the set of causal sampling policies. The optimal sampling problem for minimizing the time-average age penalty is formulated as
\begin{align}
\bar p_{\text{opt}} = \inf_{\pi\in\Pi}~&\limsup_{\Tcal\rightarrow \infty}\frac{1}{\Tcal}~\E{\int_{0}^\Tcal p(\Delta(t))\,dt}.\label{eq_problem2} \end{align}
%where $\bar p_{\text{opt}}$ denotes the optimal value of  \eqref{eq_problem2}. 
% The problem for maximizing the time-average age utility can be readily obtained from \eqref{eq_problem2} by choosing $p(\age) = - u(\age)$.

\begin{theorem}\label{ch01thm1}\cite{SunSPAWC2018,SunNonlinear2019}
If $p(\cdot)$ is non-decreasing and the service times $Y_i$ of the queue are {i.i.d.} with $0<\E{Y_i}<\infty$, then 
$(S_1(\beta),S_2(\beta),\ldots)$ with a parameter $\beta$ is an optimal solution to \eqref{eq_problem2}, where 
\begin{align}\label{ch01thm1_eq12}
S_{i+1}(\beta) &=\inf\{ t\geq  D_i(\beta): \E{p(\Delta(t + Y_{i+1}))} \geq \beta \},
\end{align}
$D_i(\beta) = S_i(\beta) +Y_i$ is the delivery time of  sample $i$, $\Delta(t) = t-S_{i}(\beta)$ is the AoI, and the threshold $\beta$ is the root  of 
\begin{align}\label{ch01thm1_eq22}
\E{\int_{D_i(\beta)}^{D_{i+1}(\beta)}\!\! p(\Delta(t))\,dt} - \beta \E{D_{i+1}(\beta)-D_i(\beta)} =0.\!\!
\end{align}
Further, $\beta$ is exactly the optimal value to \eqref{eq_problem2}, i.e., {$\beta =\bar p_{\text{opt}}$}. 
\end{theorem} 

The optimal sampling policy in \eqref{ch01thm1_eq12}-\eqref{ch01thm1_eq22} has a nice structure. Specifically,  sample $i+1$ is generated at the earliest time $t$ satisfying two conditions: \begin{enumerate}
    \item[(i)] 
 sample $i$ has already been delivered by time $t$, i.e., $t\geq D_i(\beta)$, and 
 \item[(ii)] the expected age penalty $\E{p( \Delta(t + Y_{i+1}))}$ has grown to be no smaller than a predetermined threshold $\beta$ that is equal to the optimum objective value $\bar p_{\text{opt}}$ of \eqref{eq_problem2}.
 \end{enumerate}
 One can show that the left side of \eqref{ch01thm1_eq22} is a concave and strictly decreasing function of $\beta$ \cite{Ornee2020ArXiv}. Hence, \eqref{ch01thm1_eq22} has a unique root. The root of \eqref{ch01thm1_eq22} can be calculated by using bisection search, Newton's method, and  the following fixed-point iteration method
\begin{align}
\beta_{k+1} = \frac{\Ebig{\int_{D_i(\beta_k)}^{D_{i+1}(\beta_k)} p(\Delta(t))\,dt}} {\Ebig{D_{i+1}(\beta_k)-D_i(\beta_k)}},
\end{align}
where Newton's method and fixed-point iterations converge faster than bisection search  \cite{Ornee2020ArXiv,TsaiISIT2020}.

A few variations on Theorem~\ref{ch01thm1} have appeared in other settings, including discrete-time sampling \cite{SunNonlinear2019}, sampling subject to a maximum sampling rate constraint \cite{SunNonlinear2019} and two-way communication delay \cite{TsaiISIT2020}, as well as multi-source updates for minimizing the total time-average age \cite{Ahmed2019Mobihoc} and total time-average of nonlinear age penalty functions \cite{Ahmed2020ArXiv}. 
In \cite{elif-age-online-threshold,bacinoglu-aoi-eh-finite-gnrl-pnlty}, age-optimal sampler design was obtained for an energy harvesting source with a finite battery size $B$ and zero service time $Y_i=0$. 
If energy units arrive according to a Poisson process, the optimal sampler is given by a multi-threshold sampling policy
\begin{align}\label{ch01thm1_eq12_energyharvesting}
S_{i+1}(\beta) =\inf\{ t\geq  S_i(\beta): p( \Delta(t)) \geq \beta(E(t)) \},
\end{align}
where the threshold $\beta(E(t))$ is a decreasing function of the instant battery level $E(t)\in\{0,1,\ldots,B\}$. Hence, samples are taken more frequently when the battery level is high, and less frequently when the battery level is low. %According to \eqref{ch01thm1_eq12_energyharvesting}, new samples are taken in two scenarios: (i) the AoI $\Delta(t)$ grows to be sufficiently high, or (ii) the threshold $\beta(E(t))$ drops because more energy is harvested and the battery level $E(t)$ grows. 
Interestingly, the threshold $\beta(B)$ associated to a full battery level $E(t)=B$ is equal to the optimal objective value for this setting \cite{elif-age-online-threshold,bacinoglu-aoi-eh-finite-gnrl-pnlty}.

\subsection{Sampling and Remote Estimation}
\label{sec:sampling-estimation}
%\begin{figure}[h]
%\centering
%\includegraphics[width=0.65\textwidth]{images/signal}   
%\caption{
%Signal variations during time intervals of length $T$.}\vspace{-0.0cm}
%\label{fig_signal_eg}
%\end{figure}  
The states of many cyberphysical systems, such as UAV mobility trajectory and sensor measurements, are in the form of a signal $X_t$. A natural task in these systems is to reconstruct the signal $X_t$ at the remote monitor, based on samples that are causally received from the source. This requires an extension of Nyquist sampling theory to causal sampling and reconstruction. In the non-causal case, periodic sampling can achieve perfect  reconstruction of bandlimited signals with no error; however, a non-zero reconstruction error is inevitable in causal signal processing and the design goal should be minimizing the reconstruction error. The problem of causal sampling and reconstruction was called \emph{remote estimation} in the control literature; see \cite{jog2019channels} for a recent survey on this topic.

Recently, a connection between AoI and remote estimation was revealed in \cite{SunISIT2017,SunTIT2020,Ornee2019,Ornee2020ArXiv}. To understand this connection, let $\hat X_t$ denote the MMSE estimate of $X_t$, based on the samples that have been delivered by time $t$,  so that the signal reconstruction error is 
\begin{equation}
    \err{t}=X_t-\hat X_t.
\end{equation}
Now consider the following optimal sampling problem for minimizing the mean-squared  signal reconstruction error
%As shown in Fig. \ref{fig_signal_eg}, these signals may change slowly at some time and vary more dynamically later. Hence, the time difference described by the AoI $\Delta(t) = t - u(t)$ or its non-linear functions cannot fully characterize the variation $X_t -X_{u(t)}$ of the system state. In the sequel, we will see that minimizing AoI is insufficient for optimal reconstruction of $X_t$ at the monitor. 
\begin{align}\label{eq_DPExpected}
\overline{\mathsf{mse}}_{\text{opt}}=
\inf_{\pi\in\Pi}~& \limsup_{\Tcal\rightarrow \infty}\frac{1}{\Tcal}\E{\int_0^{\Tcal} |\err{t}|^2\,dt}.
\end{align}
%where  and ${\mathsf{mse}}_{\text{opt}}$ is the optimum value of \eqref{eq_DPExpected}. 
Problem \eqref{eq_DPExpected} belongs to the class of continuous-time MDPs with continuous state space, which are usually quite challenging to solve due to  the curse of dimensionality. Nonetheless, an exact solution to \eqref{eq_DPExpected} has been found for two Gauss-Markov signals: the Wiener process and the Ornstein-Uhlenbeck process, which is the continuous-time analogue of the first-order autoregressive, i.e., AR(1), process. 

\begin{theorem} \cite{SunTIT2020,Ornee2020ArXiv} \label{ch01coro2002}
If $X_t$ is a Wiener process or an Ornstein-Uhlenbeck process, and the $Y_i$ are i.i.d. with $0<\E{Y_i} < \infty$, then $(S_1(\beta),S_2(\beta),\ldots)$ with a parameter $\beta$ is an optimal solution to \eqref{eq_DPExpected}, where
%\item[4.] \emph{Threshold policy in signal difference}: 
%The sampling times are given by
%
\begin{align}\label{eq_coro2_opt_solution}
S_{i+1} (\beta)= \inf \left\{ t \geq D_i(\beta):\! \big|\err{t}\big| \!\geq\! v(\beta)\right\},
\end{align}
%\end{itemize} 
%\begin{align}\label{eq_DPExpected}
%%\pi_{\text{opt}}=\arg
%&\min_{\pi\in\Pi}~ \limsup_{\Tcal\rightarrow \infty}\frac{1}{\Tcal}\mathbb{E}\left[\int_0^{\Tcal} (X_t - \hat X_t)^2dt\right] \\
%&~\text{s.t.}~~ \liminf_{n\rightarrow \infty} \frac{1}{n} 
%\mathbb{E}[S_n]\geq \frac{1}{f_{\max}}.\label{eq_constraint}
%\end{align
$D_i (\beta)= S_i (\beta)+ Y_i$ is the  delivery  time  of  sample $i$, and $\beta$ is the root  of 
\begin{IEEEeqnarray}{rCl}
\label{ch01coro2_eq22}
\E{\int_{\mathrlap{D_i(\beta)}}^{\mathrlap{D_{i+1}(\beta)}}~~~~~~~~ |\err{t}|^2\,dt}\!\!-\beta \E{D_{i+1}(\beta)\!-\!D_i(\beta)} &=& 0.\!\!
\IEEEeqnarraynumspace
\end{IEEEeqnarray} 
For the Wiener process, the threshold function $v(\cdot)$ is $
v(\beta)=\sqrt{3(\beta -\E{Y_i})}$;
for the Ornstein-Uhlenbeck process, $v(\cdot)$ is given by (18) of \cite{Ornee2020ArXiv}. 
Further, $\beta$ is exactly the optimal value to \eqref{eq_DPExpected}, i.e., {$\beta =\overline{\mathsf{mse}}_{\text{opt}}$}. 
\end{theorem}

The structure of the optimal sampling policy in  \eqref{eq_coro2_opt_solution}-\eqref{ch01coro2_eq22} is similar to that in \eqref{ch01thm1_eq12}-\eqref{ch01thm1_eq22}. Specifically,  sample $i+1$ is generated at the earliest time $t$ satisfying two conditions: (i) sample $i$ has already been delivered by time $t$, i.e., $t\geq D_i(\beta)$, and (ii) the instantaneous  estimation error $|\err{t}|$
%$|X_t - \hat X_t|$ 
is no smaller than a pre-determined threshold $v(\beta)$, where $\beta$ is equal to the optimum value to \eqref{eq_DPExpected} and the function $v(\cdot)$ is determined by the signal  and the service time distribution. One can add a maximum sampling rate constraint in the optimal sampling problem \eqref{eq_DPExpected} and its solution was provided in \cite{SunTIT2020,Ornee2020ArXiv}.  
Recently, remote estimation of a Wiener process with two-way random communication delay was studied in \cite{Tsai2020INFOCOM}.

In remote estimation systems, the optimal sampler is affected by the selected estimator; and conversely, the optimal estimator is also influenced by the selected sampling policy. This results in an ``chicken and egg'' dilemma. Nonetheless, one can resolve the dilemma by jointly optimizing the sampler and estimator.~In \cite{Hajek2008,Nuno2011,nayyar2013,GAO201857,ChakravortyTAC2020}, tools from majorization theory were utilized to show that threshold-type samplers remain optimal in the joint sampling and estimation problem for several discrete-time remote estimation systems. It was also pointed out in \cite[p. 619]{Hajek2008} that similar results can be also established for continuous-time systems. Recently, it was shown that the sampling policy \eqref{eq_coro2_opt_solution}-\eqref{ch01coro2_eq22} and the MMSE estimator are indeed jointly optimal for the remote estimation of a class of continuous-time Markov signals \cite{Guo2020ISIT}.\footnote{This  result was shown for a remote estimation system with zero service time, i.e., $Y_i=0$ \cite{Guo2020ISIT}. It appears that the treatment in \cite{Guo2020ISIT} also applies to the case of non-zero service time.}

%The sampling times in \eqref{ch01thm1_eq12}-\eqref{ch01thm1_eq22} are determined based on real-time knowledge of the signal value $X_t$.
\subsubsection*{AoI and Signal-agnostic Sampling}
Next, we will consider a variation of problem \eqref{eq_DPExpected} that is tightly related to the AoI. We say a sampling policy $\pi\in\Pi$ is \emph{signal-aware} (\emph{signal-agnostic}), if the sampling times $S_i$ are determined (without) using causal knowledge of the signal $X_t$.~Hence, the $S_i$'s are independent of the signal $X_t$ in signal-agnostic sampling policies.~Let $\Pi_{\text{agnostic}}\subset\Pi$ denote the set of signal-agnostic sampling policies.
%, where the sampling times $S_i$ are determined without using any knowledge of the signal value $\{X_t, t \geq 0\}$. 
For every policy $\pi\in\Pi_{\text{agnostic}}$ and time-homogeneous Markov chain $X_t$, there exists an increasing function $p(\Delta(t))$ of the AoI such that
%\fi
%the objective function in \eqref{eq_DPExpected} satisfies 
%where $p(\Delta_t)$ is an increasing function of the age $\Delta_t$, given by
%\begin{align}\label{eq_age_penalty}
%p(\Delta_t) = .
%\end{align}
\begin{align}%\label{eq_age_MSE}
%{\mathsf{mse}}_\pi =
\E{\int_0^{\Tcal} |\err{t}|^2\,dt} \!\!= \E{\int_0^{\Tcal}p(\Delta(t))\,dt}.
\end{align}
By reducing the policy space $\Pi$ to $\Pi_{\text{agnostic}}$,  \eqref{eq_DPExpected} becomes  the following signal-agnostic sampler design problem:
\begin{IEEEeqnarray}{rCl}
\label{eq_DPExpected1}
\overline{\mathsf{mse}}_{\text{age-opt}}&=&
\quad\inf_{\mathclap{\pi\in\Pi_{\text{agnostic}}}}\quad \limsup_{\Tcal\rightarrow \infty}\frac{1}{\Tcal}\E{\int_0^{\Tcal} |\err{t}|^2\,dt},\IEEEeqnarraynumspace
\end{IEEEeqnarray}
which is an instance of \eqref{eq_problem2} and hence can be solved by using Theorem \ref{ch01thm1}.
% We use $\overline{\mathsf{mse}}_{\text{age-opt}}$ to represent the optimal value to \eqref{eq_DPExpected1}. 

%If the policy space is restricted from $\Pi$ to $\Pi_{\text{signal-agnostic}}$, \eqref{eq_DPExpected} becomes

%let us consider the following optimal sampling problem for minimizing the mean-squared  signal reconstruction error: 
%As shown in Fig. \ref{fig_signal_eg}, these signals may change slowly at some time and vary more dynamically later. Hence, the time difference described by the AoI $\Delta(t) = t - u(t)$ or its non-linear functions cannot fully characterize the variation $X_t -X_{u(t)}$ of the system state. In the sequel, we will see that minimizing AoI is insufficient for optimal reconstruction of $X_t$ at the monitor. 

\begin{corollary} \label{coro_estimation}
If $X_t$ is a continuous-time homogeneous Markov chain, and the $Y_i$ are i.i.d. with $0<\E{Y_i} < \infty$, then $(S_1(\beta),S_2(\beta),\ldots)$ with a parameter $\beta$ is an optimal solution to \eqref{eq_DPExpected1}, where
%\item[4.] \emph{Threshold policy in signal difference}: 
%The sampling times are given by
%
\begin{align}
\label{eq_coro2_opt_solution1}
S_{i+1}(\beta)= \inf \left\{ t \geq D_{i}(\beta):\! \Ebig{|\err{t+Y_{i+1}}|^2}
%(X_{t+Y_{i+1}} - \hat X_{t+Y_{i+1}})^2} 
\!\geq\! \beta\right\},
\end{align}
% \begin{align}\label{eq_coro2_opt_solution1}
% S_{i+1} (\beta)= \inf \left\{ t \geq D_i(\beta):\! \Ebig{(X_{t+Y_{i+1}} - \hat X_{t+Y_{i+1}})^2} \!\geq\! \beta\right\},
% \end{align}
%\end{itemize} 
%\begin{align}\label{eq_DPExpected}
%%\pi_{\text{opt}}=\arg
%&\min_{\pi\in\Pi}~ \limsup_{\Tcal\rightarrow \infty}\frac{1}{\Tcal}\mathbb{E}\left[\int_0^{\Tcal} (X_t - \hat X_t)^2dt\right] \\
%&~\text{s.t.}~~ \liminf_{n\rightarrow \infty} \frac{1}{n} 
%\mathbb{E}[S_n]\geq \frac{1}{f_{\max}}.\label{eq_constraint}
%\end{align
$D_i (\beta)= S_i (\beta)+ Y_i$ is the  delivery  time  of  sample $i$, and the threshold $\beta$ is the root  of 
\begin{IEEEeqnarray}{rCl}
\label{ch01coro2_eq221}
\E{\int_{\mathrlap{D_{i}(\beta)}}^{\mathrlap{D_{i+1}(\beta)}}~~~~~~~~|\err{t}|^2\,dt} 
%(X_t-\hat X_t)^2\,dt}
\!\!-\beta \E{D_{i+1}(\beta)\!-\!D_{i}(\beta)} = 0.\IEEEeqnarraynumspace
\end{IEEEeqnarray}
Further, $\beta=\overline{\mathsf{mse}}_{\text{age-opt}}$, the optimal value to \eqref{eq_DPExpected1}.
\end{corollary}

Let us compare the optimal designs of signal-aware and signal-agnostic samplers: In the signal-aware sampling policy \eqref{eq_coro2_opt_solution}, the sampling time is determined by the \emph{instantaneous} estimation error
$|\err{t}|$,
%$\big|X_t - \hat X_t\big|$, 
and the threshold function $v(\cdot)$ varies with the signal model. 
In the signal-agnostic sampling policy \eqref{eq_coro2_opt_solution1}, the sampling time is determined by the \emph{expected} estimation error 
$\Ebig{|\err{t+Y_{i+1}}|^2}$
%$\Ebig{(X_{t+Y_{i+1}}\! -\! \hat X_{t+Y_{i+1}})^2}$ 
at time $t+Y_{i+1}$. If $t =S_{i+1}(\beta)$, then $t+Y_{i+1}=S_{i+1}(\beta)+Y_{i+1} = D_{i+1}(\beta)$ is the  delivery time of the new sample. Hence, \eqref{eq_coro2_opt_solution1} requires that the expected estimation error upon the delivery of the new sample is no less than $\beta$. In both cases, the parameter $\beta$ is equal to the optimal objective value. 

Age-based sampler design is of particular interest in discrete-time feedback control systems \cite{Champati2019,Markus2019}. According to \cite[Theorem 3]{SunNonlinear2019}, the optimal signal-agnostic sampler for such discrete-time systems can be obtained by (i) adding another requirement that $t$ in \eqref{eq_coro2_opt_solution1} is a discrete time, i.e., $t\in\{0, T_s, 2 T_s,\ldots\}$, and (ii) changing the integral in \eqref{ch01coro2_eq221} as a summation over discrete time. These results are not limited to the queue model; in fact, the 
optimal signal-agnostic sampler for remote estimation over a packet erasure channel has a quite similar structure, as  reported independently in \cite{Markus2019}. The scheduling of channel resources among multiple feedback control loops for reducing state estimation error was considered in \cite{AyanCPS2019,AyanCCNC2020}. When the AoI is high, it is difficult to maintain a small control error. The optimal tradeoff between the AoI and control performance was studied in \cite{SoleymaniCDC2019}.

\section{Wireless Networks}
\label{sec:wireless}
In many applications,  timely information is disseminated through wireless networks, where interference is  one of the primary limitations to system performance. In this section, we start with an overview of recent contributions to AoI in wireless networks, followed in Section~\ref{sec:wireless-channel} by works on updates over erasure channels that use ARQ/HARQ. Further in Section~\ref{sec:wireless-mac} we discuss AoI results for conventional distributed multiaccess protocols. Section~\ref{sec2.Model} then reviews the main results from \cite{igorTON18} on AoI optimization in broadcast wireless networks, followed in Section~\ref{sec:wireless-general} by results from \cite{talak18_Mobihoc} on wireless networks under general interference constraints. 

\subsection{Overview}
Over the past few years, there has been a growing body of work on AoI minimization in wireless settings.  at the physical layer, updates through channels with bit erasures have been studied in~\cite{Parag-TC-wcnc2017realtime,Najm-YS-isit2017,Yates-NSZ-isit2017,Ceran-GG-wcnc2018,Sac-BUBD-spawc2018}.  The problem of scheduling finite number of update packets under physical interference constraint for age minimization was shown to be NP-hard in, e.g., \cite{2016Ep_WiOpt,HeTIT2018}. The impact of information freshness on collision avoidance in a network of UAVs was studied in~\cite{rajat16cdc}.  Age for a wireless network where only a single link can be activated at any time was studied in, e.g., \cite{Kadota-UBSM-Allerton2016, 2017ISIT_YuPin,Jiang-KZN-itc2018,Jiang-KZZN-isit2018,Sun-Jiang-KZN-TCOM2020}, and index policies were proposed.  Threshold policies were proven to be optimal in, e.g., \cite{CeranTWC2019,ZhouGLOBECOM2018}.  

Various works have considered decentralized access of a shared medium by nodes sending updates. In~\cite{Kaul-GRK-secon2011, Kaul-Yates-isit2017, Ali2019ArXivCSMA, AoI_adhoc, yates2020age, Chen-Gatsis-2019a} authors analyzed ALOHA and CSMA like random access. 
%Repeated in next subsection:
%In~\cite{Jiang-KZN-itc2018, Sun-Jiang-KZN-TCOM2020} the age minimization problem is viewed as a restless multi-armed bandit problem, the Whittle's index is derived and used to propose a decentralized index prioritized random access. 
In~\cite{Jiang-KZZN-iot2019} the authors showed asymptotic optimality of a decentralized round robin scheduling policy. In~\cite{Chen-Yifan-2020a} the authors considered multiaccess in which a node chooses its access probability as a function of the age of its updates. In~\cite{Bedewy2020MobiHoc} a sleep-wake strategy is designed for a network of low-powered battery constrained nodes, which use carrier sensing based access.

%AoI optimized designs of slotted ALOHA and carrier sensing random access (CSMA)  protocols were considered in, e.g., \cite{Kaul-Yates-isit2017,Ali2019ArXivCSMA,Bedewy2020MobiHoc}.

In \cite{igorTON18}, the authors addressed the problem of scheduling transmissions in broadcast wireless networks in order to minimize AoI. Age minimization with minimum throughput requirements was studied in, e.g., \cite{BinLi18,igorINFOCOM}. Stochastic arrival processes were considered in \cite{igorMobiHoc}. In \cite{JooTON2018,talak18_Mobihoc}, the authors considered the problem of optimizing AoI in wireless networks under general interference constraints. Several extensions have recently appeared in~\cite{Talak-KM-allerton2017, talak18_greece, talak18_WiOpt,talak18_ISIT}.  The case of multi-source multi-hop wireless networks was considered in, e.g., \cite{Talak-KM-allerton2017,farazi2018SPAWC}. 
In~\cite{talak18_greece}, the authors developed distributed policies for age minimization, and in \cite{talak18_WiOpt} and~\cite{talak18_ISIT}  the authors proposed age-based and virtual-queue-based policies for age minimization. In~\cite{talak18_WiOpt}, the authors showed that using the current channel state information can result in significant age improvement. Scaling of age as a function of the number of nodes in a large multi-hop wireless network has been studied in \cite{Buyukates18c, Buyukates19b, Buyukates20d,farazi2019JCN}. 

\subsection{Updates Through Erasure Channels}
\label{sec:wireless-channel}
In~\cite{Yates-NSZ-isit2017} age is analyzed for different methods of introducing coded redundancy. While increasing redundancy improves the probability that an update will be successfully delivered, it comes at the expense of longer update transmission times. The authors analyze finite redundancy (FR), wherein a $k$ symbol update is transmitted as a fixed number $n$ of coded symbols, and infinite incremental redundancy (IIR), wherein coded symbols are sent till $k$ are successfully received. In~\cite{Najm-YS-isit2017} the HARQ protocols of FR and IIR are considered. The HARQ FR is different from FR in~\cite{Yates-NSZ-isit2017} in that it retransmits until an update is successfully transmitted. Further, a fresh update is generated at the beginning of a transmission, unlike~\cite{Yates-NSZ-isit2017} where updates arrive as a Poisson process. A new arrival must either preempt the currently transmitting update or be discarded. The latter is found  to be the better strategy. 

In~\cite{Sac-BUBD-spawc2018} ARQ and HARQ are analyzed for when an update is coded as $n$ symbols for transmission. Fresh updates arrive to a FCFS queue as a Poisson process. The codeword length $n$ that optimizes age is of interest. In~\cite{Ceran-GG-wcnc2018, Ceran-2019TWC} ARQ and HARQ are studied with the goal of minimizing age when there is a constraint on the time-average of the number of transmissions at the source. Last but not the least, in~\cite{Parag-TC-wcnc2017realtime} HARQ and a transmit scheme that encodes an update into $n$ symbols but transmits it only once are compared. Random linear codes are assumed for forward error correction.

\subsection{Decentralized Multiaccess}
\label{sec:wireless-mac}
In a vehicular network setting where each vehicle sent updates using carrier sense based $802.11$ medium access,  different packet management strategies were simulated and a heuristic gradient-descent based algorithm was empirically shown to minimize the average age ~\cite{Kaul-GRK-secon2011}.    In~\cite{Kaul-Yates-isit2017}, the authors compared ALOHA with scheduled access over unreliable channels. When all network nodes desired the same average age, it was shown that ALOHA led to age that was worse than scheduled access by a factor of $2e$. 

In~\cite{Ali2019ArXivCSMA} nodes use CSMA to access the channel. The authors suitably approximate practical CSMA and come up with a SHS model and the corresponding expression for average age. The age is optimized over the back-off rates of the nodes. They show that these rates are independent of the arrivals rates of fresh updates at the nodes. In~\cite{Chen-Gatsis-2019a} it is shown that one could achieve a reduction in average AoI by a factor of $2$ in comparison to the minimum that can be achieved when using typical slotted ALOHA. The authors propose using ALOHA, however, using large offered loads together with a load thinning mechanism, which discards fresh packets at nodes that on transmission would lead to a reduction in age smaller than a threshold. The challenge is then to find such a threshold. % and the authors propose an adaptive method of updating the threshold and also an offline way of calculating a fixed threshold.

All the above works above consider slotted access. Works~\cite{Kaul-GRK-secon2011} and~\cite{Ali2019ArXivCSMA} also assume carrier sensing. Unlike these,~\cite{yates2020age} considers a network of transmit-only network of nodes. The setting is unslotted, precludes carrier sensing, and allows for channel error. Updates are assumed to arrive as a Poisson process and transmission times are exponentially distributed; the  SHS approach is used to derive the average AoI.

In~\cite{Jiang-KZZN-isit2018, Jiang-KZZN-iot2019} the authors consider policies that schedule transmissions in a network of nodes and provide packet management at each node. The transmissions must be scheduled in a non-interfering manner. Also, it should be possible to implement the policies in a decentralized manner. The authors show that the RR-ONE policy, which enforces a round robin schedule and only retains the latest packet at any node is asymptotically optimal (minimizes average age) among all policies. Specifically, it achieves an optimal asymptotic scaling factor (average age of a node normalized by the number of nodes) of $0.5$. In comparison, this scaling factor of CSMA is shown to be at least $1$. While asymptotically optimal, RR-ONE may not perform well with fewer numbers of nodes and low packet arrival rates.

In~\cite{Jiang-KZN-itc2018, Sun-Jiang-KZN-TCOM2020} the authors formulate age minimization as a restless multi-armed bandit problem. They derive the Whittle index and show that each source can calculate its own index independently of the others. In a centralized setup, access to indices of all nodes is available and the resulting index policy has near-optimal performance. They propose a decentralized scheme that has nodes access the channel with probabilities that are a function of their index.

In~\cite{Chen-Yifan-2020a} the nodes don't transmit if the age of their update is smaller than a certain threshold. All nodes whose age exceeds the threshold access the medium with the same probability. The average age for the network is derived for such a policy and the same is optimized over the choices of threshold and access probability.

Unlike all the above works on shared medium access,~\cite{Bedewy2020MobiHoc} considers a network of low-powered battery constrained nodes. The goal is to design a sleep-wake strategy that minimizes the weighted average peak age while ensuring that the energy constraints are satisfied.

\subsection{Broadcast Wireless Networks}
\label{sec2.Model}
\begin{figure}[t]
\begin{center}
\includegraphics[height=4cm]{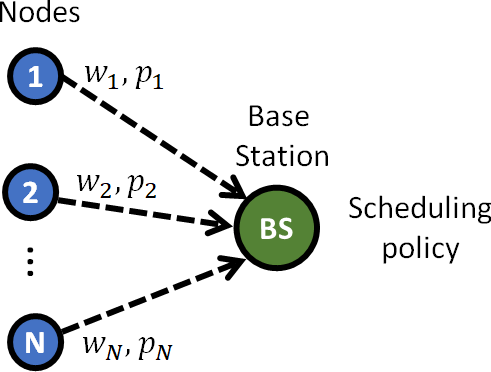}
\end{center}
\caption{Illustration of the single-hop wireless network. On the left, we have $N$ nodes. In the center, we have $N$ links with their associated priority (or weight) $w_i$ and probability of a successful packet transmission $p_i$. On the right, we have the base station running a transmission scheduling policy.}\label{fig2.Network}
\vspace{-5mm}
\end{figure}

Consider a single-hop wireless network with $N$ nodes sharing time-sensitive information through unreliable communication links to  a base station (BS), as illustrated in Fig.~\ref{fig2.Network}. Let the time be slotted, with slot duration normalized to unity and slot index $t\in\{1,2,\cdots,T\}$, where $T$ is the time-horizon of this discrete-time system. The broadcast wireless channel allows at most one packet transmission per slot. In each time-slot $t$, the BS either idles or schedules a transmission in a selected link $i\in\{1,2,\cdots,N\}$. Let $\schedulingDecision_i(t)\in\{0,1\}$ be the indicator function that is equal to $1$ when the BS selects link $i$ during slot $t$, and $\schedulingDecision_i(t)=0$ otherwise. \emph{When $\schedulingDecision_i(t)=1$ the corresponding source samples fresh information, generates a new packet and transmits this packet over link $i$.} Notice that packets are not enqueued. Since the BS can select at most one link at any given slot $t$, we have
\begin{equation}\label{eq2.interference_contraint}
\sum_{i=1}^N \schedulingDecision_{i}(t)\leq 1, \qquad t \in \{1,\dots,T\}.
\end{equation}
The transmission scheduling policy governs the sequence of decisions $\{\schedulingDecision_{i}(t)\}_{i=1}^N$ of the BS over time.

Let $c_i(t)\in\{0,1\}$ represent the channel state associated with link $i$ during slot $t$. When the channel is \emph{ON}, we have $c_i(t)=1$, and when the channel is \emph{OFF}, we have $c_i(t)=0$. The channel state process is assumed i.i.d.~over time and independent across different links, with $\mathbb{P}(c_i(t)=1)=p_i,\forall i,t$.
Let $d_i(t)\in\{0,1\}$ be the indicator function that equals $1$ when the transmission in link $i$ during slot $t$ is successful; otherwise, $d_i(t)=0$. A successful transmission occurs when a link is selected and the associated channel is ON, implying that $d_i(t)=c_i(t)\schedulingDecision_i(t),\forall i,t$. %Moreover, since the BS does not know the channel states prior to making scheduling decisions, $\schedulingDecision_i(t)$ and $c_i(t)$ are independent, and
%\begin{equation}\label{eq2.iterated_expectations}
%\mathbb{E}\left[d_i(t)\right]=p_i\mathbb{E}\left[\schedulingDecision_i(t)\right], \forall i,t \; .
%\end{equation}

In \cite{igorTON18}, the authors consider four low-complexity scheduling policies, namely Maximum Age First, Stationary Randomized, Max-Weight and Whittle's Index, and derive performance guarantees for each of them as a function of the network configuration.

%The performance of an arbitrary admissible policy $\pi\in\Pi$ is given by $\mathbb{E}[J^\pi]=\lim_{T\rightarrow\infty}\mathbb{E}[J_T^\pi]$ from \eqref{eq2.EWSAoI} and the optimal performance is $\mathbb{E}[J^*]$ from \eqref{eq2.Objective1_infinite}. We define the optimality ratio\footnote{Optimality Ratio is also known as Approximation Ratio.}
%$\rho^\pi\geq\mathbb{E}\left[J^{\pi}\right] / \mathbb{E}\left[J^*\right]$ and say that policy $\pi$ is $\rho^\pi$-optimal. Naturally, the closer the optimality ratio $\rho^\pi$ is to unity, the better is the performance of policy $\pi$ in terms of AoI. 

\paragraph{Maximum Age First (MAF)}  In each slot $t$, select  the link $i$ with highest value of $\AoI_i(t)$, with ties being broken arbitrarily.

It was shown, using stochastic ordering arguments, that the MAF policy minimizes the average AoI under symmetric conditions ($p_i=p \in (0,1]$ and weights $w_i=w > 0$ for all $i$.) Note that MAF leverages the knowledge of $\AoI_i(t)$, but disregards the values of $w_i$ and $p_i$. However, when these parameters are not symmetric, MAF age performance can be arbitrarily poor.  Next we discuss randomized scheduling policies which use the knowledge of $w_i$ and $p_i$ to achieve good performance in arbitrary settings. The randomized policy is based on a set of  positive fixed values of $\{\beta_1,\ldots,\beta_N\}$.
%\subsection{Stationary Randomized Policy}\label{sec2.Randomized}
%Consider the class of Stationary Randomized policies in which scheduling decisions are made randomly, according to fixed probabilities $\beta_i/\sum_{j=1}^N\beta_j$ for positive fixed values of $\{\beta_i\}_{i=1}^N$. 
%\begin{framed}
%\begin{definition}[Randomized policy]

\paragraph{Randomized} In each slot $t$, select  link $i$ with probability $\beta_i/\sum_{j=1}^N\beta_j$. 

%\end{definition}
%\end{framed}
%The BS transmits the packet if the selected link has an undelivered packet and idles otherwise.
%\end{policy}
%
%Denote the Randomized policy as $R$.
Although this simple policy uses no information from current or past states of the network, it was shown in \cite{igorTON18} that the  Randomized policy with $\beta_i=\sqrt{w_i/p_i}$  \emph{achieves 2-optimal performance in all network configurations} $(N,p_i,w_i)$. 

Next, we describe a Max-Weight policy that leverages the knowledge of $w_i$, $p_i$ and the age vector $\agevec(t)=[\age_1(t),\cdots,\age_N(t)]$ in making scheduling decisions. Specifically, the max-weight policy is based on the linear Lyapunov function
\begin{equation}\label{eq2.Lyapunov_function}
%L(\vec{\AoI}(t))=\frac{1}{N}\sum_{i=1}^Nw_i \AoI_i(t)^2 \; , 
L(\agevec(t))=\frac{1}{N}\sum_{i=1}^N\tilde{\alpha}_i \AoI_i(t). 
\end{equation}
where $\tilde{\alpha}_i>0$ are auxiliary parameters used to tune the performance of the MW policy.

\paragraph{Max-Weight} In each slot $t$, the Max-Weight policy minimizes the one-slot Lyapunov drift 
\begin{equation}\label{eq2.def_Lyapunov_drift}
\lyapunovDrift(\agevec(t))=\mathbb{E}\left[\left. L(\agevec(t+1))-L(\agevec(t))\right|\agevec(t)\right]
\end{equation}
by selecting the link $i$ with highest value of $p_i\tilde{\alpha}_i \AoI_i(t)$, with ties broken arbitrarily. 

Similar to  the randomized policy, it was shown \cite{igorTON18} that the max-weight policy also achieves  2-optimal performance under all network configurations.  However, in practice, the Max-Weight policy achieves much better average performance than the simple randomized policy, as shown through simulations.  

%\begin{framed}
%\begin{theorem}[Performance of Max-Weight policy]\label{theo2.performance_MaxWeight}
%Consider a broadcast wireless network with parameters $(N,p_i,w_i)$ and an infinite time-horizon. The Max-Weight policy with $\tilde{\alpha}_i=\sqrt{w_i/p_i}$ has $\rho^{MW}<2$ for all network configurations $(N,p_i,w_i)$.
%\end{theorem}
%\end{framed}

The choice in \eqref{eq2.Lyapunov_function} of a linear Lyapunov function with auxiliary parameters $\tilde{\alpha}_i$ resulted in a  performance guarantee of 2-optimality. Choosing a different Lyapunov function yields a different Max-Weight policy with a  different performance guarantee.  For example, a quadratic Lyapunov function resulted in a performance guarantee of 4-optimality \cite{igorTON18}.

Finally we consider the AoI minimization problem from a different perspective and propose an Index policy \cite{RMAB}, also known as Whittle's Index policy. This policy is surprisingly similar to the Max-Weight policy and also yields strong performance. 
%by leveraging the knowledge of $w_i$, $p_i$ and $\AoI_i(t)$ in making decisions. 
Whittle's Index policy is the optimal solution to a relaxation of the Restless Multi-Armed Bandit (RMAB) problem. This low-complexity heuristic policy has been extensively used in the literature \cite{index_regularity,index_schedule,index_myopic} and is known to have a strong performance in a range of applications \cite{index_multichannelaccess,index_assympt}. The challenge associated with this approach is that the Index policy is only defined for problems that are \emph{indexable}, a condition which is often difficult to establish. A detailed introduction to the relaxed RMAB problem can be found in \cite{RMAB,RMAB_book}.  In \cite{igorTON18}  it was shown that the AoI minimization problem is indeed indexable, and  the Whittle's Index is given by
\begin{equation}\label{eq2.Index}
C_i(\AoI_i(t))=\frac{w_ip_i}{2}\AoI_i(t) \left[ \AoI_i(t) + \frac{2}{p_i}-1 \right] .
\end{equation}
\paragraph{Whittle Index} The Whittle's Index policy selects, in each slot $t$, the link $i$ with highest value of $C_i(\AoI_i(t))$, with ties being broken arbitrarily. 

Notice that the Whittle's Index policy is similar to the Max-Weight policy despite the fact that they were developed using entirely different approaches. %Whittle's Index and Max-Weight\footnote{Max-Weight with $\tilde{\alpha}_i=\sqrt{w_i/p_i}$ prioritize link $i$ according to $\tilde{\alpha}_ip_i \AoI_i(t)$, which is equivalent to $w_ip_i \AoI_i^2(t)$.} policies prioritize link $i$ according to 
%$$
%w_ip_i \AoI_i^2(t)+w_ip_i\left(\frac{2}{p_i}-1 \right) \quad \mbox{and} \quad w_ip_i \AoI_i^2(t) \; \mbox{, respectively.}
%$$
Moreover, both are equivalent to the MAF policy when the network is symmetric, implying that both are AoI-optimal when $w_i=w$ and $p_i=p$.  It was shown in  \cite{igorTON18}  that the Whittle index policy is 8-optimal in the worst case.  However, its strong performance is demonstrated through simulation.  

%Next, we derive the performance guarantee $\rho^{WI}$ for the Whittle's Index policy in general networks, with possibly different values of $w_i$ and $p_i$.
%\begin{framed}
%\begin{theorem}[Performance of Whittle's Index policy]\label{theo2.performance_Whittle}
%Consider a broadcast wireless network with parameters $(N,p_i,w_i)$ and an infinite time-horizon. The Whittle's Index policy is $\rho^{WI}$-optimal, where
%\begin{equation}\label{eq2.performance_Whittle}
%\rho^{WI}=4 \frac{\displaystyle\left(\sum_{i=1}^{N}\sqrt{\frac{w_i}{p_i}}\left(\frac{\sqrt{2}}{p_i}+\frac{1}{\sqrt{2}}\right)\right)^2}{\displaystyle\left(\sum_{i=1}^{N}\sqrt{\frac{w_i}{p_i}}\right)^2+\left(\sum_{i=1}^Nw_i\right)} \; ,
%\rho^{WI}=4\frac{\displaystyle\left(\bar{\mathbb{M}}\left[\sqrt{\frac{w_i}{p_i}}\left(\frac{\sqrt{2}}{p_i}+\frac{1}{\sqrt{2}}\right)\right]\right)^2}{\displaystyle\left(\bar{\mathbb{M}}\left[\sqrt{\frac{w_i}{p_i}}\right]\right)^2+\frac{1}{N}\bar{\mathbb{M}}\left[w_i\right]}\; .
%\end{equation}
%and 
%\begin{equation}
%\mathbf{\widetilde{w}_i} = w_i\left(\frac{\sqrt{2}}{p_i}+\frac{1}{\sqrt{2}}\right)^2\; .
%\end{equation}
%\end{theorem}
%\end{framed}
%The complete proof of Theorem~\ref{theo2.performance_Whittle} is provided in \cite{igorTON18}. Next, we evaluate the performance of the low-complexity scheduling policies discussed in this section using MATLAB simulations.

\begin{figure}[t]
\begin{center}
\includegraphics[width=0.85\columnwidth]{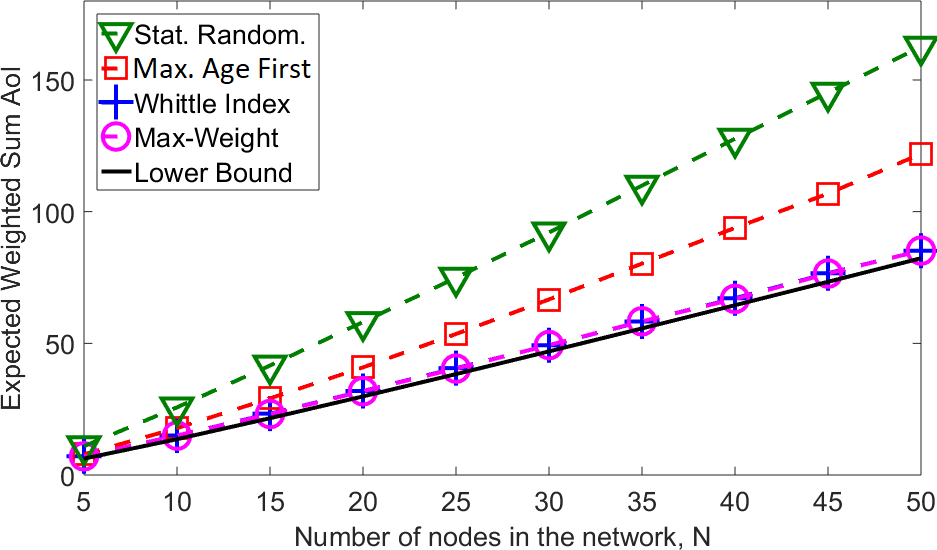}
\end{center}
\caption{Network with $T=100,000, w_i=1, p_i=i/N$. The simulation result for each policy and for each value of $N$ is an average over $10$ runs.}\label{fig2.Gen_client}
\end{figure}

%\subsubsection{Simulation Results}\label{sec2.Simulation}
In  Fig.~\ref{fig2.Gen_client}, we compare the performance of the scheduling policies in terms of the Expected Weighted Sum Age of Information.   The results show that the performances of the Max-Weight and Whittle Index policies are close to optimal and generally outperform Maximum Age First and Randomized policies.

%In Fig.~\ref{fig2.Gen_index}, we consider a network with $N=4$ links, $T=2$ slots in a slot, a total of $K=10,000$ slots, link weights $w_1=10, w_2=7, w_3=4, w_4=1$ and channel reliabilities $\{p_i\}_{i=1}^N$ chosen uniformly at random in the interval $(0.1)$. A total of $2,000$ different choices of $\{p_i\}_{i=1}^N$ are considered. Network setups are displayed in ascending order of $L_B$.

%\begin{figure}[b!]
%\begin{center}
%\includegraphics[height=5cm]{\images/2_broadcast/FIG7}
%\end{center}
%\caption{Networks with $N=4, T=2, K=10,000, w_1=10, w_2=7, w_3=4, w_4=1$ and different channel reliabilities $p_i$. For each network setup, the values of $p_i$ are sampled uniformly at random from the range $(0,1)$. For a given network setup, the performance of the policies is an average over $10$ runs. For the sake of clarity, we display $250$ out of the $2,000$ network setups by keeping only every $8$th data point.}\label{fig2.Gen_index}
%\end{figure}

%Beginning of RAJAT's chapter
\subsection{General Wireless Networks}\label{sec:wireless-general}

\newcommand{\A}{\Delta}
\newcommand{\U}{v}
\newcommand{\St}{c}

In the  previous section, we discussed several scheduling policies for minimizing age in a broadcast network. However, many of the applications in which age is an important metric involve wireless networks with more general interference constraints. 
In \cite{talak18_Mobihoc} the authors  design scheduling policies to minimize peak and average age for a wireless network with time-varying links and general interference constraints. The wireless network consists of a set of source-destination pairs, each connected by a wireless link. Each source generates information updates, which are to be sent to its destination.

The wireless  network is modeled as a graph $G = (V,E)$, where $V$ is the set of nodes and $E$ is the set of communication links between the nodes in the network. Each communication link $e \in E$ is a source-destination pair in the network. The source generates information updates that need to be communicated to the destination. Time is slotted and the duration of each slot is normalized to unity.

Wireless interference constraints limit the set of links that can be activated simultaneously.  We call a set $m \subset E$ to be a \emph{feasible activation set} if all links in $m$ can be activated simultaneously without interference, and denote by $\mathcal{A}$ the collection of all feasible activation sets. We call this the \emph{general interference model}, as it incorporates several popular interference models such as 1-hop interference, $k$-hop interference, and protocol interference models.

A non-interfering transmission over link $e$ does not always succeed due to channel errors. We let $\St_{e}(t) \in \{1,0\}$ denote the channel error process for link $e$, where $\St_{e}(t) = 1$ if a non-interfering transmission over link $e$ succeeds and $\St_{e}(t) = 0$ otherwise. We assume $\St_{e}(t)$ to be independent across links, and i.i.d. across time with $\gamma_e = \pr{\St_{e}(t) = 1} > 0$, for all $e \in E$. We assume that the channel success probabilities $\gamma_e$ are known.

We first consider  \emph{active sources} that can generate a new update packet at the beginning of each slot for transmission, while discarding old update packets that were not transmitted. Thus, for an active source, a transmitted packet always contains fresh information.  In \cite{talak18_Mobihoc} the authors design  randomized scheduling policies to minimize peak and average age. 
In the previous section, we saw a special case of a  distributed randomized stationary policy in which a single link link $e$ was activated with probability $p_e > 0$, independent across
% links and 
time slots. 

In the context of general interference, a randomized policy assigns a probability distribution $\mathbf{x} \in \mathbb{R}^{|\mathcal{A}|}$ over the collection of feasible activation sets, $\mathcal{A}$. Then, in each slot, activate the set $m \in \mathcal{A}$ with probability $x_m$, independent across time.  For this policy, the  activation frequency (probability) for link $e$ is given by $f_{e} = \sum_{m: e \in m} x_m$.%

It was shown \cite{talak18_Mobihoc}    for any stationary randomized policy that the average and peak age both equal  $\sum_{e \in E} \frac{w_e}{\gamma_e f_e}$.  Thus, the age minimization problem can be written as
\begin{align}\label{eq:opt_peak_age_problem}
\begin{aligned}
& \underset{\mathbf{x} \in \mathbb{R}^{|\mathcal{A}|}}{\text{Minimize}}
 ~~~\sum_{e \in E} \frac{w_e}{\gamma_e f_e} \\
& \text{subject to}~~ \mathbf{1}^{T}\mathbf{x} \leq 1,~\mathbf{x} \geq 0.
\end{aligned}
\end{align}
%where $f_{e}(t) = \sum_{m: e \in m} x_m$.
%RY f_e defined just above the problem
Note that the optimization is over $\mathbf{x}$, the activation probabilities of feasible activation sets $m \in \mathcal{A}$. This is because the link activation frequencies $f_e$ get completely determined by $\mathbf{x}$. The problem~\eqref{eq:opt_peak_age_problem} is a convex optimization problem in standard form~\cite{Boyd04}. The solution to it is a vector $\mathbf{x} \in \mathbb{R}^{|\mathcal{A}|}$ that defines a probability distribution over link activation sets $\mathcal{A}$, and determines a centralized stationary policy that minimizes average and peak age over all randomized policies. It was also shown in  \cite{talak18_Mobihoc} that this policy minimized peak age over all policies (not just randomized) and the average age for this policy is within a factor of $2$ from the optimal average age over all policies.

The optimization problem~\eqref{eq:opt_peak_age_problem}, although convex, has a variable space that is $|\mathcal{A}|$-dimensional, and thus, its computational complexity increases exponentially in $|V|$ and $|E|$. It is, however, possible to obtain the solution efficiently in certain specific cases. For example, under single-hop interference, where links interfere with one another if they share a node, every feasible activation set is a matching on $G$, and therefore, $\mathcal{A}$ is a collection of all matchings in $G$. As a result, the constraint set in~\eqref{eq:opt_peak_age_problem} is equal to the matching polytope~\cite{comb_opt_book}. The problem of finding an optimal schedule reduces to solving a convex optimization problem~\eqref{eq:opt_peak_age_problem} over a matching polytope. This can be efficiently solved (i.e., in polynomial time) by using the Frank-Wolfe algorithm~\cite{Garber16_FrankWolfe}, and the separation oracle for matching polytope developed in~\cite{Hajek_Sasaki_OptScheduling}.

%\subsubsection{Extensions}

Next we discuss a number of interesting extensions.

\paragraph{Buffered Sources}
In some situations it is not possible for the source to generate packets on demand, and instead each source generates update packets at random. The generated packets get queued at the MAC layer FCFS queue for transmission. In this setting, the AoI minimization problem involves optimization of both the packet generation rate and the link activation frequencies.   It was shown in  \cite{talak18_Mobihoc} that when the packets arrive according to an independent Bernoulli process, the problem of optimizing AoI can be decoupled, where the link scheduling is done according to the same randomized policy that minimized AoI in the active source case, and the packet generation rate is given by the solution to a simple optimization problem.

\paragraph{Channel State Information}
So far, we have considered error-prone wireless links where a transmission may fail due to channel errors.  In some situations, it is possible to observe the channel in advance, and decide when to transmit based on the channel conditions.  Typically, the channel is modeled as either being in either  the ``on" or  ``off" state, and a transmission is successful if it takes place when the channel is on.  The availability of channel state information (CSI) can be used to increase the achievable rate region of wireless systems.  In  \cite{talak18_WiOpt} and~\cite{talak18_ISIT}  the authors consider the impact of CSI on the age minimization problem and propose age-based scheduling policies that take CSI into account.  They show that using CSI can significantly improve average AoI. Another angle on the interplay between AoI and CSI, i.e., the age of channel state information, was considered in \cite{Farazi-KB-icassp2016,Farazi-KB-icccn2017,Klein-FHB-TW2017}. These studies recognized the importance of timely CSI in wireless networks and that CSI can be learned by direct channel estimation from a status update as well as indirectly through the contents of a status update. This work derived fundamental bounds and efficient schedules to minimize the age of global CSI in wireless networks with reciprocal channels.
%RY unfinished thought:
%Moreover, the authors 
 
\paragraph{Multi-Hop Networks} 
The multi-hop setting was first studied in the context of vehicular networks where vehicles ``piggyback'' status updates over multiple hops \cite{Kaul-YG-globecom2011piggybacking}. The multi-hop setting is interesting in that it typically removes the abstraction of the queue and explicitly considers the effect of the network topology and link contention in the analysis. The analysis in \cite{Kaul-YG-globecom2011piggybacking} and the subsequent studies in \cite{Selen-NAV-springer2013, Bedewy-SS-isit2016,asil2016farazi,arafa-age-2hop,yates2018infocom,Yates-IT2020} considered age of information in specific multi-hop network structures, e.g., line, ring, and/or two-hop networks. In~\cite{Banik-KS-ISIT2019}, sources are polled for updates by a gateway that aggregates the updates and sends them to a monitor. A general multi-hop network setting where a single-source disseminates status updates through a gateway to the network was considered in \cite{Bedewy-SS-isit2017, BedewyJournal20172}. Another general multi-hop setting in which each node is both a source and a monitor of information was considered in \cite{farazi2018SPAWC,farazi2019JCN,farazi2019INFOCOM,farazi2019ASILOMAR1,farazi2019ASILOMAR2}. These studies led to the formulation of fundamental limits and schedules that were shown to achieve an average age close to these limits. A practical age control protocol to improve AoI in multi-hop IP networks was proposed in \cite{Shreedhar2019AnInternet-of-things}. The distribution of AoI for general networks with time-invariant erasure probabilities on each link was derived in \cite{ayan2020NL}.

%\bibliographystyle{IEEEtran}
%\bibliography{aoi-bib-rajat,igor_references,ch01_sampling,ch02_scheduling} %This one works!

% In~\cite{Jiang-KZN-itc2018, Sun-Jiang-KZN-TCOM2020} the age minimization problem is viewed as a restless multi-armed bandit problem, the Whittle's index is derived and used to propose a decentralized index prioritized random access. In~\cite{Jiang-KZZN-isit2018, Jiang-KZZN-iot2019} the authors considered scheduling policies that can be implemented in a decentralized manner and are asymptotically optimal. In~\cite{Chen-Yifan-2020a} the authors considered multiaccess in which a node chooses its access probability as a function of the age of its updates. In~\cite{Bedewy2020MobiHoc} a sleep-wake strategy is designed for a network of low-powered battery constrained nodes, which use carrier sensing based access.

% !TEX root = aoisurvey.tex
\section{Applications}
\label{sec:apps}
In addition to queue/network focused analyses,  AoI has also appeared in various application areas, including  dissemination of channel state information \cite{Costa-VE-icc2015,Klein-FHB-TW2017,Farazi-KB-icccn2017,Farazi-KB-icassp2016}, timely updates via replicated servers \cite{Sang-LJ-globecom2017,Zhong-YS-aoi2018} including multi-casting networks \cite{Zhong-YS-allerton2017, Zhong-YS-spawc2018, Buyukates18, Buyukates19, Buyukates18b}, timely source coding \cite{Zhong-Yates-dcc2016lossless, Zhong-YS-isit2017, Mayekar-PT-isit2018lossless, MelihBatu1, MelihBatu2, MelihBatu4, Bastopcu20},  differential encoding of temporally correlated updates \cite{Bhambay-PP-wcnc2017}, correlated updates from multiple cameras \cite{HeFD-aoi2018}, periodic updates from correlated IoT sources \cite{Hribar-CKD-globecom2017}, mobile cloud gaming \cite{Yates-THR-infocom2017}, computation-intensive updating where updates are generated after processing raw data \cite{Bastopcu19, Bastopcu20b, Buyukates19c}, game-theoretic approaches to network resource allocation for updating sources  \cite{Nguyen-KKWE-wiopt2017,Xiao-Sun-aoi2018,garnaev2019,gac2018,nguyen2018,Gopal-Kaul-aoi2018,SGAoI2019,SGAoI2020,SGAoI2020Non,Kumar-Vaze-2020,zheng2019,ning2020}, and timely updating of researchers' citations in Google Scholar \cite{Bastopcu20a}.

\subsection{Age and Games}
 Game theory has been applied  in various settings where one or more players value timeliness \cite{Nguyen-KKWE-wiopt2017,Xiao-Sun-aoi2018,garnaev2019,gac2018,nguyen2018,Gopal-Kaul-aoi2018,SGAoI2019,SGAoI2020,SGAoI2020Non,Kumar-Vaze-2020,zheng2019,ning2020}.  This includes adversarial settings where one player aims to maintain the freshness of information updates while the other player aims to prevent this \cite{Nguyen-KKWE-wiopt2017, Xiao-Sun-aoi2018,gac2018,garnaev2019}. 
 
 In~\cite{Nguyen-KKWE-wiopt2017}, the authors formulated a two-player game between  a transmitter that aims to establish a connection to its receiver and an interferer that attempts to disrupt the connection. The players choose power levels as strategies and it is shown that both players have the same strategy at the Nash Equilibrium. It also shown that the Stackelberg strategy, when led by the interferer, dominates the Nash strategy.
 
 In~\cite{Xiao-Sun-aoi2018}, a dynamic game between a real-time monitoring system that cares about the timeliness of status updates sent over a wireless channel and an attacker that jams the channel to delay the status updates was considered. The authors proved the existence of a unique stationary equilibrium in the game and characterized the equilibrium analytically. They showed that the attacker chooses a jamming time distribution with high variance, while the system chooses a sampling policy that results in a low variance in the time between the reception of two consecutive updates. 
 
 In~\cite{garnaev2019}, authors modeled the interaction between a UAV transmitter and an adversarial interferer and showed that there exists a unique Nash equilibrium but multiple Stackelberg equilibria. In~\cite{gac2018}, the  authors considered a communication scheduling and remote estimation problem in the presence of an adversary and obtained a Nash equilibrium for the non-zero-sum dynamic game. 
 
 In~\cite{nguyen2018}, the authors formulated a two-player game to model the interaction between two transmitter-receiver pairs over an interference channel. The transmitters desire freshness of their updates at their receivers and can choose their transmit power levels. The Nash and Stackelberg strategies are derived and it is shown that the Stackelberg strategy dominates the Nash strategy.

In~\cite{Gopal-Kaul-aoi2018,SGAoI2019,SGAoI2020} authors studied age in the context of spectrum sharing between networks.  A game-theoretic approach was proposed in~\cite{Gopal-Kaul-aoi2018} to study the coexistence of Dedicated Short Range Communication (DSRC) and WiFi;  the DSRC network desires to minimize the average age of information and the WiFi network aims to maximize the average throughput. For the one-shot game,  the Nash and Stackelberg equilibrium strategies were evaluated.  The DSRC-WiFi coexistence problem from~\cite{Gopal-Kaul-aoi2018} was generalized in~\cite{SGAoI2019} to the coexistence of age and throughput optimizing networks. A repeated game approach was employed to capture the interaction of the networks over time.
This line of work was further extended \cite{SGAoI2020} to explore the possibility of cooperation between age and throughput optimizing networks using a randomized signaling device. It was shown that networks choose to cooperate only when they consist of a sufficiently small number of nodes, but  otherwise they prefer to compete. 

In~\cite{SGAoI2020Non} and \cite{Kumar-Vaze-2020}, authors studied the coexistence of nodes that value timeliness of their information at others and provided insights into how competing nodes would coexist. In~\cite{SGAoI2020Non}, authors proposed a one-shot multiple access game with nodes as players, where each node shares the spectrum using a CSMA/CA based access mechanism. Authors investigated the equilibrium strategies of nodes in each CSMA/CA slot when collision slots are shorter than successful transmissions, as well as when they are longer. They showed that when collisions are shorter, transmitting is a weakly dominant strategy. However,  when collisions are longer, no weakly dominant strategy exists and a mixed strategy Nash equilibrium was derived. In~\cite{Kumar-Vaze-2020}, authors considered a distributed competition mode where each node wants to minimize a function of its age and transmission cost but information such as the number of nodes in the network and their  strategies is not available. A learning strategy  was proposed for each node to use its current empirical average of age and transmission cost to determine its transmit probability in each slot.  They showed that for a certain set of parameters the proposed strategy converges to an equilibrium that is identified as the Nash equilibrium for a suitable virtual game.

In~\cite{zheng2019}, authors proposed a Stackelberg game between an access point and a set of  helpers that contribute toward charging a sensor via wireless power transfer. The access point would like to minimize a utility that includes the age of information from the sensor, the power transferred by it to charge the sensor, and the payments it makes to the helpers. The helpers benefit from the payments but  bear costs of  transferring power. In~\cite{ning2020}, authors designed a mobile edge computing enabled $5$G health monitoring system for wireless body area networks (WBANs) in the Internet of Medical Things (IoMT).The goal was  to minimize system-wide costs that depend on medical criticality, age of information, and energy consumption of health monitoring packets. The IoMT was divided into two sub-networks, namely, intra-WBANs and beyond-WBANs. For the intra-WBANs, the authors formulated a cooperative game to minimize the cost per patient. For the beyond-WBANs, where patients can choose to analyze the information either at local devices or at edge servers, the authors formulated a non-cooperative game and analyzed the Nash equilibrium.

\subsection{Age and Learning}
Data driven model-free learning techniques have been applied to optimizing age-of-information for different network assumptions and applications~\cite{Ceran-GG-wcnc2018, CeranTWC2019, ceran2018reinforcement, yin2020application, elgabli2019reinforcement, age_multiple_flows, AoI_adhoc, abedin2020data,yi2020deep,Elmagid2019Deep}. In addition, age-based computation distribution to servers in machine learning applications with straggling servers has been studied in \cite{Buyukates20c}. In \cite{Buyukates20c}, already aged parts of the computations are assigned to servers that are less likely to straggle in the next computation cycle.

In~\cite{Ceran-GG-wcnc2018, CeranTWC2019, ceran2018reinforcement, yin2020application, elgabli2019reinforcement, age_multiple_flows}, the goal is to  learn policies that schedule updates from one or more sources with the goal of minimizing their age at a monitor. Authors in ~\cite{Ceran-GG-wcnc2018, CeranTWC2019} aim to schedule updates of a source over an error-prone channel under different feedback mechanisms including ARQ and Hybrid ARQ. They also proposed an on-policy SARSA algorithm~\cite{sutton2018reinforcement}, with the policy modeled by a soft-max distribution, for when the channel error statistics are not known. The authors consider the case where a source may send updates to multiple users in~\cite{ceran2018reinforcement}. In~\cite{yin2020application}, the authors consider a network of devices whose updates are delivered to applications via an IoT hub. An application is updated at the end of a frame only if updates from all devices relevant to the application are scheduled for transmission during the frame. The authors use the options framework (hierarchical reinforcement learning)~\cite{sutton1999between} to model the scheduling policy as composed of two policies, the first which chooses the application that must be updated and the other that chooses how devices corresponding to the application must be scheduled. A deep Q network based architecture is proposed to learn the scheduling policy. 

In~\cite{elgabli2019reinforcement}, the authors consider a network of sensors sending their updates to a remote monitor. The network would like to optimize a weighted sum of the average age of information and the probabilities that the ages of sensors' updates at the monitor exceed a predefined threshold. The probabilities model reliability for ultra reliable low latency communication (URLLC) systems. The actor-critic~\cite{sutton2018reinforcement} algorithm A3C is used to learn a policy that chooses the sensor that must send its update. In~\cite{age_multiple_flows}, the authors empirically compare the age of information achieved using the reinforcement learning algorithms of Deep Q-Learning and Policy Gradient for scheduling packets from multiple (four) flows that arrive at a single queue single server facility. They also compare against the often analyzed maximum age first scheduling policy.

In~\cite{AoI_adhoc} the authors consider a fully connected ad hoc network in which nodes send updates to each other. They use deep Q-learning in a multi-agent setting to learn the policy that every node must use. The policy chooses the transmit power for a node at every time slot.

Works~\cite{abedin2020data,yi2020deep,Elmagid2019Deep} learn trajectories of one or more Unmanned Aerial Vehicles (UAVs) that collect information from sensors. In~\cite{abedin2020data}, mobile UAV base stations plan trajectories to obtain updates from IoT devices that must be delivered to a ground station. The UAV trajectories must ensure that average AoI at the ground station is below a certain threshold. However, one would like to choose trajectories such that energy efficiency is maximized. In~\cite{yi2020deep}, deep reinforcement learning is used to find the trajectory a UAV must take, during a fixed duration and under the constraint that it must not run out of energy, through the sensors and back to its base to minimize the weighted sum AoI of the sensor updates at the base. In a similar but different problem formulation,  the authors of~\cite{Elmagid2019Deep} propose a deep reinforcement learning algorithm to optimize (i) the trajectory a UAV must take to obtain updates from sensors on the ground and (ii) the scheduling of updates at the sensors with the goal of minimizing the sum AoI at the UAV.

\subsection{Caching}
Gao et al.~\cite{Gao2012} were the first to introduce the concept of cache freshness in opportunistic mobile networks. They assume a single data source updating multiple caching servers (nodes), wherein each caching node may further update other nodes. 
The authors in \cite{Yates-CWY-isit2017cache, Kam2017, Tang2019} use age of information together with the popularity of content for cache management. In \cite{Yates-CWY-isit2017cache}, the authors design a popularity weighted AoI metric based update policy for updating a local cache connected to a remote server via a limited capacity link. The authors show that the update rate of content in the local cache should  be proportional to the square root of its popularity.
In \cite{Tang2019} the authors propose an update policy that minimizes the average AoI of the files with respect to a given popularity distribution in a single-server single-cache system. The authors consider AoI-dependent update durations and compare their policy with the square root policy proposed in \cite{Yates-CWY-isit2017cache}.

In \cite{Kam2017}, the authors propose a caching model for requesting content based on the popularity (history of requests) and freshness (age), captured using \textit{effective age}. They propose an optimal policy that minimizes the number of missed requests while managing the contents of the cache when the cache is full and new content arrives. 
In \cite{Zhang2018}, the authors propose a cache-assisted lazy update and delivery (CALUD) to jointly optimize content freshness and service latency in vehicular networks. They consider multiple data sources periodically updating a single roadside unit (RSU).   

In \cite{Yang2019} the authors discuss the real-time performance of a cache enabled network with AoI as a freshness metric. They propose a random caching framework and show that it performs better than the most popular content based and uniform caching strategies with respect to minimizing the peak AoI. 
In \cite{Bastopcu20e} the authors consider AoI in serial and multi-access connected cache networks, and analyze age at intermediate caches and the end-users. Extending \cite{Bastopcu20e},  the authors in \cite{Bastopcu20c} consider the trade-off between obtaining a file from the source and obtaining it from the cache. While the former could entail larger file transmission times, albeit of the freshest copy, the latter may result in an older copy obtained quickly. They consider whether a file must be stored in the cache and if so then the rate at which it must be updated.

In \cite{Zhang2020} the authors propose a freshness-aware refreshing scheme that minimizes the average service delay while guaranteeing content freshness. The cached contents are refreshed on user requests if their AoI exceeds a certain threshold (refreshing window). A small window increases service delays as the latest version is fetched more often. On the other hand, a large window increases age but keeps service delays small as the cached version is sent to the user.

In~\cite{AtakanStalenessMLModels2020} the authors use the Q-learning algorithm to achieve staleness control of machine learning models available at the edge for data analytics. They consider six different performance metrics including the age of information and the value of information updates.

%In \cite{Zhang2020} too the authors use a threshold based policy for refreshing a cached content when requested by a user. A ``refreshing window" is chosen for an item so as to balance the average delay in servicing the user and content freshness. A small window increases service delays as the latest version is fetched more often. On the other hand, a large window increases age but keeps service delays small as the cached version is sent to the user.
%The authors also investigate how the refreshing window size impacts the trade-off between system delay and age. %with respect to the refreshing window size. 

%SIGMETRICS, Sandeep's work

\subsection{Protocols}
In \cite{Sonmez2018Age-of-InformationLTE} the authors measure the age of information at a remote server when using TCP/IP over a selection of access networks including WiFi, LTE, 2G/3G and Ethernet. Experiments are conducted for various application  update rates,  both in an emulated testbed and over the Internet.  In \cite{Beytur2019MeasuringConnections, Beytur2020TowardsSystems} authors analyzed age of information over real networks using different devices, protocols and networks (wired/wireless). The authors discuss the challenges associated in measuring AoI in real networks, including synchronization, selection of hardware, and choice of transport protocol and draw insights for AoI aware transmission protocols. 
In \cite{Shreedhar2018PosterInternet, Shreedhar2019AnInternet-of-things,shreedhar2018acpArXiv} the authors propose the Age Control Protocol,  an end-to-end transport protocol that minimizes the AoI of the delivered update at the monitor in a network-transparent manner. The authors detail the protocol algorithm and empirically demonstrate its efficacy using real-world experiments that have one or more devices sending updates to a server in the cloud.

%!TEX root = aoisurvey.tex
\section{Conclusion}
\label{sec:conclusion}
Age of information has emerged as an end-to-end performance metric for systems that employ status update messages. In this survey, we have seen that the concept of age is so general that age-based optimization problems can be found almost everywhere, arising in all network layers and in all system components that can be viewed as communicating via updates. A source can optimize the rate at which it submits updates or the source may adopt an  age-optimizing update transmission policy. 
%This policy may or may not employ knowledge of the state of the system.  
A service facility that processes updates can employ scheduling policies to maximize the timeliness of delivered updates. A base station can use age to schedule downlink transmissions to its users. On the uplink, users can use age-based policies to access the channel. These policies differ substantially from rate/throughput maximization, particularly for energy harvesting sources. Moreover, in these network settings, age-based update policies are practical because senders, servers and routers understand packet time-stamps, as opposed to  application-specific measurements. 

This AoI introduction has described the development of new analytical models for freshness  and tools for age analysis. These new methods have been used to characterize age in relatively abstract models of sensors, networks, and service facilities. Some key ideas have emerged: The updating process should be matched to the system that is handling the updates, neither underloading nor overloading the system. The system should aim to process new updates rather than  old. The system should avoid processing updates that lack sufficient novelty.  These observations  suggest that the introduction of timeliness requirements results in systems that work smarter not harder. 

We also have seen that AoI metrics are beginning to be applied in abstract models of various application domains ranging from UAV trajectory guidance to mobile gaming. However, it should be clear that  there remain many unanswered questions, open problems, and unexplored application areas.  

%RY
A promising area is edge cloud processing for low-latency edge-assisted applications. The system model is that the delivery of an update requires (timely) computation in the cloud. In this setting, mechanisms like service preemption  that were introduced here in the context of queues and communication links are in fact more applicable in the context of edge-cloud processing; killing a job  on a processor is a lot simpler than killing a packet in a network.

Often these edge-cloud computations will involve machine learning. The intersection of predictive machine learning and AoI could become important. Using machine learning to solve scheduling problems for reduced AoI is an obvious opportunity; but using AoI to accelerate real-time learning algorithms through prioritization of fresh data is also feasible. 
% one can also try to interpret and exploit AoI in real-time machine learning applications.

With respect to networking, an interesting open problem is how to enable ``age-optimality'' as a network service, which caters to a host of cyber-physical system applications. In such a world, all a CPS application (and the many sources and monitors that underpin its functionality) would do is open an ``age-socket.'' The network would then manage the end-to-end connections, including polling/sampling, and managing application queues. The networking stack of course performs some of these functions already, though not for timeliness.

We have already seen that age is applicable to various estimation problems. It should be apparent that functions of AoI can play an important role in scheduling feedback for control systems.  How this should be done yields problems at the intersection of sensors, processing, and communications, where there is a tradeoff between processing (e.g., compression) and communication latency.  More compression takes time and processing capacity, but can reduce communication latency.
%RY: I started working on the compression part of this with Jing Zhong in 2016. There lots to do here.

Age/distortion tradeoffs represent another aspect of these types of problems, For example, suppose on can get a low resolution image with small age or a higher resolution image with increased age. In what settings would one be preferable?  How should these tradeoffs be optimized in different settings? 
% There was an arxiv paper about this in Dec 2019 called "Energy Harvesting Powered Sensing in IoT: Timeliness Versus Distortion".

For remote estimation and control, a (big) open question is how to design sampling, quantization, coding, and control for multi-dimensional signals that could be correlated and non-Markovian. Of course, this will be  related to age/distortion tradeoffs.

In each of these potential problem areas, there are likely to be a host of abstract but informative problem formulations. The solutions will perhaps answer the question of when  or whether AoI based problem formulations will become  a useful practical mechanism for optimizing the design and operation of widely deployed cyberphysical systems.

% For now, let's use alphabetical citation references. It will help us keep track of duplication.
\begin{singlespace}
\bibliographystyle{IEEEtran}
\bibliography{AOI-2020-05, aoi-bib-rajat,igor_references,ch01_sampling,%
ch02_scheduling,apps-sanjit,aoi_eh_summary,aoi_eh,lib_v1_melih,multihop}

\begin{thebibliography}{100}
\providecommand{\url}[1]{#1}
\csname url@samestyle\endcsname
\providecommand{\newblock}{\relax}
\providecommand{\bibinfo}[2]{#2}
\providecommand{\BIBentrySTDinterwordspacing}{\spaceskip=0pt\relax}
\providecommand{\BIBentryALTinterwordstretchfactor}{4}
\providecommand{\BIBentryALTinterwordspacing}{\spaceskip=\fontdimen2\font plus
\BIBentryALTinterwordstretchfactor\fontdimen3\font minus
  \fontdimen4\font\relax}
\providecommand{\BIBforeignlanguage}[2]{{%
\expandafter\ifx\csname l@#1\endcsname\relax
\typeout{** WARNING: IEEEtran.bst: No hyphenation pattern has been}%
\typeout{** loaded for the language `#1'. Using the pattern for}%
\typeout{** the default language instead.}%
\else
\language=\csname l@#1\endcsname
\fi
#2}}
\providecommand{\BIBdecl}{\relax}
\BIBdecl

\bibitem{Parvez-RGSD-commsurveys2018}
I.~{Parvez}, A.~{Rahmati}, I.~{Guvenc}, A.~I. {Sarwat}, and H.~{Dai}, ``A
  survey on low latency towards {5G}: {RAN}, core network and caching
  solutions,'' \emph{IEEE Communications Surveys Tutorials}, vol.~20, no.~4,
  pp. 3098--3130, 2018.

\bibitem{Popovski-SNCATB-comm2019}
P.~{Popovski}, {\v C}.~{Stefanovic}, J.~J. {Nielsen}, E.~{de Carvalho},
  M.~{Angjelichinoski}, K.~F. {Trillingsgaard}, and A.~{Bana}, ``Wireless
  access in ultra-reliable low-latency communication {(URLLC)},'' \emph{IEEE
  Transactions on Communications}, vol.~67, no.~8, pp. 5783--5801, 2019.

\bibitem{Sachs-AACLRSW-ieeeproc2019}
J.~{Sachs}, L.~A.~A. {Andersson}, J.~{Araujo}, C.~{Curescu}, J.~{Lundsjo},
  G.~{Rune}, E.~{Steinbach}, and G.~{Wikstrom}, ``Adaptive {5G} low-latency
  communication for tactile internet services,'' \emph{Proceedings of the
  IEEE}, vol. 107, no.~2, pp. 325--349, 2019.

\bibitem{Kaul-YG-infocom2012}
S.~Kaul, R.~Yates, and M.~Gruteser, ``Real-time status: How often should one
  update?'' in \emph{Proc. IEEE INFOCOM}, March 2012, pp. 2731--2735.

\bibitem{Karakasidis-VP-iqis2005etl}
A.~Karakasidis, P.~Vassiliadis, and E.~Pitoura, ``{ETL} queues for active data
  warehousing,'' in \emph{Proc. 2nd international workshop on Information
  quality in information systems (IQIS)}, 2005, pp. 28--39.

\bibitem{Yu-BS-sigcomm1999scalable}
H.~Yu, L.~Breslau, and S.~Shenker, ``A scalable web cache consistency
  architecture,'' \emph{{SIGCOMM} Comput. Commun. Rev.}, vol.~29, no.~4, pp.
  163--174, Aug. 1999.

\bibitem{Chen1998}
{Shigang Chen} and K.~{Nahrstedt}, ``Distributed {QoS} routing with imprecise
  state information,'' in \emph{Proc. 7th International Conference on Computer
  Communications and Networks}, 1998, pp. 614--621.

\bibitem{Hu-Johnson-pomc2002ensuring}
Y.~C. Hu and D.~B. Johnson, ``Ensuring cache freshness in on-demand ad hoc
  network routing protocols,'' in \emph{{ACM} international workshop on
  Principles of Mobile Computing (POMC)}, 2002, pp. 25--30.

\bibitem{Giruka-Singhal-wowmom2005hello}
V.~C. Giruka and M.~Singhal, ``Hello protocols for ad-hoc networks: overhead
  and accuracy tradeoffs,'' in \emph{World of Wireless Mobile and Multimedia
  Networks, {WoWMoM}, Sixth {IEEE} International Symposium on}.\hskip 1em plus
  0.5em minus 0.4em\relax {IEEE}, Jun. 2005, pp. 354--361.

\bibitem{Song-Liu-ISCSA1990}
X.~Song and J.~W.-S. Liu, ``Performance of multiversion concurrency control
  algorithms in maintaining temporal consistency,'' in \emph{Proceedings.,
  Fourteenth Annual International Computer Software and Applications
  Conference}.\hskip 1em plus 0.5em minus 0.4em\relax IEEE, 1990, pp. 132--139.

\bibitem{Xiong-Ramamritham-real1999deriving}
M.~Xiong and K.~Ramamritham, ``Deriving deadlines and periods for real-time
  update transactions,'' in \emph{The 20th {IEEE} {Real-Time} Systems
  Symposium}.\hskip 1em plus 0.5em minus 0.4em\relax {IEEE}, 1999, pp. 32--43.

\bibitem{Cho-GM-tods2003effective}
J.~Cho and H.~Garcia-Molina, ``Effective page refresh policies for web
  crawlers,'' \emph{ACM Transactions on Database Systems (TODS)}, vol.~28,
  no.~4, pp. 390--426, 2003.

\bibitem{Ioannidis2009}
S.~Ioannidis, A.~Chaintreau, and L.~Massoulie, ``Optimal and scalable
  distribution of content updates over a mobile social network,'' in
  \emph{Proc. IEEE INFOCOM}, 2009.

\bibitem{Kaul-GRK-secon2011}
S.~Kaul, M.~Gruteser, V.~Rai, and J.~Kenney, ``Minimizing age of information in
  vehicular networks,'' in \emph{{IEEE} Conference on Sensor, Mesh and Ad Hoc
  Communications and Networks {(SECON)}}, 2011.

\bibitem{Kaul-YG-globecom2011piggybacking}
S.~K. Kaul, R.~D. Yates, and M.~Gruteser, ``On piggybacking in vehicular
  networks,'' in \emph{{IEEE} Global Telecommunications Conference, {GLOBECOM}
  2011}, Dec. 2011.

\bibitem{Ross1996stochastic}
S.~M. Ross, \emph{Stochastic Processes}, 2nd~ed.\hskip 1em plus 0.5em minus
  0.4em\relax John Wiley \& Sons, 1996.

\bibitem{Gallager2013stochastic}
R.~G. Gallager, \emph{Stochastic processes: theory for applications}.\hskip 1em
  plus 0.5em minus 0.4em\relax Cambridge University Press, 2013.

\bibitem{Costa-CE-IT2016management}
M.~Costa, M.~Codreanu, and A.~Ephremides, ``On the age of information in status
  update systems with packet management,'' \emph{IEEE Trans. Info. Theory},
  vol.~62, no.~4, pp. 1897--1910, April 2016.

\bibitem{Costa-CE-isit2014}
------, ``Age of information with packet management,'' in \emph{Proc. IEEE
  Int'l. Symp. Info. Theory (ISIT)}, June 2014, pp. 1583--1587.

\bibitem{Inoue-MTT-IT2019}
Y.~{Inoue}, H.~{Masuyama}, T.~{Takine}, and T.~{Tanaka}, ``A general formula
  for the stationary distribution of the age of information and its application
  to single-server queues,'' \emph{IEEE Trans. Info. Theory}, vol.~65, no.~12,
  pp. 8305--8324, 2019.

\bibitem{Hespanha-2006modelling}
J.~Hespanha, ``Modelling and analysis of stochastic hybrid systems,'' \emph{IEE
  Proceedings-Control Theory and Applications}, vol. 153, no.~5, pp. 520--535,
  2006.

\bibitem{Yates-Kaul-IT2019}
R.~D. Yates and S.~K. Kaul, ``The age of information: Real-time status updating
  by multiple sources,'' \emph{IEEE Trans. Info. Theory}, vol.~65, no.~3, pp.
  1807--1827, March 2019.

\bibitem{Vermes-1980}
D.~Vermes, ``Optimal dynamic control of a useful class of randomly jumping
  processes,'' International Institute for Applied Systems Analysis, Tech. Rep.
  PP-80-015, 1980.

\bibitem{Davis-1984}
M.~H.~A. Davis, ``Piecewise-deterministic {Markov} processes: a general class
  of nondiffusion stochastic models,'' \emph{J. Roy. Statist. Soc.}, vol.~46,
  pp. 353--388, 1984.

\bibitem{Deville-DDZ-siam2016moment}
L.~DeVille, S.~Dhople, A.~D. Dom{\'\i}nguez-Garc{\'\i}a, and J.~Zhang, ``Moment
  closure and finite-time blowup for piecewise deterministic {Markov}
  processes,'' \emph{SIAM Journal on Applied Dynamical Systems}, vol.~15,
  no.~1, pp. 526--556, 2016.

\bibitem{Yates-IT2020}
R.~D. Yates, ``The age of information in networks: Moments, distributions, and
  sampling,'' \emph{IEEE Trans. Info. Theory}, 2020.

\bibitem{Ornee2019}
T.~Z. Ornee and Y.~Sun, ``Sampling for remote estimation through queues: Age of
  information and beyond,'' in \emph{Proc.~IEEE/IFIP WiOpt}, 2019.

\bibitem{Ornee2020ArXiv}
------, ``Sampling for remote estimation through queues: Age of information and
  beyond,'' 2020, submitted to \emph{IEEE/ACM Trans. Netw.},
  https://arxiv.org/abs/1902.03552.

\bibitem{Champati2019}
J.~P. Champati, M.~H. Mamduhi, K.~H. Johansson, and J.~Gross, ``Performance
  characterization using {AoI} in a single-loop networked control system,'' in
  \emph{Proc.~IEEE INFOCOM Age of Information Workshop}, 2019.

\bibitem{Markus2019}
M.~Kl{\"u}gel, M.~H. Mamduhi, S.~Hirche, and W.~Kellerer, ``{AoI}-penalty
  minimization for networked control systems with packet loss,'' in
  \emph{Proc.~IEEE INFOCOM Age of Information Workshop}, 2019.

\bibitem{SunInfocom2016}
Y.~{Sun}, E.~{Uysal-Biyikoglu}, R.~{Yates}, C.~E. {Koksal}, and N.~B. {Shroff},
  ``Update or wait: How to keep your data fresh,'' in \emph{IEEE INFOCOM},
  2016, pp. 1--9.

\bibitem{SunJournal2017}
------, ``Update or wait: How to keep your data fresh,'' \emph{IEEE Trans.~Inf.
  Theory}, vol.~63, no.~11, pp. 7492--7508, Nov. 2017.

\bibitem{Kosta2017}
A.~Kosta, N.~Pappas, A.~Ephremides, and V.~Angelakis, ``Age and value of
  information: Non-linear age case,'' in \emph{Proc. IEEE Int’l. Symp. Info.
  Theory (ISIT)}, June 2017, pp. 326--330.

\bibitem{SunNonlinear2019}
Y.~Sun and B.~Cyr, ``Sampling for data freshness optimization: Non-linear age
  functions,'' \emph{J. Commun. Netw.}, vol.~21, no.~3, pp. 204--219, 2019.

\bibitem{KostaTCOM2020}
A.~{Kosta}, N.~{Pappas}, A.~{Ephremides}, and V.~{Angelakis}, ``The cost of
  delay in status updates and their value: Non-linear ageing,'' \emph{IEEE
  Trans. Commun., in press}, 2020.

\bibitem{Shapiro1999}
C.~Shapiro and H.~Varian, \emph{Information Rules: A Strategic Guide to the
  Network Economy}.\hskip 1em plus 0.5em minus 0.4em\relax Harvard Business
  Press, 1999.

\bibitem{Even:2007}
A.~Even and G.~Shankaranarayanan, ``Utility-driven assessment of data
  quality,'' \emph{SIGMIS Database}, vol.~38, no.~2, pp. 75--93, May 2007.

\bibitem{Heinrich:2009}
B.~Heinrich, M.~Klier, and M.~Kaiser, ``A procedure to develop metrics for
  currency and its application in {CRM},'' \emph{J. Data and Information
  Quality}, vol.~1, no.~1, pp. 5:1--5:28, 2009.

\bibitem{Altman2011}
E.~Altman, R.~El-Azouzi, D.~S. Menasche, and Y.~Xu, ``Forever young: Aging
  control for smartphones in hybrid networks,'' 2010,
  https://arxiv.org/abs/1009.4733.

\bibitem{Razniewski:2016}
S.~Razniewski, ``Optimizing update frequencies for decaying information,'' in
  \emph{Proceedings of the 25th ACM International on Conference on Information
  and Knowledge Management}, 2016, pp. 1191--1200.

\bibitem{Bedewy-SS-isit2016}
A.~M. Bedewy, Y.~Sun, and N.~B. Shroff, ``Optimizing data freshness,
  throughput, and delay in multi-server information-update systems,'' in
  \emph{Proc. IEEE Int'l. Symp. Info. Theory (ISIT)}, 2016, pp. 2569--2574.

\bibitem{Bedewy-SS-isit2017}
------, ``Age-optimal information updates in multihop networks,'' in
  \emph{Proc. IEEE Int'l. Symp. Info. Theory (ISIT)}, June 2017, pp. 576--580.

\bibitem{BedewyJournal2017}
------, ``Minimizing the age of information through queues,'' \emph{IEEE
  Trans.~Inf. Theory}, vol.~65, no.~8, pp. 5215--5232, Aug. 2019.

\bibitem{BedewyJournal20172}
------, ``The age of information in multihop networks,'' \emph{ACM/IEEE
  Trans.~Netw.}, vol.~27, no.~3, pp. 1248 -- 1257, Jun. 2019.

\bibitem{Sun-UBK-aoi2018}
Y.~Sun, E.~Uysal-Biyikoglu, and S.~Kompella, ``Age-optimal updates of multiple
  information flows,'' in \emph{IEEE Conference on Computer Communications
  (INFOCOM) Workshops}, April 2018, pp. 136--141.

\bibitem{maatouk2020status}
A.~Maatouk, Y.~Sun, A.~Ephremides, and M.~Assaad, ``Status updates with
  priorities: Lexicographic optimality,'' in \emph{IEEE/IFIP WiOpt}, 2020.

\bibitem{Truong2013}
K.~T. Truong and R.~W. Heath, ``Effects of channel aging in massive {MIMO}
  systems,'' \emph{J. Commun. Netw.}, vol.~15, no.~4, pp. 338--351, Aug 2013.

\bibitem{SunISIT2017}
Y.~Sun, Y.~Polyanskiy, and E.~Uysal-Biyikoglu, ``Remote estimation of the
  {Wiener} process over a channel with random delay,'' in \emph{Proc. IEEE
  Int’l. Symp. Info. Theory (ISIT)}, 2017.

\bibitem{SunTIT2020}
Y.~{Sun}, Y.~{Polyanskiy}, and E.~{Uysal}, ``Sampling of the {Wiener} process
  for remote estimation over a channel with random delay,'' \emph{IEEE
  Trans.~Inf. Theory}, vol.~66, no.~2, pp. 1118--1135, Feb 2020.

\bibitem{SoleymaniArXiv2018}
T.~Soleymani, J.~S. Baras, and K.~H. Johansson, ``Stochastic control with stale
  information--part i: Fully observable systems,'' 2018,
  https://arxiv.org/abs/1810.10983.

\bibitem{SunSPAWC2018}
Y.~Sun and B.~Cyr, ``Information aging through queues: A mutual information
  perspective,'' in \emph{Proc. IEEE SPAWC Workshop}, 2018.

\bibitem{Cover}
T.~Cover and J.~Thomas, \emph{Elements of Information Theory}.\hskip 1em plus
  0.5em minus 0.4em\relax John Wiley and Sons, 1991.

\bibitem{Soleymani2016-1}
T.~Soleymani, S.~Hirche, and J.~S. Baras, ``Optimal self-driven sampling for
  estimation based on value of information,'' in \emph{Proceedings of the 13th
  International Workshop on Discrete Event Systems (WODES)}, 2016.

\bibitem{Soleymani2016-2}
------, ``Maximization of information in energy-limited directed
  communication,'' in \emph{Proc. European Control Conference (ECC)}, 2016.

\bibitem{Soleymani2016-3}
------, ``Optimal stationary self-triggered sampling for estimation,'' in
  \emph{Proc. IEEE CDC}, 2016.

\bibitem{Bastopcu18}
M.~Bastopcu and S.~Ulukus, ``Age of information with soft updates,'' in
  \emph{Allerton Conference}, October 2018.

\bibitem{Bastopcu19b}
------, ``Minimizing age of information with soft updates,'' \emph{Journal of
  Communications and Networks, special issue on Age of Information}, vol.~21,
  no.~3, pp. 233--243, July 2019.

\bibitem{StochasticOrderBook}
M.~Shaked and J.~G. Shanthikumar, \emph{Stochastic Orders}.\hskip 1em plus
  0.5em minus 0.4em\relax Springer, 2007.

\bibitem{Champati-AG-aoi2018}
J.~P. Champati, H.~Al-Zubaidy, and J.~Gross, ``Statistical guarantee
  optimization for age of information for the {D/G/1} queue,'' in \emph{IEEE
  Conference on Computer Communications (INFOCOM) Workshops}, April 2018, pp.
  130--135.

\bibitem{soysal18}
A.~Soysal and S.~Ulukus, ``Age of information in {G/G/1/1} systems,'' in
  \emph{Asilomar Conference}, November 2019.

\bibitem{soysal19}
------, ``Age of information in {G/G/1/1} systems: Age expressions, bounds,
  special cases, and optimization,'' May 2019, available on arXiv: 1905.13743.

\bibitem{Buyukates20a}
B.~Buyukates and S.~Ulukus, ``Age of information with {Gilbert-Elliot} servers
  and samplers,'' in \emph{CISS}, March 2020.

\bibitem{Kam-KNWE-isit2016deadline}
C.~Kam, S.~Kompella, G.~D. Nguyen, J.~Wieselthier, and A.~Ephremides, ``Age of
  information with a packet deadline,'' in \emph{Proc. IEEE Int'l. Symp. Info.
  Theory (ISIT)}, 2016, pp. 2564--2568.

\bibitem{Wang-FY-spawc2018}
B.~{Wang}, S.~{Feng}, and J.~{Yang}, ``To skip or to switch? {Minimizing} age
  of information under link capacity constraint,'' in \emph{2018 IEEE 19th
  International Workshop on Signal Processing Advances in Wireless
  Communications (SPAWC)}, June 2018, pp. 1--5.

\bibitem{Chen-Huang-isit2016}
K.~Chen and L.~Huang, ``Age-of-information in the presence of error,'' in
  \emph{Proc. IEEE Int'l. Symp. Info. Theory (ISIT)}, 2016, pp. 2579--2584.

\bibitem{Huang-Qian-globecom2017}
L.~Huang and L.~P. Qian, ``Age of information for transmissions over {Markov}
  channels,'' in \emph{IEEE Global Communications Conference (GLOBECOM)}, Dec
  2017.

\bibitem{Kaul-YG-ciss2012}
S.~Kaul, R.~Yates, and M.~Gruteser, ``Status updates through queues,'' in
  \emph{Conf. on Information Sciences and Systems (CISS)}, Mar. 2012.

\bibitem{Yates-Kaul-isit2012}
R.~Yates and S.~Kaul, ``Real-time status updating: Multiple sources,'' in
  \emph{Proc. IEEE Int'l. Symp. Info. Theory (ISIT)}, Jul. 2012.

\bibitem{Huang-Modiano-isit2015}
L.~Huang and E.~Modiano, ``Optimizing age-of-information in a multi-class
  queueing system,'' in \emph{Proc. IEEE Int'l. Symp. Info. Theory (ISIT)},
  Jun. 2015.

\bibitem{Najm-Telatar-aoi2018}
E.~Najm and E.~Telatar, ``Status updates in a multi-stream {M/G/1/1} preemptive
  queue,'' in \emph{IEEE Conference on Computer Communications (INFOCOM)
  Workshops}, April 2018, pp. 124--129.

\bibitem{Moltafet-LC-comm2020}
M.~{Moltafet}, M.~{Leinonen}, and M.~{Codreanu}, ``On the age of information in
  multi-source queueing models,'' \emph{IEEE Transactions on Communications},
  pp. 1--1, 2020, {IEEE} Early Access.

\bibitem{Najm-YS-isit2017}
E.~Najm, R.~Yates, and E.~Soljanin, ``Status updates through {M/G/1/1} queues
  with {HARQ},'' in \emph{Proc. IEEE Int'l. Symp. Info. Theory (ISIT)}, Jun.
  2017, pp. 131--135.

\bibitem{Kaul-Yates-ciss2020}
S.~K. {Kaul} and R.~D. {Yates}, ``Timely updates by multiple sources: The
  {M/M/1} queue revisited,'' in \emph{54th Annual Conference on Information
  Sciences and Systems (CISS)}, 2020, pp. 1--6.

\bibitem{Najm-NT-IT2020}
E.~{Najm}, R.~{Nasser}, and E.~{Telatar}, ``Content based status updates,''
  \emph{IEEE Trans. Info. Theory}, vol.~66, no.~6, pp. 3846--3863, 2020.

\bibitem{Kaul-Yates-isit2018priority}
S.~Kaul and R.~Yates, ``Age of information: Updates with priority,'' in
  \emph{Proc. IEEE Int'l. Symp. Info. Theory (ISIT)}, Jun. 2018, pp.
  2644--2648.

\bibitem{MaatoukISIT2019}
A.~{Maatouk}, M.~{Assaad}, and A.~{Ephremides}, ``Age of information with
  prioritized streams: When to buffer preempted packets?'' in \emph{Proc. IEEE
  Int’l. Symp. Info. Theory (ISIT)}, July 2019, pp. 325--329.

\bibitem{Kam-KE-isit2014diversity}
C.~Kam, S.~Kompella, and A.~Ephremides, ``Effect of message transmission
  diversity on status age,'' in \emph{Proc. IEEE Int'l. Symp. Info. Theory
  (ISIT)}, June 2014, pp. 2411--2415.

\bibitem{Kam-KE-isit2013random}
------, ``Age of information under random updates,'' in \emph{Proc. IEEE Int'l.
  Symp. Info. Theory (ISIT)}, 2013, pp. 66--70.

\bibitem{Kam-KNE-IT2016diversity}
C.~Kam, S.~Kompella, G.~D. Nguyen, and A.~Ephremides, ``Effect of message
  transmission path diversity on status age,'' \emph{IEEE Trans. Info. Theory},
  vol.~62, no.~3, pp. 1360--1374, Mar. 2016.

\bibitem{Yates-isit2018}
R.~D. Yates, ``Status updates through networks of parallel servers,'' in
  \emph{Proc. IEEE Int'l. Symp. Info. Theory (ISIT)}, Jun. 2018, pp.
  2281--2285.

\bibitem{Tripathi-Moharir-globecom2017}
V.~Tripathi and S.~Moharir, ``Age of information in multi-source systems,'' in
  \emph{IEEE Global Communications Conference (GLOBECOM)}, Dec 2017.

\bibitem{Bruce2005}
B.~Hajek and P.~Seri, ``Lex-optimal online multiclass scheduling with hard
  deadlines,'' \emph{Mathematics of Operations Research}, vol.~30, no.~3, pp.
  562--596, 2005.

\bibitem{jing-age-online}
X.~Wu, J.~Yang, and J.~Wu, ``Optimal status update for age of information
  minimization with an energy harvesting source,'' \emph{IEEE Trans. Green
  Commun. Netw.}, vol.~2, no.~1, pp. 193--204, March 2018.

\bibitem{arafa-age-online-finite}
A.~Arafa, J.~Yang, S.~Ulukus, and H.~V. Poor, ``Age-minimal transmission for
  energy harvesting sensors with finite batteries: Online policies,''
  \emph{IEEE Trans. Inf. Theory}, vol.~66, no.~1, pp. 534--556, January 2020.

\bibitem{jing-age-erasures-infinite-jour}
S.~Feng and J.~Yang, ``Age of information minimization for an energy harvesting
  source with updating erasures: Without and with feedback,'' available Online:
  ar{X}iv:1808.05141.

\bibitem{arafa-age-erasure-no-fb}
A.~Arafa, J.~Yang, S.~Ulukus, and H.~V. Poor, ``Online timely status updates
  with erasures for energy harvesting sensors,'' in \emph{Proc. Allerton},
  October 2018.

\bibitem{arafa-age-erasure-fb}
------, ``Using erasure feedback for online timely updating with an energy
  harvesting sensor,'' in \emph{Proc. IEEE ISIT}, July 2019.

\bibitem{Yates-isit2015}
R.~Yates, ``Lazy is timely: Status updates by an energy harvesting source,'' in
  \emph{Proc. IEEE Int'l. Symp. Info. Theory (ISIT)}, June 2015, pp.
  3008--3012.

\bibitem{farazi-K-B-ISIT2018-aoi-eh-preempt}
S.~Farazi, A.~G. Klein, and D.~R.~B. III, ``Age of information in energy
  harvesting status update systems: When to preempt in service?'' in
  \emph{Proc. IEEE ISIT}, June 2018.

\bibitem{Farazi-KB-aoi2018}
S.~Farazi, A.~G. Klein, and D.~R. Brown, ``Average age of information for
  status update systems with an energy harvesting server,'' in \emph{IEEE
  Conference on Computer Communications (INFOCOM) Workshops}, April 2018, pp.
  112--117.

\bibitem{elif_age_eh}
B.~T. Bacinoglu, E.~T. Ceran, and E.~Uysal-Biyikoglu, ``Age of information
  under energy replenishment constraints,'' in \emph{Proc. ITA}, February 2015.

\bibitem{liu-age-eh-sensing}
W.~Liu, X.~Zhou, S.~Durrani, H.~Mehrpouyan, and S.~D. Blostein, ``Energy
  harvesting wireless sensor networks: Delay analysis considering energy costs
  of sensing and transmission,'' \emph{IEEE Trans. Wireless Commun.}, vol.~15,
  no.~7, pp. 4635--4650, July 2016.

\bibitem{arafa-age-2hop}
A.~Arafa and S.~Ulukus, ``Age-minimal transmission in energy harvesting two-hop
  networks,'' in \emph{Proc. IEEE Globecom}, December 2017.

\bibitem{arafa-aoi-eh-2hop-inf-battery}
------, ``Timely updates in energy harvesting two-hop networks: Offline and
  online policies,'' \emph{IEEE Trans. Wireless Commun.}, vol.~18, no.~8, pp.
  4017--4030, August 2019.

\bibitem{arafa-age-var-serv}
------, ``Age minimization in energy harvesting communications:
  Energy-controlled delays,'' in \emph{Proc. Asilomar}, October 2017.

\bibitem{elif-age-Emax}
B.~T. Bacinoglu and E.~Uysal-Biyikoglu, ``Scheduling status updates to minimize
  age of information with an energy harvesting sensor,'' in \emph{Proc. IEEE
  ISIT}, June 2017.

\bibitem{jing-age-error-infinite-no-fb}
S.~Feng and J.~Yang, ``Optimal status updating for an energy harvesting sensor
  with a noisy channel,'' in \emph{Proc. IEEE Infocom}, April 2018.

\bibitem{jing-age-error-infinite-w-fb}
------, ``Minimizing age of information for an energy harvesting source with
  updating failures,'' in \emph{Proc. IEEE ISIT}, June 2018.

\bibitem{baknina-age-coding}
A.~Baknina and S.~Ulukus, ``Coded status updates in an energy harvesting
  erasure channel,'' in \emph{Proc. CISS}, March 2018.

\bibitem{ZhengTWC2019}
X.~{Zheng}, S.~{Zhou}, Z.~{Jiang}, and Z.~{Niu}, ``Closed-form analysis of
  non-linear age of information in status updates with an energy harvesting
  transmitter,'' \emph{IEEE Trans. Wireless Commun.}, vol.~18, no.~8, pp.
  4129--4142, Aug 2019.

\bibitem{baknina-updt-info}
A.~Baknina, O.~Ozel, J.~Yang, S.~Ulukus, and A.~Yener, ``Sending information
  through status updates,'' in \emph{Proc. IEEE ISIT}, June 2018.

\bibitem{arafa-age-rbr}
A.~Arafa, J.~Yang, and S.~Ulukus, ``Age-minimal online policies for energy
  harvesting sensors with random battery recharges,'' in \emph{Proc. IEEE ICC},
  May 2018.

\bibitem{arafa-age-sgl}
A.~Arafa, J.~Yang, S.~Ulukus, and H.~V. Poor, ``Age-minimal online policies for
  energy harvesting sensors with incremental battery recharges,'' in
  \emph{Proc. ITA}, February 2018.

\bibitem{elif-age-online-threshold}
B.~T. Bacinoglu, Y.~Sun, E.~Uysal-Biyikoglu, and V.~Mutlu, ``Achieving the
  age-energy tradeoff with a finite-battery energy harvesting source,'' in
  \emph{Proc. IEEE ISIT}, June 2018.

\bibitem{bacinoglu-aoi-eh-finite-gnrl-pnlty}
------, ``Optimal status updating with a finite-battery energy harvesting
  source,'' \emph{J. Commun. Netw.}, vol.~21, no.~3, pp. 280--294, June 2019.

\bibitem{krikidis-aoi-wpt}
I.~Krikidis, ``Average age of information in wireless powered sensor
  networks,'' \emph{IEEE Wireless Commun. Lett.}, vol.~8, no.~2, pp. 628--631,
  April 2019.

\bibitem{chen-aoi-eh-mac-het}
Z.~Chen, N.~Pappas, E.~Bjornson, and E.~G. Larsson, ``Age of information in a
  multiple access channel with heterogeneous traffic and an energy harvesting
  node,'' in \emph{Proc. IEEE Infocom}, May 2019.

\bibitem{Leng-Yener-2019TCCN}
S.~{Leng} and A.~{Yener}, ``Age of information minimization for an energy
  harvesting cognitive radio,'' \emph{IEEE Transactions on Cognitive
  Communications and Networking}, vol.~5, no.~2, pp. 427--439, 2019.

\bibitem{stamatakis-aoi-eh-alarm}
G.~Stamatakis, N.~Pappas, and A.~Traganitis, ``Control of status updates for
  energy harvesting devices that monitor processes with alarms,'' available
  Online: ar{X}iv:1907.03826.

\bibitem{rafiee-aoi-eh-drop}
P.~Rafiee and O.~Ozel, ``Active status update packet drop control in an energy
  harvesting node,'' available Online: ar{X}iv:1911.01407.

\bibitem{ozel-aoi-eh-sensing}
O.~Ozel, ``Timely status updating through intermittent sensing and
  transmission,'' available Online: ar{X}iv:2001.01122.

\bibitem{dong-aoi-mse}
Y.~Dong, P.~Fan, and K.~B. Letaief, ``Energy harvesting powered sensing in
  {IoT}: Timeliness versus distortion,'' available Online: ar{X}iv:1912.12427.

\bibitem{Tunc-Panwar-2019}
C.~Tunc and S.~Panwar, ``Optimal transmission policies for energy harvesting
  age of information systems with battery recovery,'' in \emph{2019 53rd
  Asilomar Conference on Signals, Systems, and Computers}.\hskip 1em plus 0.5em
  minus 0.4em\relax IEEE, 2019, pp. 2012--2016.

\bibitem{zheng-aoi-eh-queue}
X.~Zheng, S.~Zhou, Z.~Jiang, and Z.~Niu, ``Closed-form analysis of non-linear
  age-of-information in status updates with an energy harvesting transmitter,''
  \emph{IEEE Trans. Wireless Commun.}, vol.~18, no.~8, pp. 4129--4142, August
  2019.

\bibitem{Gu-CZLV2019}
Y.~Gu, H.~Chen, Y.~Zhou, Y.~Li, and B.~Vucetic, ``Timely status update in
  internet of things monitoring systems: An age-energy tradeoff,'' \emph{IEEE
  Internet of Things Journal}, vol.~6, no.~3, pp. 5324--5335, 2019.

\bibitem{Ceran-GG-wcnc2018}
E.~T. Ceran, D.~G{\"u}nd{\"u}z, and A.~Gy{\"o}rgy, ``Average age of information
  with hybrid {ARQ} under a resource constraint,'' in \emph{2018 IEEE Wireless
  Communications and Networking Conference (WCNC)}, April 2018.

\bibitem{CeranTWC2019}
E.~T. {Ceran}, D.~{G\"und\"uz}, and A.~{Gy\"orgy}, ``Average age of information
  with hybrid {ARQ} under a resource constraint,'' \emph{IEEE Trans. Wireless
  Commun.}, vol.~18, no.~3, pp. 1900--1913, March 2019.

\bibitem{Abd-Elmagid-Dhillon-vt2019}
M.~A. {Abd-Elmagid} and H.~S. {Dhillon}, ``Average peak age-of-information
  minimization in uav-assisted iot networks,'' \emph{IEEE Transactions on
  Vehicular Technology}, vol.~68, no.~2, pp. 2003--2008, 2019.

\bibitem{Hu-Dong-JCN-2019}
C.~Hu and Y.~Dong, ``Age of information of two-way data exchanging systems with
  power-splitting,'' \emph{Journal of Communications and Networks}, vol.~21,
  no.~3, pp. 295--306, 2019.

\bibitem{Gindullina-Badia-Gunduz-2020}
E.~Gindullina, L.~Badia, and D.~G{\"u}nd{\"u}z, ``Age-of-information with
  information source diversity in an energy harvesting system,'' \emph{arXiv
  preprint arXiv:2004.11135}, 2020.

\bibitem{TsaiISIT2020}
C.-H. Tsai and C.-C. Wang, ``Age-of-information revisited: Two-way delay and
  distribution-oblivious online algorithm,'' in \emph{Proc. IEEE Int’l. Symp.
  Info. Theory (ISIT)}, June 2020.

\bibitem{Ahmed2019Mobihoc}
A.~M. Bedewy, Y.~Sun, S.~Kompella, and N.~B. Shroff, ``Age-optimal sampling and
  transmission scheduling in multi-source systems,'' in \emph{ACM MobiHoc},
  2019, p. 121–130.

\bibitem{Ahmed2020ArXiv}
------, ``Optimal sampling and scheduling for timely status updates in
  multi-source networks,'' 2020, https://arxiv.org/abs/2001.09863.

\bibitem{jog2019channels}
V.~Jog, R.~J. La, and N.~C. Martins, ``Channels, learning, queueing and remote
  estimation systems with a utilization-dependent component,'' 2019, coRR,
  abs/1905.04362.

\bibitem{Tsai2020INFOCOM}
C.-H. Tsai and C.-C. Wang, ``Unifying {AoI} minimization and remote estimation
  — optimal sensor/controller coordination with random two-way delay,'' in
  \emph{Proc. IEEE INFOCOM}, 2020.

\bibitem{Hajek2008}
B.~Hajek, K.~Mitzel, and S.~Yang, ``Paging and registration in cellular
  networks: Jointly optimal policies and an iterative algorithm,'' \emph{IEEE
  Trans.~Inf. Theory}, vol.~54, no.~2, pp. 608--622, Feb 2008.

\bibitem{Nuno2011}
G.~M. Lipsa and N.~C. Martins, ``Remote state estimation with communication
  costs for first-order {LTI} systems,'' \emph{IEEE Trans.~Auto.~Control},
  vol.~56, no.~9, pp. 2013--2025, Sept. 2011.

\bibitem{nayyar2013}
A.~Nayyar, T.~Ba?ar, D.~Teneketzis, and V.~V. Veeravalli, ``Optimal strategies
  for communication and remote estimation with an energy harvesting sensor,''
  \emph{IEEE Trans.~Auto.~Control}, vol.~58, no.~9, pp. 2246--2260, Sept. 2013.

\bibitem{GAO201857}
X.~Gao, E.~Akyol, and T.~Ba\c{s}ar, ``Optimal communication scheduling and
  remote estimation over an additive noise channel,'' \emph{Automatica},
  vol.~88, pp. 57 -- 69, 2018.

\bibitem{ChakravortyTAC2020}
J.~{Chakravorty} and A.~{Mahajan}, ``Remote estimation over a packet-drop
  channel with {Markovian} state,'' \emph{IEEE Trans. Auto. Control}, vol.~65,
  no.~5, pp. 2016--2031, 2020.

\bibitem{Guo2020ISIT}
N.~Guo and V.~Kostina, ``Optimal causal rate-constrained sampling for a class
  of continuous {Markov} processes,'' in \emph{Proc. IEEE Int’l. Symp. Info.
  Theory (ISIT)}, 2020.

\bibitem{AyanCPS2019}
O.~Ayan, M.~Vilgelm, M.~Kl\"{u}gel, S.~Hirche, and W.~Kellerer,
  ``Age-of-information vs. value-of-information scheduling for cellular
  networked control systems,'' in \emph{Proceedings of the 10th ACM/IEEE
  International Conference on Cyber-Physical Systems}, 2019, p. 109–117.

\bibitem{AyanCCNC2020}
O.~{Ayan}, M.~{Vilgelm}, and W.~{Kellerer}, ``Optimal scheduling for discounted
  age penalty minimization in multi-loop networked control,'' in \emph{IEEE
  Annual Consumer Communications Networking Conference (CCNC)}, 2020, pp. 1--7.

\bibitem{SoleymaniCDC2019}
T.~{Soleymani}, J.~S. {Baras}, and K.~H. {Johansson}, ``Stochastic control with
  stale information–part i: Fully observable systems,'' in \emph{IEEE
  Conference on Decision and Control (CDC)}, 2019, pp. 4178--4182.

\bibitem{igorTON18}
I.~Kadota, A.~Sinha, E.~Uysal-Biyikoglu, R.~Singh, and E.~Modiano, ``Scheduling
  policies for minimizing age of information in broadcast wireless networks,''
  \emph{IEEE/ACM Trans. Netw.}, 2018.

\bibitem{talak18_Mobihoc}
R.~Talak, S.~Karaman, and E.~Modiano, ``Optimizing information freshness in
  wireless networks under general interference constraints,'' in \emph{Proc.
  ACM MobiHoc}, Jun. 2018.

\bibitem{Parag-TC-wcnc2017realtime}
P.~Parag, A.~Taghavi, and J.~Chamberland, ``On real-time status updates over
  symbol erasure channels,'' in \emph{2017 IEEE Wireless Communications and
  Networking Conference (WCNC)}, March 2017.

\bibitem{Yates-NSZ-isit2017}
R.~Yates, E.~Najm, E.~Soljanin, and J.~Zhong, ``Timely updates over an erasure
  channel,'' in \emph{Proc. IEEE Int'l. Symp. Info. Theory (ISIT)}, Jun. 2017,
  pp. 316--320.

\bibitem{Sac-BUBD-spawc2018}
H.~Sac, T.~Bacinoglu, E.~Uysal-Biyikoglu, and G.~Durisi, ``Age-optimal channel
  coding blocklength for an {M/G/1} queue with {HARQ},'' in \emph{19th
  International Workshop on Signal Processing Advances in Wireless
  Communications (SPAWC)}, June 2018, pp. 486--490.

\bibitem{2016Ep_WiOpt}
Q.~He, D.~Yuan, and A.~Ephremides, ``Optimizing freshness of information: On
  minimum age link scheduling in wireless systems,'' in \emph{Proc. IEEE/IFIP
  {W}i{O}pt}, May 2016, pp. 1--8.

\bibitem{HeTIT2018}
------, ``Optimal link scheduling for age minimization in wireless systems,''
  \emph{IEEE Trans.\ Inf.\ Theory}, vol.~64, no.~7, pp. 5381--5394, July 2018.

\bibitem{rajat16cdc}
R.~Talak, S.~Karaman, and E.~Modiano, ``Speed limits in autonomous vehicular
  networks due to communication constraints,'' in \emph{Proc. IEEE CDC}, Dec
  2016.

\bibitem{Kadota-UBSM-Allerton2016}
I.~Kadota, E.~Uysal-Biyikoglu, R.~Singh, and E.~Modiano, ``Minimizing the age
  of information in broadcast wireless networks,'' in \emph{54th Annual
  Allerton Conference on Communication, Control, and Computing (Allerton)},
  Sept 2016, pp. 844--851.

\bibitem{2017ISIT_YuPin}
Y.-P. Hsu, E.~Modiano, and L.~Duan, ``Age of information: Design and analysis
  of optimal scheduling algorithms,'' in \emph{Proc. IEEE Int’l. Symp. Info.
  Theory (ISIT)}, Jun. 2017, pp. 1--5.

\bibitem{Jiang-KZN-itc2018}
Z.~{Jiang}, B.~{Krishnamachari}, S.~{Zhou}, and Z.~{Niu}, ``Can decentralized
  status update achieve universally near-optimal age-of-information in wireless
  multiaccess channels?'' in \emph{2018 30th International Teletraffic Congress
  (ITC 30)}, vol.~01, 2018, pp. 144--152.

\bibitem{Jiang-KZZN-isit2018}
Z.~Jiang, B.~Krishnamachari, X.~Zheng, S.~Zhou, and Z.~Niu, ``Decentralized
  status update for age-of-information optimization in wireless multiaccess
  channels,'' in \emph{Proc. IEEE Int'l. Symp. Info. Theory (ISIT)}, June 2018,
  pp. 2276--2280.

\bibitem{Sun-Jiang-KZN-TCOM2020}
J.~Sun, Z.~Jiang, B.~Krishnamachari, S.~Zhou, and Z.~Niu, ``{Closed-form
  whittle's index-enabled random access for timely status update},'' \emph{IEEE
  Transactions on Communications}, vol.~68, no.~3, pp. 1538--1551, 2020.

\bibitem{ZhouGLOBECOM2018}
B.~{Zhou} and W.~{Saad}, ``Optimal sampling and updating for minimizing age of
  information in the internet of things,'' in \emph{Proc.~IEEE GLOBECOM}, Dec
  2018, pp. 1--6.

\bibitem{Kaul-Yates-isit2017}
S.~K. Kaul and R.~Yates, ``Status updates over unreliable multiaccess
  channels,'' in \emph{Proc. IEEE Int'l. Symp. Info. Theory (ISIT)}, Jun. 2017,
  pp. 331--335.

\bibitem{Ali2019ArXivCSMA}
A.~{Maatouk}, M.~{Assaad}, and A.~{Ephremides}, ``On the age of information in
  a {CSMA} environment,'' \emph{IEEE/ACM Trans. Netw.}, vol.~28, no.~2, pp.
  818--831, 2020.

\bibitem{AoI_adhoc}
S.~Leng and A.~Yener, ``Age of information minimization for wireless ad hoc
  networks: A deep reinforcement learning approach,'' in \emph{2019 IEEE Global
  Communications Conference (GLOBECOM)}.\hskip 1em plus 0.5em minus 0.4em\relax
  IEEE, 2019, pp. 1--6.

\bibitem{yates2020age}
\BIBentryALTinterwordspacing
R.~D. Yates and S.~K. Kaul, ``Age of information in uncoordinated unslotted
  updating,'' 2020. [Online]. Available: \url{https://arxiv.org/abs/2002.02026}
\BIBentrySTDinterwordspacing

\bibitem{Chen-Gatsis-2019a}
\BIBentryALTinterwordspacing
X.~Chen, K.~Gatsis, H.~Hassani, and S.~S. Bidokhti, ``{Age of Information in
  Random Access Channels},'' 2019. [Online]. Available:
  \url{http://arxiv.org/abs/1912.01473}
\BIBentrySTDinterwordspacing

\bibitem{Jiang-KZZN-iot2019}
Z.~{Jiang}, B.~{Krishnamachari}, X.~{Zheng}, S.~{Zhou}, and Z.~{Niu}, ``Timely
  status update in wireless uplinks: Analytical solutions with asymptotic
  optimality,'' \emph{IEEE Internet of Things Journal}, vol.~6, no.~2, pp.
  3885--3898, 2019.

\bibitem{Chen-Yifan-2020a}
\BIBentryALTinterwordspacing
H.~Chen, Y.~Gu, and S.-C. Liew, ``{Age-of-Information Dependent Random Access
  for Massive IoT Networks},'' jan 2020. [Online]. Available:
  \url{http://arxiv.org/abs/2001.04780}
\BIBentrySTDinterwordspacing

\bibitem{Bedewy2020MobiHoc}
A.~M. Bedewy, Y.~Sun, R.~Singh, and N.~B. Shroff, ``Optimizing information
  freshness using low-power status updates via sleep-wake scheduling,'' in
  \emph{Proc. ACM MobiHoc}, 2020.

\bibitem{BinLi18}
N.~Lu, B.~Ji, and B.~Li, ``Age-based scheduling: Improving data freshness for
  wireless real-time traffic,'' in \emph{Proc. ACM MobiHoc}, 2018.

\bibitem{igorINFOCOM}
I.~Kadota, A.~Sinha, and E.~Modiano, ``Optimizing age of information in
  wireless networks with throughput constraints,'' in \emph{Proc. IEEE
  INFOCOM}, 2018.

\bibitem{igorMobiHoc}
I.~Kadota and E.~Modiano, ``Minimizing the age of information in wireless
  networks with stochastic arrivals,'' in \emph{Proc. ACM MobiHoc}, 2019.

\bibitem{JooTON2018}
C.~Joo and A.~Eryilmaz, ``Wireless scheduling for information freshness and
  synchrony: Drift-based design and heavy-traffic analysis,'' \emph{IEEE/ACM
  Trans. Netw.}, vol.~26, no.~6, pp. 2556--2568, Dec 2018.

\bibitem{Talak-KM-allerton2017}
R.~Talak, S.~Karaman, and E.~Modiano, ``Minimizing age-of-information in
  multi-hop wireless networks,'' in \emph{55th Annual Allerton Conference on
  Communication, Control, and Computing}, Oct 2017, pp. 486--493.

\bibitem{talak18_greece}
------, ``Distributed scheduling algorithms for optimizing information
  freshness in wireless networks,'' in \emph{Proc. {SPAWC} (arXiv:1803.06469)},
  Jun. 2018.

\bibitem{talak18_WiOpt}
------, ``Optimizing age of information in wireless networks with perfect
  channel state information,'' in \emph{Proc. IEEE/IFIP {WiOpt}}, May 2018.

\bibitem{talak18_ISIT}
R.~Talak, I.~Kadota, S.~Karaman, and E.~Modiano, ``Scheduling policies for age
  minimization in wireless networks with unknown channel state,'' in
  \emph{Proc. IEEE Int’l. Symp. Info. Theory (ISIT)}, Jun. 2018.

\bibitem{farazi2018SPAWC}
S.~Farazi, A.~G. Klein, J.~A. McNeill, and D.~R. Brown, ``On the age of
  information in multi-source multi-hop wireless status update networks,'' in
  \emph{2018 IEEE 19th International Workshop on Signal Processing Advances in
  Wireless Communications (SPAWC)}.\hskip 1em plus 0.5em minus 0.4em\relax
  IEEE, 2018, pp. 1--5.

\bibitem{Buyukates18c}
B.~Buyukates, A.~Soysal, and S.~Ulukus, ``Age of information scaling in large
  networks,'' in \emph{IEEE ICC}, May 2019.

\bibitem{Buyukates19b}
------, ``Age of information scaling in large networks with hierarchical
  cooperation,'' in \emph{IEEE Globecom}, December 2019.

\bibitem{Buyukates20d}
------, ``Scaling laws for age of information in wireless networks,'' November
  2019.

\bibitem{farazi2019JCN}
S.~Farazi, A.~G. Klein, and D.~R. Brown~III, ``Fundamental bounds on the age of
  information in multi-hop global status update networks,'' \emph{Journal of
  Communications and Networks}, vol.~21, no.~3, pp. 268--279, 2019.

\bibitem{Ceran-2019TWC}
E.~T. Ceran, D.~G{\"u}nd{\"u}z, and A.~Gy{\"o}rgy, ``Average age of information
  with hybrid arq under a resource constraint,'' \emph{IEEE Transactions on
  Wireless Communications}, vol.~18, no.~3, pp. 1900--1913, 2019.

\bibitem{RMAB}
P.~Whittle, ``Restless bandits: Activity allocation in a changing world,''
  \emph{Journal of Applied Probability}, vol.~25, pp. 287--298, 1988.

\bibitem{index_regularity}
R.~Singh, X.~Guo, and P.~Kumar, ``Index policies for optimal mean-variance
  trade-off of inter-delivery times in real-time sensor networks,'' in
  \emph{Proc. IEEE INFOCOM}, Jan. 2015, pp. 505--512.

\bibitem{index_schedule}
V.~Raghunathan, V.~Borkar, M.~Cao, and P.~R. Kumar, ``Index policies for
  real-time multicast scheduling for wireless broadcast systems,'' in
  \emph{Proc. IEEE INFOCOM}, Apr. 2008.

\bibitem{index_myopic}
P.~Mansourifard, T.~Javidi, and B.~Krishnamachari, ``Optimality of myopic
  policy for a class of monotone affine restless multi-armed bandits,'' in
  \emph{Proc. IEEE CDC}, Dec. 2012, pp. 877--882.

\bibitem{index_multichannelaccess}
K.~Liu and Q.~Zhao, ``Indexability of restless bandit problems and optimality
  of {Whittle} index for dynamic multichannel access,'' \emph{IEEE Trans. Inf.
  Theory}, vol.~56, pp. 5547--5567, Nov. 2010.

\bibitem{index_assympt}
R.~R. Weber and G.~Weiss, ``On an index policy for restless bandits,''
  \emph{Journal of Applied Probability}, vol.~27, no.~3, pp. 637--648, 1990.

\bibitem{RMAB_book}
J.~Gittins, K.~Glazebrook, and R.~Weber, \emph{Multi-armed Bandit Allocation
  Indices}, 2nd~ed.\hskip 1em plus 0.5em minus 0.4em\relax Wiley, Mar. 2011.

\bibitem{Boyd04}
S.~Boyd and L.~Vandenberghe, \emph{Convex Optimization}.\hskip 1em plus 0.5em
  minus 0.4em\relax Cambridge, U.K.: Cambridge Univerisity Press, 2004.

\bibitem{comb_opt_book}
B.~Korte and J.~Vygen, \emph{Combinatorial Optimization: Theory and
  Algorithms}, 4th~ed.\hskip 1em plus 0.5em minus 0.4em\relax Springer
  Publishing Company, Incorporated, 2007.

\bibitem{Garber16_FrankWolfe}
D.~Garber and E.~Hazan, ``A linearly convergent variant of the conditional
  gradient algorithm under strong convexity, with applications to online and
  stochastic optimization,'' \emph{SIAM J. on Opt.}, vol.~26, no.~3, pp.
  1493--1528, 2016.

\bibitem{Hajek_Sasaki_OptScheduling}
B.~Hajek and G.~Sasaki, ``Link scheduling in polynomial time,'' \emph{IEEE
  Trans. Inf. Theory}, vol.~34, no.~5, pp. 910--917, Sep. 1988.

\bibitem{Farazi-KB-icassp2016}
S.~Farazi, A.~G. Klein, and D.~R. Brown, ``On the average staleness of global
  channel state information in wireless networks with random transmit node
  selection,'' in \emph{2016 IEEE International Conference on Acoustics, Speech
  and Signal Processing (ICASSP)}, March 2016, pp. 3621--3625.

\bibitem{Farazi-KB-icccn2017}
------, ``Bounds on the age of information for global channel state
  dissemination in fully-connected networks,'' in \emph{Int'l Conference on
  Computer Communication and Networks (ICCCN)}, July 2017.

\bibitem{Klein-FHB-TW2017}
A.~G. Klein, S.~Farazi, W.~He, and D.~R. Brown, ``Staleness bounds and
  efficient protocols for dissemination of global channel state information,''
  \emph{IEEE Transactions on Wireless Communications}, vol.~16, no.~9, pp.
  5732--5746, Sept 2017.

\bibitem{Selen-NAV-springer2013}
J.~Selen, H.~N. Yu, L.~L. Andrew, and H.~L. Vu, ``The age of information in
  gossip networks,'' in \emph{Analytical and Stochastic Modeling Techniques and
  Applications}.\hskip 1em plus 0.5em minus 0.4em\relax Berlin, Heidelberg:
  Springer, 2013, pp. 364--379.

\bibitem{asil2016farazi}
S.~Farazi, {D.\ R.\ Brown III}, and A.~G. Klein, ``On global channel state
  estimation and dissemination in ring networks,'' in \emph{Proc.\ Asilomar
  Conf.\ on Signals, Systems, and Computers}, Nov. 2016, pp. 1122--1127.

\bibitem{yates2018infocom}
R.~D. Yates, ``Age of information in a network of preemptive servers,'' in
  \emph{Proc. \ IEEE Intl. Conf.\ on Computer Comm.\ Workshops ({INFOCOM
  WKSHPS})}, Apr. 2018, pp. 118--123.

\bibitem{Banik-KS-ISIT2019}
\BIBentryALTinterwordspacing
S.~Banik, S.~K. Kaul, and P.~B. Sujit, ``{Minimizing Age in Gateway Based
  Update Systems},'' in \emph{IEEE International Symposium on Information
  Theory - Proceedings}, vol. 2019-July.\hskip 1em plus 0.5em minus 0.4em\relax
  Institute of Electrical and Electronics Engineers Inc., jul 2019, pp.
  1032--1036. [Online]. Available: \url{http://arxiv.org/abs/1903.07963}
\BIBentrySTDinterwordspacing

\bibitem{farazi2019INFOCOM}
S.~Farazi, A.~G. Klein, and D.~R. Brown, ``Fundamental bounds on the age of
  information in general multi-hop interference networks,'' in \emph{IEEE
  INFOCOM 2019-IEEE Conference on Computer Communications Workshops (INFOCOM
  WKSHPS)}.\hskip 1em plus 0.5em minus 0.4em\relax IEEE, 2019, pp. 96--101.

\bibitem{farazi2019ASILOMAR1}
------, ``Age of information with unreliable transmissions in multi-source
  multi-hop status update systems,'' in \emph{2019 53rd Asilomar Conference on
  Signals, Systems, and Computers}.\hskip 1em plus 0.5em minus 0.4em\relax
  IEEE, 2019, pp. 2017--2021.

\bibitem{farazi2019ASILOMAR2}
------, ``Average age of information in multi-source self-preemptive status
  update systems with packet delivery errors,'' in \emph{2019 53rd Asilomar
  Conference on Signals, Systems, and Computers}.\hskip 1em plus 0.5em minus
  0.4em\relax IEEE, 2019, pp. 396--400.

\bibitem{Shreedhar2019AnInternet-of-things}
T.~Shreedhar, S.~K. Kaul, and R.~D. Yates, ``{An age control transport protocol
  for delivering fresh updates in the internet-of-things},'' \emph{20th IEEE
  International Symposium on A World of Wireless, Mobile and Multimedia
  Networks, WoWMoM 2019}, 2019.

\bibitem{ayan2020NL}
O.~Ayan, H.~M. G{\"u}rsu, A.~Papa, and W.~Kellerer, ``Probability analysis of
  age of information in multi-hop networks,'' \emph{IEEE Networking Letters},
  vol.~2, no.~2, pp. 76--80, 2020.

\bibitem{Costa-VE-icc2015}
M.~Costa, S.~Valentin, and A.~Ephremides, ``On the age of channel information
  for a finite-state {Markov} model,'' in \emph{2015 IEEE International
  Conference on Communications (ICC)}, June 2015, pp. 4101--4106.

\bibitem{Sang-LJ-globecom2017}
Y.~Sang, B.~Li, and B.~Ji, ``The power of waiting for more than one response in
  minimizing the age-of-information,'' in \emph{IEEE Global Communications
  Conference (GLOBECOM)}, Dec 2017.

\bibitem{Zhong-YS-aoi2018}
J.~Zhong, R.~Yates, and E.~Soljanin, ``Minimizing content staleness in
  dynamo-style replicated storage systems,'' in \emph{Infocom Workshop on Age
  of Information}, Apr. 2018, arXiv preprint arXiv:1804.00742.

\bibitem{Zhong-YS-allerton2017}
------, ``Status updates through multicast networks,'' in \emph{Proc. Allerton
  Conf. on Commun., Control and Computing}, Oct. 2017, pp. 463--469.

\bibitem{Zhong-YS-spawc2018}
------, ``Multicast with prioritized delivery: How fresh is your data?'' in
  \emph{{Signal Processing Advance for Wireless Communications (SPAWC)}}, Jun.
  2018, pp. 476--480.

\bibitem{Buyukates18}
B.~Buyukates, A.~Soysal, and S.~Ulukus, ``Age of information in two-hop
  multicast networks,'' in \emph{Asilomar Conference}, October 2018.

\bibitem{Buyukates19}
------, ``Age of information in multicast networks with multiple update
  streams,'' in \emph{Asilomar Conference}, November 2019.

\bibitem{Buyukates18b}
------, ``Age of information in multihop multicast networks,'' \emph{Journal of
  Communications and Networks}, vol.~21, no.~3, pp. 256--267, July 2019.

\bibitem{Zhong-Yates-dcc2016lossless}
J.~Zhong and R.~D. Yates, ``Timeliness in lossless block coding,'' in
  \emph{2016 Data Compression Conference (DCC)}, March 2016, pp. 339--348.

\bibitem{Zhong-YS-isit2017}
J.~Zhong, R.~Yates, and E.~Soljanin, ``Backlog-adaptive compression: Age of
  information,'' in \emph{Proc. IEEE Int'l. Symp. Info. Theory (ISIT)}, Jun.
  2017, pp. 566--570.

\bibitem{Mayekar-PT-isit2018lossless}
P.~Mayekar, P.~Parag, and H.~Tyagi, ``Optimal lossless source codes for timely
  updates,'' in \emph{Proc. IEEE Int'l. Symp. Info. Theory (ISIT)}, Jun. 2018,
  pp. 1246--1250.

\bibitem{MelihBatu1}
M.~Bastopcu, B.~Buyukates, and S.~Ulukus, ``Optimal selective encoding for
  timely updates,'' in \emph{CISS}, March 2020.

\bibitem{MelihBatu2}
B.~Buyukates, M.~Bastopcu, and S.~Ulukus, ``Optimal selective encoding for
  timely updates with empty symbol,'' in \emph{IEEE ISIT}, June 2020.

\bibitem{MelihBatu4}
M.~Bastopcu, B.~Buyukates, and S.~Ulukus, ``Selective encoding policies for
  maximizing information freshness,'' April 2020, available on
  arXiv:2004.06091.

\bibitem{Bastopcu20}
M.~Bastopcu and S.~Ulukus, ``Partial updates: Losing information for
  freshness,'' in \emph{IEEE ISIT}, June 2020.

\bibitem{Bhambay-PP-wcnc2017}
S.~Bhambay, S.~Poojary, and P.~Parag, ``Differential encoding for real-time
  status updates,'' in \emph{2017 IEEE Wireless Communications and Networking
  Conference (WCNC)}, March 2017.

\bibitem{HeFD-aoi2018}
Q.~He, G.~Dan, and V.~Fodor, ``Minimizing age of correlated information for
  wireless camera networks,'' in \emph{IEEE Conference on Computer
  Communications (INFOCOM) Workshops}, April 2018, pp. 547--552.

\bibitem{Hribar-CKD-globecom2017}
J.~Hribar, M.~Costa, N.~Kaminski, and L.~A. DaSilva, ``Updating strategies in
  the internet of things by taking advantage of correlated sources,'' in
  \emph{IEEE Global Communications Conference (GLOBECOM)}, Dec 2017.

\bibitem{Yates-THR-infocom2017}
R.~Yates, M.~Tavan, Y.~Hu, and D.~Raychaudhuri, ``Timely cloud gaming,'' in
  \emph{Proc. INFOCOM}, May 2017, pp. 1--9.

\bibitem{Bastopcu19}
M.~Bastopcu and S.~Ulukus, ``Age of information for updates with distortion,''
  in \emph{IEEE ITW}, August 2019.

\bibitem{Bastopcu20b}
------, ``Age of information for updates with distortion: Constant and
  age-dependent distortion constraints,'' December 2019, available on
  arXiv:1912.13493.

\bibitem{Buyukates19c}
B.~Buyukates and S.~Ulukus, ``Timely distributed computation with stragglers,''
  \emph{IEEE Transactions on Communications}, to appear. Also available on
  arXiv: 1910.03564.

\bibitem{Nguyen-KKWE-wiopt2017}
G.~D. Nguyen, S.~Kompella, C.~Kam, J.~E. Wieselthier, and A.~Ephremides,
  ``Impact of hostile interference on information freshness: A game approach,''
  in \emph{Int'l Symposium on Modeling and Optimization in Mobile, Ad Hoc, and
  Wireless Networks (WiOpt)}, May 2017.

\bibitem{Xiao-Sun-aoi2018}
Y.~Xiao and Y.~Sun, ``A dynamic jamming game for real-time status updates,'' in
  \emph{IEEE Conference on Computer Communications (INFOCOM) Workshops}, April
  2018, pp. 354--360.

\bibitem{garnaev2019}
A.~Garnaev, W.~Zhang, J.~Zhong, and R.~D. Yates, ``Maintaining information
  freshness under jamming,'' in \emph{IEEE INFOCOM 2019-IEEE Conference on
  Computer Communications Workshops (INFOCOM WKSHPS)}.\hskip 1em plus 0.5em
  minus 0.4em\relax IEEE, 2019, pp. 90--95.

\bibitem{gac2018}
X.~Gac, E.~Akyol, and T.~Ba\c{s}ar, ``On communication scheduling and remote
  estimation in the presence of an adversary as a nonzero-sum game,'' in
  \emph{2018 IEEE Conference on Decision and Control (CDC)}.\hskip 1em plus
  0.5em minus 0.4em\relax IEEE, 2018, pp. 2710--2715.

\bibitem{nguyen2018}
G.~D. Nguyen, S.~Kompella, C.~Kam, J.~E. Wieselthier, and A.~Ephremides,
  ``Information freshness over an interference channel: A game theoretic
  view,'' in \emph{IEEE INFOCOM 2018-IEEE Conference on Computer
  Communications}.\hskip 1em plus 0.5em minus 0.4em\relax IEEE, 2018, pp.
  908--916.

\bibitem{Gopal-Kaul-aoi2018}
S.~Gopal and S.~K. Kaul, ``A game theoretic approach to {DSRC} and {WiFi}
  coexistence,'' in \emph{IEEE Conference on Computer Communications (INFOCOM)
  Workshops}, April 2018, pp. 565--570.

\bibitem{SGAoI2019}
S.~{Gopal}, S.~K. {Kaul}, and R.~{Chaturvedi}, ``Coexistence of age and
  throughput optimizing networks: A game theoretic approach,'' in \emph{2019
  IEEE 30th Annual International Symposium on Personal, Indoor and Mobile Radio
  Communications (PIMRC)}, Sep. 2019, pp. 1--6.

\bibitem{SGAoI2020}
S.~Gopal, S.~K. Kaul, R.~Chaturvedi, and S.~Roy, ``Coexistence of age and
  throughput optimizing networks: A spectrum sharing game,'' \emph{arXiv
  preprint arXiv:1909.02863}, 2019.

\bibitem{SGAoI2020Non}
------, ``{A Non-Cooperative Multiple Access Game for Timely Updates},'' in
  \emph{INFOCOM 2020 - IEEE Conference on Computer Communications Workshopss
  (INFOCOM WKSHPS)}, 2020.

\bibitem{Kumar-Vaze-2020}
K.~Saurav and R.~Vaze, ``{Game of Ages},'' in \emph{INFOCOM 2020 - IEEE
  Conference on Computer Communications Workshopss (INFOCOM WKSHPS)}, 2020.

\bibitem{zheng2019}
H.~Zheng, K.~Xiong, P.~Fan, Z.~Zhong, and K.~B. Letaief, ``Age-based utility
  maximization for wireless powered networks: A stackelberg game approach,'' in
  \emph{2019 IEEE Global Communications Conference (GLOBECOM)}.\hskip 1em plus
  0.5em minus 0.4em\relax IEEE, 2019, pp. 1--6.

\bibitem{ning2020}
Z.~Ning, P.~Dong, X.~Wang, X.~Hu, L.~Guo, B.~Hu, Y.~Guo, T.~Qiu, and R.~Kwok,
  ``Mobile edge computing enabled 5g health monitoring for internet of medical
  things: A decentralized game theoretic approach,'' \emph{IEEE J. Sel. Areas
  Commun.}, pp. 1--16, 2020.

\bibitem{Bastopcu20a}
M.~Bastopcu and S.~Ulukus, ``Who should {Google} {Scholar} update more often?''
  in \emph{IEEE Infocom}, July 2020.

\bibitem{ceran2018reinforcement}
E.~T. Ceran, D.~G{\"u}nd{\"u}z, and A.~Gy{\"o}rgy, ``A reinforcement learning
  approach to age of information in multi-user networks,'' in \emph{2018 IEEE
  29th Annual International Symposium on Personal, Indoor and Mobile Radio
  Communications (PIMRC)}.\hskip 1em plus 0.5em minus 0.4em\relax IEEE, 2018,
  pp. 1967--1971.

\bibitem{yin2020application}
B.~Yin, S.~Zhang, and Y.~Cheng, ``Application-oriented scheduling for
  optimizing the age of correlated information: A deep reinforcement learning
  based approach,'' \emph{IEEE Internet of Things Journal}, 2020.

\bibitem{elgabli2019reinforcement}
A.~Elgabli, H.~Khan, M.~Krouka, and M.~Bennis, ``Reinforcement learning based
  scheduling algorithm for optimizing age of information in ultra reliable low
  latency networks,'' in \emph{2019 IEEE Symposium on Computers and
  Communications (ISCC)}.\hskip 1em plus 0.5em minus 0.4em\relax IEEE, 2019,
  pp. 1--6.

\bibitem{age_multiple_flows}
H.~B. Beytur and E.~Uysal, ``Age minimization of multiple flows using
  reinforcement learning,'' in \emph{2019 International Conference on
  Computing, Networking and Communications (ICNC)}.\hskip 1em plus 0.5em minus
  0.4em\relax IEEE, 2019, pp. 339--343.

\bibitem{abedin2020data}
S.~F. Abedin, M.~Munir, N.~H. Tran, Z.~Han, C.~S. Hong \emph{et~al.}, ``Data
  freshness and energy-efficient uav navigation optimization: A deep
  reinforcement learning approach,'' \emph{arXiv preprint arXiv:2003.04816},
  2020.

\bibitem{yi2020deep}
M.~Yi, X.~Wang, J.~Liu, Y.~Zhang, and B.~Bai, ``Deep reinforcement learning for
  fresh data collection in uav-assisted iot networks,'' \emph{arXiv preprint
  arXiv:2003.00391}, 2020.

\bibitem{Elmagid2019Deep}
M.~A. Abd-Elmagid, A.~Ferdowsi, H.~S. Dhillon, and W.~Saad, ``Deep
  reinforcement learning for minimizing age-of-information in uav-assisted
  networks,'' \emph{arXiv preprint arXiv:1905.02993}, 2019.

\bibitem{Buyukates20c}
E.~Ozfatura, B.~Buyukates, D.~Gunduz, and S.~Ulukus, ``Age-based coded
  computation for bias reduction in distributed learning,'' June 2020,
  available on arXiv: 2006.01816.

\bibitem{sutton2018reinforcement}
R.~S. Sutton and A.~G. Barto, \emph{Reinforcement learning: An
  introduction}.\hskip 1em plus 0.5em minus 0.4em\relax MIT press, 2018.

\bibitem{sutton1999between}
R.~S. Sutton, D.~Precup, and S.~Singh, ``Between mdps and semi-mdps: A
  framework for temporal abstraction in reinforcement learning,''
  \emph{Artificial intelligence}, vol. 112, no. 1-2, pp. 181--211, 1999.

\bibitem{Gao2012}
W.~Gao, G.~Cao, M.~Srivatsa, and A.~Iyengar, ``{Distributed maintenance of
  cache freshness in opportunistic mobile networks},'' \emph{Proceedings -
  International Conference on Distributed Computing Systems}, pp. 132--141,
  2012.

\bibitem{Yates-CWY-isit2017cache}
R.~Yates, P.~Ciblat, M.~Wigger, and A.~Yener, ``Age-optimal constrained cache
  updating,'' in \emph{Proc. IEEE Int'l. Symp. Info. Theory (ISIT)}, Jun. 2017,
  pp. 141--145.

\bibitem{Kam2017}
C.~Kam, S.~Kompella, G.~D. Nguyen, J.~E. Wieselthier, and A.~Ephremides,
  ``{Information freshness and popularity in mobile caching},'' in \emph{IEEE
  International Symposium on Information Theory - Proceedings}, 2017.

\bibitem{Tang2019}
\BIBentryALTinterwordspacing
H.~Tang, P.~Ciblat, J.~Wang, M.~Wigger, and R.~Yates, ``{Age of Information
  Aware Cache Updating with File- and Age-Dependent Update Durations},'' pp.
  1--6, 2019. [Online]. Available: \url{http://arxiv.org/abs/1909.05930}
\BIBentrySTDinterwordspacing

\bibitem{Zhang2018}
S.~Zhang, J.~Li, H.~Luo, J.~Gao, L.~Zhao, and X.~S. Shen, ``{Towards Fresh and
  Low-Latency Content Delivery in Vehicular Networks: An Edge Caching
  Aspect},'' \emph{2018 10th International Conference on Wireless
  Communications and Signal Processing, WCSP 2018}, pp. 1--6, 2018.

\bibitem{Yang2019}
\BIBentryALTinterwordspacing
L.~Yang, Y.~Zhong, F.-C. Zheng, and S.~Jin, ``{Edge Caching with Real-Time
  Guarantees},'' 2019. [Online]. Available:
  \url{http://arxiv.org/abs/1912.11847}
\BIBentrySTDinterwordspacing

\bibitem{Bastopcu20e}
M.~Bastopcu and S.~Ulukus, ``Information freshness in cache updating systems,''
  April 2020, available on arXiv:2004.09475.

\bibitem{Bastopcu20c}
------, ``Information freshness in cache updating systems with limited cache
  storage capacity,'' May 2020, available on arXiv:2005.10683.

\bibitem{Zhang2020}
\BIBentryALTinterwordspacing
S.~Zhang, L.~Wang, H.~Luo, X.~Ma, and S.~Zhou, ``{AoI-Delay Tradeoff in Mobile
  Edge Caching with Freshness-Aware Content Refreshing},'' vol. 100191, pp.
  1--28, 2020. [Online]. Available: \url{http://arxiv.org/abs/2002.05868}
\BIBentrySTDinterwordspacing

\bibitem{AtakanStalenessMLModels2020}
A.~Aral, M.~Erol-Kantarci, and I.~Brandi\'{c}, ``Staleness control for edge
  data analytics,'' \emph{Proc. ACM Meas. Anal. Comput. Syst.}, vol.~4, no.~2,
  Jun. 2020.

\bibitem{Sonmez2018Age-of-InformationLTE}
C.~S{\"{o}}nmez, S.~Baghaee, A.~Ergi{\c{s}}i, and E.~Uysal-Biyikoglu,
  ``{Age-of-Information in Practice: Status Age Measured Over TCP/IP
  Connections Through WiFi, Ethernet and LTE},'' \emph{2018 IEEE International
  Black Sea Conference on Communications and Networking, BlackSeaCom 2018}, pp.
  1--5, 2018.

\bibitem{Beytur2019MeasuringConnections}
H.~B. Beytur, S.~Baghaee, and E.~Uysal, ``{Measuring age of information on
  real-life connections},'' \emph{27th Signal Processing and Communications
  Applications Conference, SIU 2019}, 2019.

\bibitem{Beytur2020TowardsSystems}
------, ``{Towards AoI-aware Smart IoT Systems},'' \emph{2020 International
  Conference on Computing, Networking and Communications, ICNC 2020}, pp.
  353--357, 2020.

\bibitem{Shreedhar2018PosterInternet}
T.~Shreedhar, S.~K. Kaul, and R.~D. Yates, ``Poster: {ACP}: {Age Control
  Protocol} for minimizing age of information over the internet,'' in
  \emph{Proceedings of the 24th Annual International Conference on Mobile
  Computing and Networking}, ser. MobiCom ’18.\hskip 1em plus 0.5em minus
  0.4em\relax ACM, 2018, p. 699–701.

\bibitem{shreedhar2018acpArXiv}
------, ``{ACP}: An end-to-end transport protocol for delivering fresh updates
  in the internet-of-things,'' \emph{arXiv preprint arXiv:1811.03353}, 2018.

\end{thebibliography}
\end{singlespace}
\end{document}